\documentclass[
 reprint,
 superscriptaddress,
 nofootinbib,
 amsmath,amssymb,
 aps,
]{revtex4-2}

\usepackage{graphicx}
\usepackage{dcolumn}
\usepackage{bm}
\usepackage{mathtools}
\usepackage{mathrsfs}
\usepackage{siunitx}

\makeatletter
\let\MYcaption\@makecaption
\makeatother

\usepackage{subcaption}

\makeatletter
\let\@makecaption\MYcaption
\makeatother

\begin{document}

\title{Finding Optimal Pathways in Chemical Reaction Networks Using Ising Machines}

\author{Yuta Mizuno}
  \email{mizuno@es.hokudai.ac.jp}
  \affiliation{Research Institute for Electronic Science, Hokkaido University,
               Sapporo, Hokkaido 001-0020, Japan}
  \affiliation{Institute for Chemical Reaction Design and Discovery
                 (WPI-ICReDD), Hokkaido University, Sapporo, Hokkaido 001-0021, Japan}
  \affiliation{Graduate School of Chemical Sciences and Engineering,
               Hokkaido University, Sapporo, Hokkaido 060-8628, Japan}
\author{Tamiki Komatsuzaki}
  \affiliation{Research Institute for Electronic Science, Hokkaido University,
               Sapporo, Hokkaido 001-0020, Japan}
  \affiliation{Institute for Chemical Reaction Design and Discovery
                 (WPI-ICReDD), Hokkaido University, Sapporo, Hokkaido 001-0021, Japan}
  \affiliation{Graduate School of Chemical Sciences and Engineering,
               Hokkaido University, Sapporo, Hokkaido 060-8628, Japan}
  \affiliation{The Institute of Scientific and Industrial Research,
               Osaka University, Ibaraki, Osaka 567-0047, Japan}

\date{\today}

\begin{abstract}
  Finding optimal pathways in chemical reaction networks is essential for
  elucidating and designing chemical processes, with significant applications
  such as synthesis planning and metabolic pathway analysis. Such a chemical
  pathway-finding problem can be formulated as a constrained combinatorial
  optimization problem, aiming to find an optimal combination of chemical
  reactions connecting starting materials to target materials in a given
  network. Due to combinatorial explosion, the computation time required
  to find an optimal pathway increases exponentially with the network size.
  Ising machines, including quantum and simulated annealing devices, are
  promising novel computers dedicated to such hard combinatorial optimization.
  However, to the best of our knowledge, there has yet to be an attempt to
  apply Ising machines to chemical pathway-finding problems. In this article,
  we present the first Ising/quantum computing application for chemical
  pathway-finding problems. The Ising model, translated from a chemical
  pathway-finding problem, involves several types of penalty terms for
  violating constraints. It is not obvious how to set appropriate penalty
  strengths of different types. To address this challenge, we employ Bayesian
  optimization for parameter tuning. Furthermore, we introduce a novel technique
  that enhances tuning performance by grouping penalty terms according to
  the underlying problem structure. The performance evaluation and analysis
  of the proposed algorithm were conducted using a D-Wave Advantage system
  and simulated annealing. The benchmark results reveal challenges in finding
  exact optimal pathways. Concurrently, the results indicate the feasibility
  of finding approximate optimal pathways, provided that a certain degree of
  relative error in cost value is acceptable.
\end{abstract}

\maketitle

\section{Introduction} \label{sec:introduction}

Since the inception of modern chemistry in the 18th century,
scientists have designed and discovered numerous chemical reactions.
These discoveries are recorded in chemical databases such as
Reaxys \cite{Reaxys} and SciFinder\textsuperscript{n} \cite{SciFinder}. 
The vast collections of chemical reactions stored in these databases are
big data in chemistry and have experienced rapid growth. For instance,
Reaxys contains more than 60 million chemical reactions as of 2023.
The number of chemical reactions added to Reaxys each year is more than
hundreds of thousands, which has more than doubled between 2000 and 2015
\cite{Jacob2017}. Additionally, emerging technologies for automatic reaction
exploration, such as chemical synthesis robots \cite{Granda2018, Coley2019},
artificial intelligence and machine learning \cite{Segler2018, Tsuji2023},
and quantum chemical calculation \cite{Maeda2021}, will further accelerate
the growth of chemical databases.

A chemical reaction database can be represented by a gigantic chemical
reaction network. A chemical reaction network (CRN) is a graph-theoretical
representation of a collection of chemical reactions, modeled as a directed
bipartite graph with two types of nodes: chemical reactions and species
\cite{Temkin1996, Szymkuc2016}. This representation is used in chemistry and
systems biology to visualize, elucidate, and design the mechanisms of
complex chemical processes.

Finding optimal pathways in CRNs is a static analysis of CRNs,
which has significant applications such as synthesis planning
\cite{Szymkuc2016, Kowalik2012, Shibukawa2020, Gao2021} and metabolic pathway
analysis \cite{Andersen2012, Andersen2017}. Such a chemical pathway-finding
problem can be formulated as a combinatorial optimization problem to find
an optimal combination of chemical reactions connecting starting materials
to target materials in a given CRN. In chemical synthesis, chemists aim to find
synthesis pathways from readily available materials to target compound(s).
Moreover, they seek synthesis pathways with better properties, such as
low monetary costs, short execution times, low risks, and eco-friendliness.
In systems biology, identifying metabolic pathways is fundamental to
characterizing the function and mechanism of a living system. Furthermore,
metabolic engineers aim to design artificial metabolic pathways to maximize
the target metabolite production of cells. There are several algorithms for
these pathway-finding problems, including recursive network search algorithms
\cite{Kowalik2012, Shibukawa2020}, optimization/enumeration algorithms using
a mixed integer programming solver \cite{Gao2021, Andersen2017}, and
a simulated annealing algorithm designed for synthesis planning \cite{Kowalik2012}.

However, finding optimal pathways in CRNs is NP-hard in general
\cite{Andersen2012}; the computation time to find an optimal solution
increases exponentially as the network size increases due to combinatorial
explosion. Consequently, the continuous improvement of software and hardware
for pathway finding in CRNs is crucial to address the rapidly increasing
computational demands associated with growing data size in chemistry.

In recent years, Ising machines have gained considerable attention as
next-generation hardware devices dedicated to combinatorial optimization
\cite{Mohseni2022}. Ising machines are specially designed to find the ground
state(s) of an Ising model characterized by the Hamiltonian
\begin{equation}
H_\mathrm{Ising}(\bm{\sigma})
= -\sum_{i=1}^{N_\sigma-1} \sum_{j=i+1}^{N_\sigma} J_{ij} \sigma_i \sigma_j
  -\sum_{i=1}^{N_\sigma} h_i \sigma_i.
\label{eq:ising-model}
\end{equation}
Here, $\sigma_i \in \{-1, +1\}$, $\bm{\sigma}$, $N_\sigma$, $J_{ij}$, 
and $h_i$ denote, respectively, a spin variable, the spin configuration
$(\sigma_1, \sigma_2, \dots, \sigma_{N_\sigma})$, the number of spins,
the interaction coefficient between two spins $\sigma_i$ and $\sigma_j$,
and the local field interacting with $\sigma_i$. Ising machines include
quantum annealing devices developed by D-Wave systems
\cite{Johnson2011, McGeoch2022} and classical simulated annealing devices
\cite{Yamaoka2015, Okuyama2019, Matsubara2020, Yamamoto2020, Kawamura2023}.
Many combinatorial optimization problems are efficiently reducible to
the ground state finding problems of Ising models \cite{Lucas2014, Yarkoni2022}.
If chemical pathway-finding problems are also reducible to Ising model
problems, these physics-inspired, novel computers may potentially offer
a solution to the challenges posed by the accelerating growth in chemical
data and the concurrent slowdown in performance growth for conventional
von-Neumann computers due to the end of Moore's law \cite{Patterson2020}.
However, to the best of our knowledge, there has yet to be an attempt
to apply Ising machines to chemical pathway-finding problems.

Pathway-finding problems in CRNs, as potential application targets
of Ising machines and quantum annealing, exhibit the following notable
features: (1) The available data size is enormous and increasing rapidly.
(2) Various problem settings exist, including multi-objective and
mixed integer programming formulation of synthesis planning \cite{Gao2021}
and exhaustive enumeration of possible metabolic pathways \cite{Andersen2017}.
(3) The bipartite graph representation of a CRN can be viewed as
a Petri net \cite{Chaouiya2007}, a graphical, mathematical modeling tool
for discrete event systems. Hence, chemical pathway-finding problems
can be model cases for developing various general-purpose algorithms
using Ising machines, which are anticipated to be transferable
to other systems.

In this article, we present an Ising computing framework for pathway finding
in CRNs. In Sec.~\ref{sec:problem}, we formulate a synthesis-planning problem
as a typical chemical pathway-finding problem for algorithm development.
In Sec.~\ref{sec:algorithm}, we detail our proposed algorithm, which includes
a translation procedure of a chemical pathway-finding problem into
an Ising model problem and postprocessing methods for enhancing
the feasibility and optimality of solutions. Additionally, to address
the challenge of tuning penalty strengths in the translated Ising model,
we introduce Bayesian optimization and a novel technique to enhance
tuning performance by reducing the search complexity in Sec.~\ref{sec:tuning}.
In Sec.~\ref{sec:evaluation}, we evaluate the performance of the proposed algorithm
using a D-Wave Advantage quantum annealing machine \cite{McGeoch2022} and
simulated annealing on a classical computer. Finally, we summarize
the present study and remark on possible directions for the future
development of chemical pathway-finding algorithms using Ising machines
in Sec.~\ref{sec:conclusion}.

\section{Problem Formulation} \label{sec:problem}

\subsection{Chemical reaction networks and pathways}

\begin{figure*}[tb]
  \centering
  \includegraphics[width=0.7\linewidth]{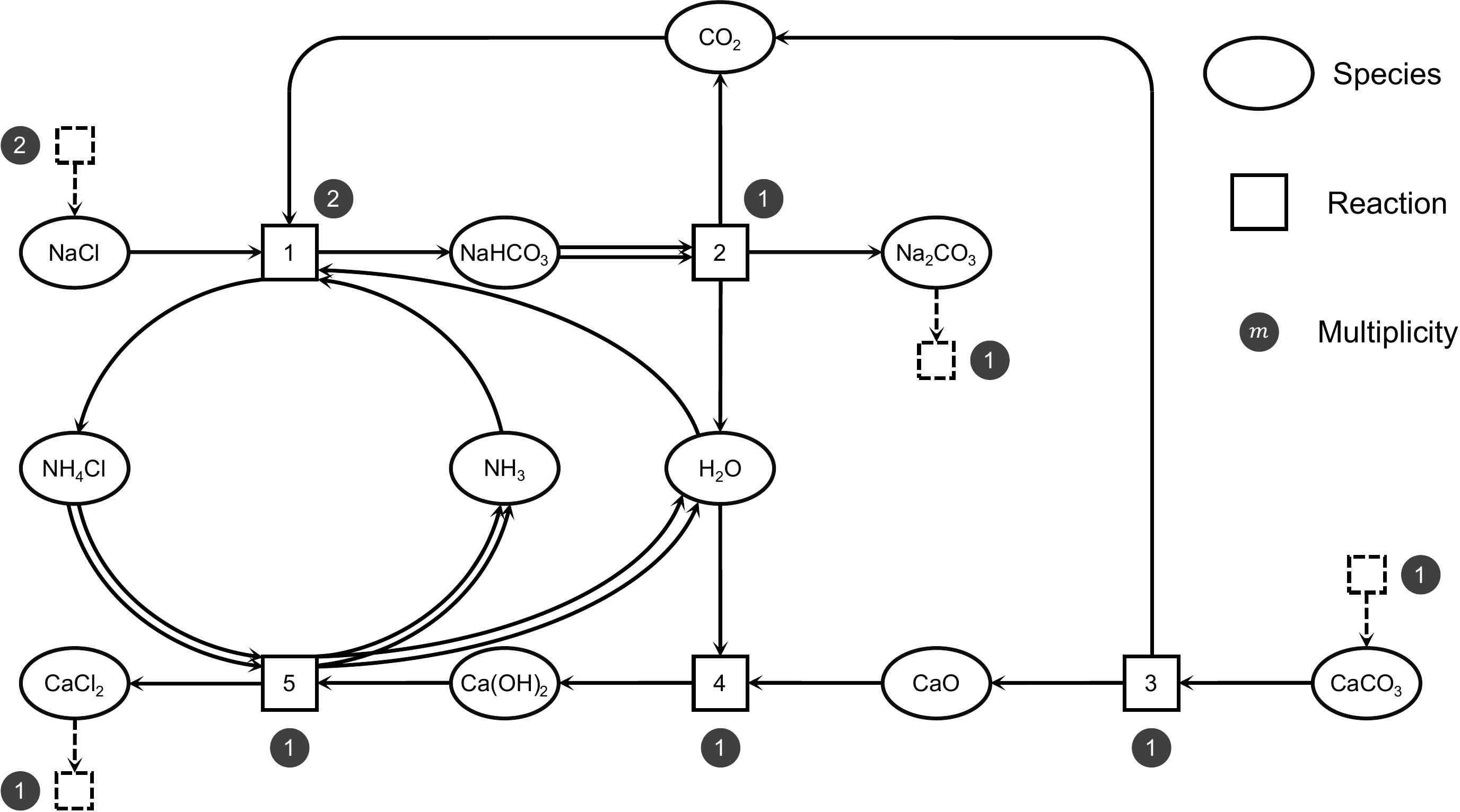}
  \caption{
    An example of a chemical reaction network and a pathway:
    the Solvay process.
   }
  \label{fig:solvay}
\end{figure*}

Figure~\ref{fig:solvay} shows an example of chemical reaction networks (CRNs)
in the bipartite graph representation \cite{Temkin1996, Szymkuc2016}, also known
as the Petri net representation \cite{Chaouiya2007}. The network illustrates
the Solvay process, an industrial chemical production process of soda ash
(Na\textsubscript{2}CO\textsubscript{3}). The overall process is represented as
2NaCl + CaCO\textsubscript{3} → Na\textsubscript{2}CO\textsubscript{3}
+ CaCl\textsubscript{2}, which consists of five chemical reactions:
\begin{enumerate}
  \item NaCl + CO\textsubscript{2} + NH\textsubscript{3}+ H\textsubscript{2}O
        → NaHCO\textsubscript{3} + NH\textsubscript{4}Cl
  \item 2NaHCO\textsubscript{3}
        → Na\textsubscript{2}CO\textsubscript{3} + CO\textsubscript{2} + H\textsubscript{2}O
  \item CaCO\textsubscript{3} → CaO + CO\textsubscript{2}
  \item CaO + H\textsubscript{2}O → Ca(OH)\textsubscript{2}
  \item Ca(OH)\textsubscript{2} + 2NH\textsubscript{4}Cl
        → CaCl\textsubscript{2} + 2NH\textsubscript{3} + 2H\textsubscript{2}O
\end{enumerate}
In the graph, these five chemical reactions involved in the Solvay process
are depicted as squares, and 11 chemical species are represented as ellipses.
Each directed edge (arrow) connects a reaction with one of its reactants or
products. The edge direction is either from a reactant species to a reaction
or from a reaction to a product species. The number of directed edges between
a reaction and a species indicates the stoichiometric coefficient, that is,
the coefficient of the species in the chemical equation of the reaction.

A CRN may include inflow and outflow \textit{dummy} reactions representing
mass exchange between a chemical system and its environment.
The network depicted in Fig.~\ref{fig:solvay} includes four dummy reactions
represented as dashed squares: the inflow reactions of NaCl and
CaCO\textsubscript{3} (the purchase of the starting materials);
the outflow reaction of Na\textsubscript{2}CO\textsubscript{3}
(the shipping of the chemical product); and the outflow reaction
of CaCl\textsubscript{2} (either the shipping or disposal of the byproduct).
Dummy reactions are useful for modeling pathway-finding problems,
as detailed below. Henceforth, we will use the term``reaction" to refer
to both chemical and dummy reactions.

We define a \textit{pathway} in a CRN as a combination of chemical and inflow/outflow
reactions, taking into account reaction multiplicity. Reaction multiplicity,
which indicates the number of occurrences of a reaction, is essential for
a quantitative description of complex reaction processes
\cite{Gao2021, Andersen2017, Horiuti1953}. For instance, in Fig.~\ref{fig:solvay},
the numbers in black circles indicate the multiplicity of each reaction
of the Solvay process. During a single cycle of the overall reaction,
chemical reaction 1 occurs twice, while the other chemical reactions
2-5 occur once each. This specific combination of the multiplicities of
the five chemical reactions enables the reuse of ammonia and carbon dioxide,
resulting in an economical chemical transformation. In addition,
the multiplicity of each inflow or outflow reaction corresponds to
the consumption of the starting material or production of the (by)product
during a single overall process, respectively. It is important to mention,
however, that the definition of pathways here does not consider the order
of occurrence of chemical reactions, as in \cite{Gao2021, Andersen2017}. 

Physically-feasible pathways must satisfy the mass balance equations
for all chemical species in a network. The mass balance equation
for species $s$ is given by $\sum_{r \in R}\nu_{sr}m_r = 0$,
where $R$, $m_r$, and $\nu_{sr}$ are the set of all reactions in the network, 
the multiplicity of reaction $r$, and the \textit{signed} stoichiometric coefficient
of species $s$ in reaction $r$, respectively. The sign of $\nu_{sr}$ is defined
as positive if $r$ is a chemical reaction producing $s$ or an inflow reaction
supplying $s$ into the system; conversely, it is defined as negative if $r$ is
a chemical reaction consuming $s$ or an outflow reaction exporting $s$ from
the system; otherwise, it is zero. The mass balance equation states that
the sum of the total production and input equals the sum of the total
consumption and output. The mass balance constraints ensure that no species
are created from nothing nor annihilated to nothing during the chemical
process, in accordance with the law of conservation of mass.

\subsection{Synthesis-planning problem}

Let us formulate a synthesis-planning problem to find the minimum monetary
cost synthesis pathway from commercially available substrates to target
compound(s).

Let $R$ be a set of chemical and inflow/outflow reactions potentially used
for synthesizing the target species. Let $S$ be the set of all chemical
species participating in chemical reactions of $R$. Here, the candidate
reactions in $R$ and species in $S$ can be listed in advance while
the multiplicities of reactions of the desired synthesis pathway are unknown.
For example, the chemical reactions in $R$ can be either extracted from
a chemical reaction database (see Appendix~\ref{appx:synthesis-relevant-network})
or generated by computer-aided retrosynthesis \cite{Gao2021}. Only commercially
available species have inflow reactions in $R$. All target species must have
outflow reactions, while others may also have outflow reactions representing
disposal.

The decision variables in the synthesis-planning problem are the multiplicities
of each reaction in $R$. Let $x_r$ denote the variable representing
the multiplicity of $r \in R$. In this article, we assume each variable $x_r$
is a non-negative integer variable with lower bound $l_r \ (\ge 0)$ and upper
bound $u_r$. Since feasible synthesis pathways must ensure the positive outflow
of each target species, the lower bound of the outflow reaction multiplicity
for each target species must be positive; in this way, positive-outflow
constraints specify which species are targets of the chemical synthesis.

The monetary costs can be categorized into two types: variable costs and
fixed costs. Variable costs vary with changes in the multiplicities of
reactions. For example, the costs of substrate purchase and byproduct
disposal are proportional to the substrate inflows and the byproduct
outflows, respectively. The simplest model of such variable costs can be
represented as $c_r^\mathrm{unit} x_r$, where $c_r^\mathrm{unit}$ is
the unit cost of reaction $r$. On the other hand, fixed costs depend
only on whether reactions are carried out. For example, the costs of
equipment preparation for reactions are overhead costs for carrying out
the reactions and are independent of the amounts of the reactions.
These fixed costs can be modeled as $c_r^\mathrm{fixed} \chi_+(x_r)$,
where $c_r^\mathrm{fixed}$ is the fixed cost of reaction $r$ and
$\chi_+$ denotes the positivity indicator function defined as
$\chi_+(x)=0$ for $x=0$ and $\chi_+(x)=1$ for $x>0$.

In summary, the synthesis-planning problem is formulated as follows:
\begin{equation}
\begin{alignedat}{2}
&\operatorname*{minimize}_{\bm{x} \in \mathbb{N}_0^{|R|}}& \quad
&\sum_{r \in R} c_r^\mathrm{unit} x_r
+\sum_{r \in R} c_r^\mathrm{fixed} \chi_+(x_r), \\
&\operatorname{subject\ to}& \quad
&\forall s \in S, \ \sum_{r \in R} \nu_{sr} x_r = 0, \\
&&
&\forall r \in R, \ l_r \le x_r \le u_r.
\end{alignedat}
\label{eq:synthesis-planning-problem}
\end{equation}
Here, $\bm{x}$ denotes the vector representation of the reaction
multiplicity variables $\{x_r\}$.

Note that a gap still exists between the Ising model (Eq.~\ref{eq:ising-model})
and the pathway-finding problem (Eq.~\ref{eq:synthesis-planning-problem}).
In the next section, we will present a connection between them.

\section{Proposed Algorithm} \label{sec:algorithm}

\begin{figure}
  \centering
  \includegraphics[width=1.0\linewidth]{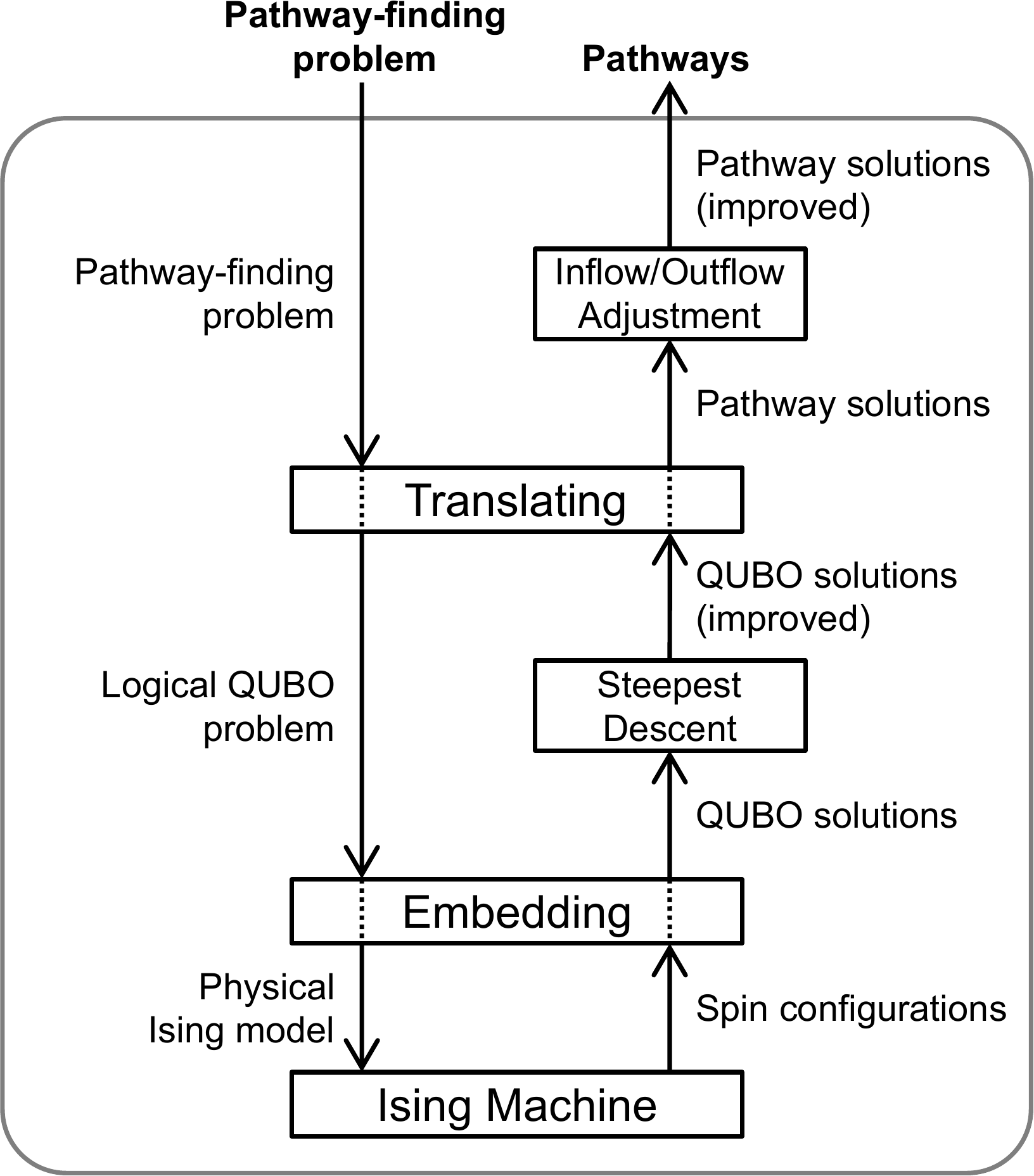}
  \caption{
    The overview of the proposed algorithm for finding pathways in chemical
    reaction networks using Ising machines.
  }
  \label{fig:algorithm-structure}
\end{figure}

This section presents an algorithm for finding pathways in chemical reaction
networks using Ising machines. Figure~\ref{fig:algorithm-structure} illustrates
the overview of the proposed algorithm. This algorithm first translates
a chemical pathway-finding problem into a quadratic unconstrained binary
optimization (QUBO) problem, mathematically equivalent to an Ising model
problem. Then, the translated problem is solved by an Ising machine. Here,
the algorithm embeds the \textit{logical} QUBO problem into a \textit{physical}
Ising model of the device if the device hardware has limited spin connectivity.
Finally, the algorithm decodes binary solutions returned from the Ising
machine and applies postprocessing methods to enhance the feasibility
and optimality of the solutions.

In the following subsections, we will elaborate on (1) the translating
procedure, (2) the postprocessing methods, and (3) the automatic tuning
of parameters in the QUBO/Ising form of a pathway-finding problem.
In addition, we review Ising machines and the embedding process
in Appendix~\ref{appx:Ising-machine} for readers unfamiliar with
Ising computing.

\subsection{Translating into QUBO} \label{sec:translating}

We translate our pathway-finding problem (Eq.~\ref{eq:synthesis-planning-problem})
into a quadratic unconstrained binary optimization (QUBO) problem. This QUBO
problem is mathematically equivalent to the energy minimization problem of
an Ising model. QUBO is defined as
\begin{equation}
\operatorname*{minimize}_{\bm{q} \in \{0, 1\}^{N_q}} \quad
\sum_{i=1}^{N_q} \sum_{j=i}^{N_q} Q_{ij} q_i q_j,
\end{equation}
where $q_i \in \{0, 1\}$, $\bm{q}$, $N_q$, and $Q_{ij}$ denote, respectively,
a binary variable, the binary variables vector $(q_1, q_2, \dots, q_{N_q})$,
the number of binary variables, and a constant associated with $q_i$ and $q_j$.
The binary variables are interconvertible with the Ising spin variables as
$q_i = (1-\sigma_i)/2$. Note that the diagonal term $Q_{ii} q_i^2$ is
essentially linear due to the idempotent law
$q_i^2 = q_i$ for $q_i \in \{0, 1\}$.

First, we translate the objective function into a quadratic function.
To achieve this, we devised the following mapping from the positivity
indicator function into a quadratic expression. The positivity indicator
function of non-negative integer variable $x$ can be expressed as follows:
\begin{equation}
\chi_+(x) = \min_{y \in \{0, 1\}} \left[ y + (1-y)x \right].
\end{equation}
Therefore, we can replace $\chi_+(x_r)$ in the original cost function
with $[y_r + (1-y_r)x_r]$, adding an auxiliary binary variable
$y_r \in \{0, 1\}$.

Next, instead of imposing the mass balance constraints directly,
we add the penalty term
\begin{equation}
\sum_{s \in S} M_s \left( \sum_{r \in R} \nu_{sr} x_r \right)^2 
\end{equation}
to the QUBO objective function. Here, $M_s$ is a positive constant. This term
equals zero if solution $\bm{x}$ satisfies all the mass balance constraints
and a positive value otherwise. Thus, this term penalizes solutions violating
the mass balance constraints. The penalty strength is determined by $M_s$.
In theory, large-enough penalty strength ensures that the optimal solution
of this unconstrained problem equals that of the original problem.
However, the penalty strength requires careful tuning in practice,
as extremely large penalty strength can impair the annealing performance.
Sec.~\ref{sec:tuning} describes the penalty strength parameter tuning in detail.

Finally, we encode integer variables into binary variables. We employ
four types of integer-to-binary encoding methods: unary, order, log, and
one-hot \cite{Zaman2022}. Table~\ref{tbl:encoding-methods} summarizes
these encoding methods. In the table, an integer variable $x \in [l, u]$
is expressed by a linear expression $x(\bm{q})$ of $n$ binary variables
$\bm{q} \in \{0, 1\}^n$. The number $n$ of required binary variables is
$d \ (\coloneqq u-l)$ for the unary and order encoding methods,
$\lfloor \log_2 d \rfloor + 1$ for the log encoding method, and $d+1$ for
the one-hot encoding method. In addition, the order and one-hot encoding
methods require a penalty term in the QUBO objective function.
The additional penalty in the order encoding method corresponds to constraints
that $q_k = 0 \Rightarrow q_{k+1} = 0$ for $k = 1, \dots, d-1$. Note that
the order constraints are not necessarily hard constraints; even if the order
constraints are not satisfied, the integer variable expression $x(\bm{q})$
takes an integer value in $[l, u]$. On the other hand, the additional penalty
in the one-hot encoding method corresponds to a constraint that only one binary
variable takes the value one and all the others zero. In contrast to the order
encoding method, if the one-hot constraint is violated, $x(\bm{q})$ may take
an infeasible value of $x$; thus, the one-hot constraint is a hard constraint.
We will compare the performance of these four encoding methods in
Sec.~\ref{sec:results-discussion}.

\begin{table*}[tb]
  \centering
  \caption{
    Integer-to-binary encoding methods.
    Here, an integer variable $x \in [l, u]$ is expressed by $n$ binary
    variables $\bm{q} \in \{0, 1\}^n$. The number $n$ of required binary
    variables is $d \ (\coloneqq u-l)$ for the unary and order encoding methods,
    $K + 1 \ (K \coloneqq \lfloor\log_2d\rfloor)$ for the log encoding method,
    and $d+1$ for the one-hot encoding method.
  }
  \begin{tabular}{lll}
    \hline
    Type & Integer variable expression $x(\bm{q})$ & Additional penalty $P_x(\bm{q})$ \\
    \hline \hline
    Unary & $l+\sum_{k=1}^{d}q_k$ & $0$ \\
    Order & $l+\sum_{k=1}^{d}q_k$ & $\sum_{k=1}^{d-1}q_{k+1}(1-q_k)$ \\
    Log & $l+\sum_{k=0}^{K-1}2^kq_k+[d-(2^K-1)]q_K$ & $0$ \\
    One-hot & $l+\sum_{k=0}^{d}k q_k$ & $\left(\sum_{k=0}^{d}q_k-1\right)^2$ \\
    \hline
    \end{tabular}
  \label{tbl:encoding-methods}
\end{table*}

In summary, the synthesis-planning problem (Eq.~\ref{eq:synthesis-planning-problem})
can be represented in QUBO form as follows:
\begin{equation}
\begin{alignedat}{2}
&\operatorname*{minimize}_{\bm{q} \in \{0, 1\}^{N_q}}& \quad
&\sum_{r \in R} c_r^\mathrm{unit} x_r(\bm{q}_r)
+\sum_{r \in R} c_r^\mathrm{fixed} \left[ y_r + (1-y_r)x_r(\bm{q}_r) \right] \\
&& +
&\sum_{s \in S} M_s \left( \sum_{r \in R} \nu_{sr} x_r(\bm{q}_r) \right)^2
+\sum_{r \in R} L_r P_{x_r}(\bm{q}_r).
\end{alignedat}
\label{eq:synthesis-planning-problem-QUBO}
\end{equation}
Here, each $x_r$ is expressed by binary variables $\bm{q}_r$ by one of
the four encoding methods listed in Table~\ref{tbl:encoding-methods}.
$P_{x_r}(\bm{q}_r)$ and $L_r\ (>0)$ are the additional penalty term
and its strength parameter for the encoding of $x_r$, respectively.
Each $y_r \ (\in \{0, 1\})$ is an auxiliary binary variable
for translating the positivity indicator function $\chi_+(x_r)$.
Finally, the symbol $\bm{q}$ collectively denotes all binary variables
$\{\bm{q}_r \mid r \in R\} \cup \{y_r \mid r \in R\}$, and $N_q$ is
the total number of the binary variables.

\subsection{Postprocessing} \label{sec:postprocessing}

Ising machines often return non-optimal or infeasible solutions.
To improve the quality of solutions, we employ two postprocessing methods:
steepest descent and inflow/outflow adjustment.

Steepest descent is a standard postprocessing method in Ising computing.
This method iteratively descends the energy landscape of an Ising model
by single-spin flips. At each step, it performs the spin flip that reduces
the energy the most among all possible single-spin flips. This postprocessing
method is implemented in the D-Wave Ocean Software \cite{DWaveOcean}. We employ
it to enhance the optimality and feasibility of Ising/QUBO solutions
returned by an Ising machine.

\begin{figure}[tb]
  \centering
  \includegraphics[width=1\linewidth]{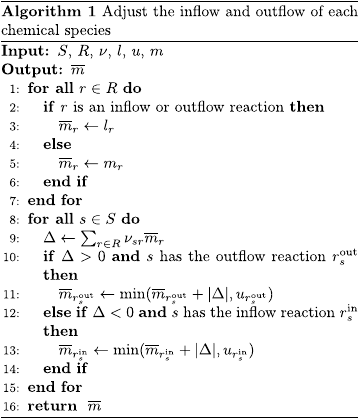}
  \caption{
    Inflow/outflow adjustment algorithm.
    The input arguments are a set of chemical species $S$,
    a set of chemical and inflow/outflow reactions $R$,
    a signed stoichiometric coefficient function
    $\nu \colon S \times R \to \mathbb{Z}; (s, r) \mapsto \nu_{sr}$,
    a lower bound function $l \colon R \to \mathbb{N}_0; r \mapsto l_r$,
    an upper bound function $u \colon R \to \mathbb{N}_0; r \mapsto u_r$,
    and the multiplicity function representing a pathway
    $m \colon R \to \mathbb{N}_0; r \mapsto m_r$.
    This algorithm, first, resets the multiplicity of each inflow/outflow
    reaction to its lower bound. Then, it increases the outflow of overproduced
    species ($\Delta > 0$) and the inflow of overconsumed species ($\Delta < 0$).
    Finally, it returns the multiplicity function of the improved pathway
    $\overline{m} \colon R \to \mathbb{N}_0; r \mapsto \overline{m}_r$.
    Note that this postprocessing does not ensure that a post-processed pathway
    satisfies all mass balance constraints. For example, if a species is
    overconsumed, but the associated inflow is unavailable, this method cannot
    adjust the inflow/outflow to make the mass balance zero.
  }
  \label{fig:ioflow-adjustment-algorithm}
\end{figure}

Inflow/outflow adjustment, which we developed, is aimed at adjusting
the inflow and outflow of each species so that the mass balance equals zero.
Specifically, this method (1) sets the inflow and outflow of each species to
their minimum values and (2) increases either the inflow or outflow as much as
necessary to satisfy the mass balance constraint. The pseudo-code is given in
Fig.~\ref{fig:ioflow-adjustment-algorithm}. We apply this postprocessing method
at the end of the workflow to enhance the feasibility of solutions.

\subsection{Penalty strength tuning} \label{sec:tuning}

The QUBO form of the pathway-finding problem
(Eq.~\ref{eq:synthesis-planning-problem-QUBO}) has penalty strength parameters.
Moreover, when the QUBO problem is embedded into a physical Ising machine
with limited connectivity of spins, the physical Ising model has chain
strength parameters (see Appendix~\ref{appx:embedding}).
In the following discussion, we consider parameter tuning to improve
algorithm performance.

Parameter tuning aims to find parameter values that maximize the quality of
solutions returned by the Ising computing algorithm. The parameter tuning
problem is formulated as
\begin{equation}
\operatorname*{minimize}_{\bm{\lambda} \in D} f(\bm{\lambda}),
\end{equation}
where $\bm{\lambda}$ is the parameter vector to be tuned (in this study,
penalty and chain strength parameters), $D$ is the domain of $\bm{\lambda}$,
and $f \colon D \to \mathbb{R}$ is a function measuring the quality of
solutions (smaller is better).

Let us define the parameter search space $D$. In the present case,
we can estimate an upper bound of penalty strength parameters, which is
large enough to ensure the equivalence between the original and QUBO problems.
A simple upper bound is the maximum cost among all feasible and infeasible
pathways:
\begin{equation}
\overline{C} = \sum_{r \in R} (c_r^\mathrm{unit} u_r + c_r^\mathrm{fixed}).
\end{equation}
If all penalty strength parameters (i.e., $\{M_s\}$ and $\{L_r\}$) equal
$\overline{C}$, the QUBO objective function value for any feasible solution is
less than $\overline{C}$
\footnote{
  If solution $x_r = u_r \ (\forall r \in R)$ is feasible, this statement
  is false; in that case, a value greater than $\overline{C}$ can be employed
  as the upper bound of the parameter values. However, in every benchmark
  problem used in the present article, the solution $x_r = u_r$ is confirmed
  to be infeasible.
}; in contrast, that for any infeasible solution
is greater than or equals to $\overline{C}$. Therefore, the penalty strength
$\overline{C}$ ensures the equivalence between the original and QUBO problems.
This upper bound estimation is also valid for chain strength parameters;
if the chain strength equals $\overline{C}$, a spin configuration with chain
breaks has an energy greater than any feasible solution without chain breaks.
Therefore, we define the search space as
\begin{equation}
D
=\left[0, \overline{C}\right]^{N_\mathrm{params}}
\subset \mathbb{R}^{N_\mathrm{params}},
\end{equation}
where $N_\mathrm{params}$ is the total number of penalty and chain strength
parameters to be tuned.

Next, we define the evaluation function $f$. Multiple runs of the algorithm
produce a set of synthesis pathways. This set contains feasible and infeasible
pathways with various synthesis costs. Our goal here is (1) to get many
feasible solutions and (2) to get low costs solutions. Therefore, we use
the following function to evaluate a single solution:
\begin{equation}
E(\bm{x})
=\sum_{r \in R} \left(c_r^\mathrm{unit} x_r
+c_r^\mathrm{fixed} \chi_+(x_r) \right)
+\overline{C}\sum_{s \in S} \left( \sum_{r \in R} \nu_{sr} x_r \right)^2.
\end{equation}
Here, the first term evaluates the total costs of synthesis, and the second
is the penalty for infeasible solutions. Since the penalty strength equals
$\overline{C}$ in the formula, all infeasible solutions have a higher value
of $E$ than any feasible solution. Using this measure of single solution
quality, we define $f(\bm{\lambda})$ as the expected value of $E(\bm{x})$,
\begin{equation}
\langle E \rangle_{\bm{\lambda}}
= \sum_{\bm{x} \in \mathbb{N}_0^{|R|}} E(\bm{x})p(\bm{x}|\bm{\lambda}),
\label{eq:mean-standard-energy}
\end{equation}
where $p(\bm{x}|\bm{\lambda})$ is the probability that the algorithm returns
solution $\bm{x}$ when using parameters $\bm{\lambda}$. In practice,
$\langle E \rangle_{\bm{\lambda}}$ is evaluated by a sample mean of
$E(\bm{x})$ from a finite set of samples of $\bm{x}$.

We adopt Bayesian optimization (BO) to minimize
$\langle E \rangle_{\bm{\lambda}}$ over parameters $\bm{\lambda}$.
Specifically, we employ a BO algorithm implemented in the Optuna library
\cite{Akiba2019, Optuna}. BO is a sequential black-box optimization technique
that selects points to be evaluated in an iteration step based on
a surrogate model of $f$ constructed with previous evaluation results.
The Optuna's algorithm uses the tree-structured Parzen estimator (TPE)
\cite{Bergstra2011} as a surrogate model. BO-TPE is known to be effective
in a wide range of hyperparameter tuning in machine learning and
has less computational time complexity than standard BO algorithms
using the Gaussian process \cite{Yang2020}.

However, the high dimensionality of the search space $D$ may impair
the efficiency of BO. Although BO is applicable to parameter tuning
in a complex and high-dimensional search space, its efficiency decreases
rapidly as the dimension of the search space increases \cite{Yang2020}.
The dimension of $D$ is $N_\mathrm{params}$, which is on the same order
of magnitude as the total numbers of species and reactions. Therefore,
reducing the dimensionality of $D$ is crucial for efficient parameter tuning.

To address the issue of high dimensionality in $D$, we devised parameter
grouping methods. In these grouping methods, penalty and chain strength
parameters are grouped according to the associated constraint type,
and all parameters in each group are constrained to take the same value.
We first classify parameters into three major categories:
(1) penalty strength parameters associated with mass balance constraints,
(2) penalty strength parameters related to integer-to-binary encoding,
and (3) chain strength parameters for embedding. In this article,
we further consider four types of subdivisions of the mass balance constraints
group according to chemical species classifications listed in
Table~\ref{tbl:grouping-methods} and detailed below. For simplicity,
we leave the other groups related to encoding and embedding intact,
though these groups can be further subdivided according to reaction
classifications.

\begin{table*}[tbh]
  \centering
  \caption{
    Grouping methods for penalty strength parameters associated with
    mass balance constraints. The right column lists species groups corresponding
    to mass balance constraint groups. See the text for details on the chemical
    species classifications.
  }
  \begin{tabular}{ll}
    \hline
    Type & Groups \\
    \hline \hline
    Unified & all species \\
    Degree & degree-1 species, degree-2 species, ... \\
    Depth & targets, depth-1 precursors, depth-2 precursors, ..., byproducts \\
    Category & targets, substrates, intermediates, substrates/intermediates, byproducts \\
    \hline
  \end{tabular}
  \label{tbl:grouping-methods}
\end{table*}

The details of the four grouping methods for mass balance constraints are
as follows. The first method, `unified,' consolidates all mass balance
constraints into one group without any further subdivision. The second method,
`degree,' classifies mass balance constraints by the associated species' node
degree. Here, the node degree is defined as the number of reactions involving
the species. The third and fourth methods have been designed based on
the specific structure of synthesis-planning problems. The third method,
named `depth,' organizes mass balance constraints by the associated species'
depth. The depth of a precursor species is defined as the minimum number of
synthesis steps from the species to one of the target species, as detailed in
Appendix~\ref{appx:synthesis-relevant-network}. The fourth method, termed
`category,' categorizes mass balance constraints by the associated species'
category: `targets,' `substrates' (species that are exclusively starting
materials and not synthesized from others), `intermediates' (species that
need to be synthesized from others), `substrates/intermediates' (species that
can be either starting materials or synthesized from others), and `byproducts.'
We will compare the performance of parameter tuning using these grouping
methods in Sec.~\ref{sec:results-discussion}.

We note that the Optuna library has already been used in parameter tuning
for Ising computing in \cite{Ayodele2022, Goh2022}. In \cite{Ayodele2022}, Optuna
was applied to tuning parameters of the Digital Annealer, not penalty strength
parameters. In \cite{Goh2022}, all penalty strength parameters are consolidated
into one group, and the single, unified penalty strength parameter was tuned
by Optuna. In contrast to these studies, the present study systematically
elucidates the effects of several different parameter grouping methods on
tuning performance. In the next section, we will demonstrate that appropriate
choices of parameter grouping methods significantly contribute to
performance improvement.

\section{Performance Evaluation} \label{sec:evaluation}

In this section, we present a performance evaluation of the proposed
algorithm using a D-Wave Advantage quantum annealing machine and
a simulated annealing software. The primary objective of these experiments
is to assess the feasibility and efficacy of the proposed algorithm.

\subsection{Experimental setting} \label{sec:experimental-setting}

We generated 100 benchmark problems of synthesis planning based on
the USPTO dataset of chemical reactions \cite{Lowe2017}. In our benchmark
problems, fixed reaction costs and substrate purchase costs are randomly
assigned to all chemical reactions and commercially available species,
respectively. The benchmark problem generation process is detailed in
Appendix~\ref{appx:benchmark-problems}. These benchmark problems have been
designed to assess the feasibility of the proposed algorithm using a D-Wave
Advantage quantum annealer with a limited number of qubits. Thus, they are
relatively small-scale problems: the number of integer variables $N_x$,
equivalent to the number of chemical and inflow/outflow reactions $|R|$,
ranges from 3 to 116; the number of chemical species $|S|$, ranges from
2 to 80. All of these problems can be embedded into the D-Wave Advantage's
Pegasus topology QPU \cite{Boothby2019}, with the number of required
physical qubits ranging from 18 to 3752 when utilizing the unary or order
encoding method. Refer to Appendix~\ref{appx:embedding-performance} for
the details of the embedding results.

We solved the benchmark problems using our proposed algorithm. The Ising
machines employed were (1) the D-Wave Advantage system 4.1 quantum annealer
\cite{McGeoch2022} and (2) the \texttt{SimulatedAnnealingSampler} of
the D-Wave Ocean Software \cite{DWaveOcean}. We used
the \texttt{minorminer.find\_embedding} function of the D-Wave Ocean Software
to find embeddings of QUBO problems into the D-Wave Advantage system 4.1.
Note that the simulated annealing software does not require the embedding
process, as it can solve Ising models with any spin connectivity. We employed
the \texttt{SteepestDescentSolver} of the D-Wave Ocean Software for the steepest
descent postprocessing. We implemented the other components of the algorithm,
that is, the translation process based on SymPy \cite{Meurer2017} and
the inflow/outflow adjustment postprocessing, in Python. We integrated
all these components into the workflow illustrated in
Fig.~\ref{fig:algorithm-structure}. We utilized the multivariate version of
the \texttt{TPESampler} of the Optuna library \cite{Optuna} to tune penalty
and chain strength parameters. Unless otherwise noted, we used default settings
for the functions from the D-Wave Ocean Software and the Optuna library
in the benchmarking. The simulated annealing, embedding finding, translating,
postprocessing, and parameter tuning were executed on a local Linux machine
with Intel Xeon Gold 6248R processors (3.0 GHz, 24 cores $\times$ 2). Our local
machine in Hokkaido, Japan, communicated with the D-Wave Advantage system 4.1
in British Columbia, Canada, via the internet, using the Leap quantum cloud
service.

For performance comparison, we solved the benchmark problems also with
the Gurobi Optimizer version 9.5.1 \cite{Gurobi}. The Gurobi Optimizer is
a commercial state-of-the-art solver for mathematical programming,
including integer linear programming, and was employed in previous studies
on chemical pathway-finding problems \cite{Gao2021, Andersen2017}. To solve
the synthesis planning problems with the Gurobi Optimizer, we translated
them into integer linear programming by replacing the positivity indicator
function in a fixed cost term, $\chi_+(x_r)$, with an auxiliary binary
variable $y_r \in \{0, 1\}$ satisfying $x_r \le u_r y_r$. The chemical
pathway-finding algorithm with the Gurobi Optimizer was executed on
our local machine using up to 32 threads.

\subsection{Results and discussion} \label{sec:results-discussion}

\subsubsection{Best choice of parameters}

\begin{figure*}
  \centering
  \includegraphics{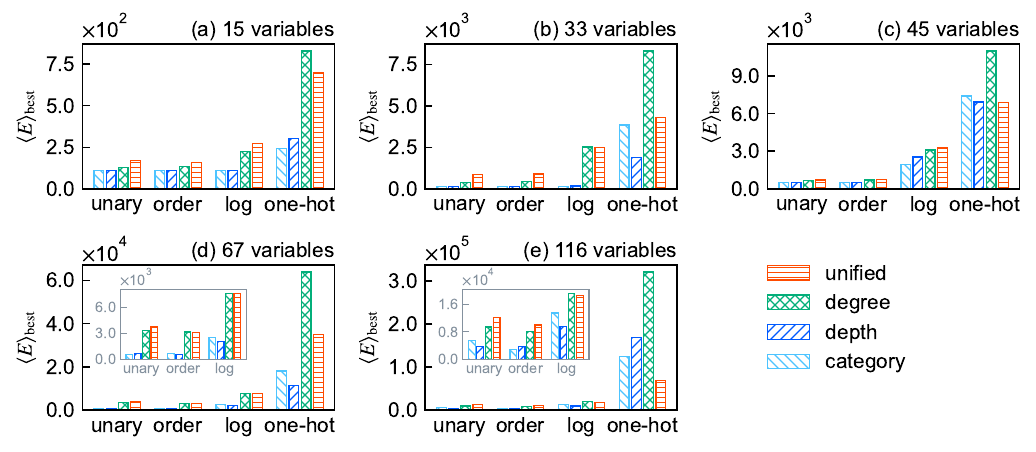}
  \caption{
    Performance comparison of different combinations of an integer-to-binary
    encoding method (unary, order, log, or one-hot) and a parameter grouping
    method (unified, degree, depth, or category) in the proposed algorithm
    using the D-Wave Advantage system 4.1. Each panel displays the results for
    (a) the 20th, (b) the 40th, (c) the 60th, (d) the 80th, and (e) the 100th
    from the 100 benchmark problems sorted in ascending order by the number of
    integer variables, respectively. The performance measure
    $\langle E \rangle_\mathrm{best}$ is the best score of the tuning objective
    function, Eq.~\ref{eq:mean-standard-energy}, during 300 iterations of Bayesian
    optimization (smaller is better). The insets in panels (d) and (e) provide
    magnified views.
  }
  \label{fig:encoding-grouping-performance-qa}
\end{figure*}

\begin{figure*}
  \centering
  \includegraphics{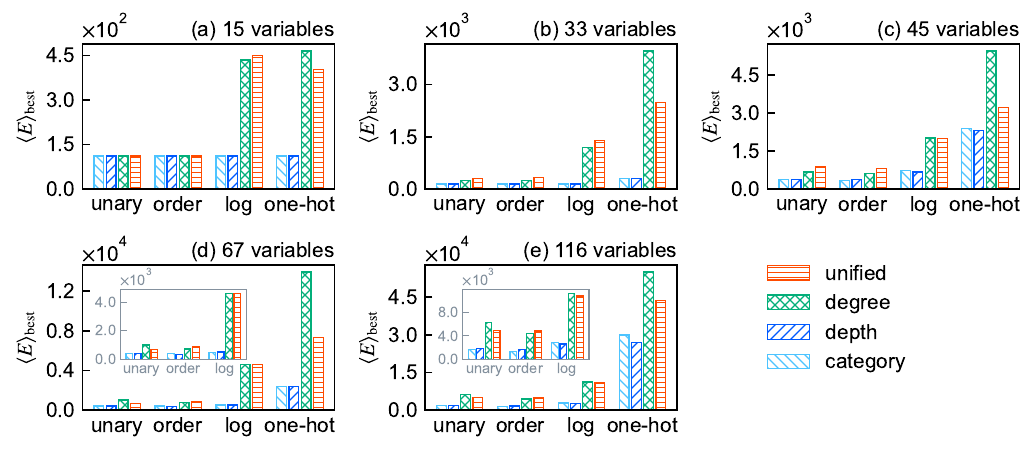}
  \caption{
    Performance comparison of different combinations of an integer-to-binary
    encoding method (unary, order, log, or one-hot) and a parameter grouping
    method (unified, degree, depth, or category) in the proposed algorithm
    using simulated annealing. The notation is the same as that in
    Fig.~\ref{fig:encoding-grouping-performance-qa}.
  }
  \label{fig:encoding-grouping-performance-sa}
\end{figure*}

We first investigated the best combination of an integer-to-binary
encoding method (unary, order, log, or one-hot) and a parameter
grouping method (unified, degree, depth, or category). This investigation
was conducted on five representative problems selected from the 100 benchmark
problems to cover a wide range of problem sizes. The problems selected were
(a) the 20th, (b) the 40th, (c) the 60th, (d) the 80th, and (e) the 100th
in ascending order of the number of integer variables. For each combination
of encoding and grouping methods, we tuned penalty and chain strength
parameters for the five representative problems. During the tuning processes,
the annealing time for quantum annealing (QA) was fixed to 20 \si{\micro s},
and the number of sweeps (corresponding to the annealing time) for simulated
annealing (SA) was fixed to 1000. These are the default values for the D-Wave
Advantage system and the SA program we used. We set the number of Bayesian
optimization iterations to 300, which we consider sufficient to check for
convergence trends. We sampled 200 solutions per tuning iteration to compute
the tuning score function, Eq.~\ref{eq:mean-standard-energy}, with balancing
statistical uncertainty and computational cost. We measured the performance
of each combination of encoding and grouping methods by the best score of
the tuning objective function over the 300 tuning trials. The performance
comparisons for the D-wave Advantage system and SA are shown in
Fig.~\ref{fig:encoding-grouping-performance-qa} and
Fig.~\ref{fig:encoding-grouping-performance-sa}, respectively.

Concerning the encoding methods, the unary and order encoding methods
demonstrate the highest performance. Subsequently, the log encoding method
also yields satisfactory results, whereas the one-hot encoding method exhibits
the poorest performance. These results on the performance of the encoding
methods are consistent with a previous study \cite{Tamura2021}.

Among the four grouping methods, the depth and category grouping methods
outperform the others when employing the unary or order encoding methods
for both QA and SA. The performance difference between these two methods and
the others is prominent, especially for large problems (d) and (e).
Furthermore, the degree grouping method exhibits the least performance
in many cases, although the degree grouping method offers more degrees of
freedom for parameter tuning due to the subdivision of mass balance
constraints than the unified grouping method. These findings suggest that
the appropriate design of parameter groups based on the specific structure
of problems is crucial for performance improvement.

Next, we examined the optimized penalty and chain strength values
for the depth and category grouping methods to clarify the reason for
the advantage of these grouping methods as well as the trends in penalty
and chain strength parameter values of each group with respect to the problem
size. This examination was conducted on all 100 benchmark problems and employed
the unary and order encoding methods. The settings for the tuning processes
were the same as those described above.

\begin{figure*}
  \centering
  \includegraphics{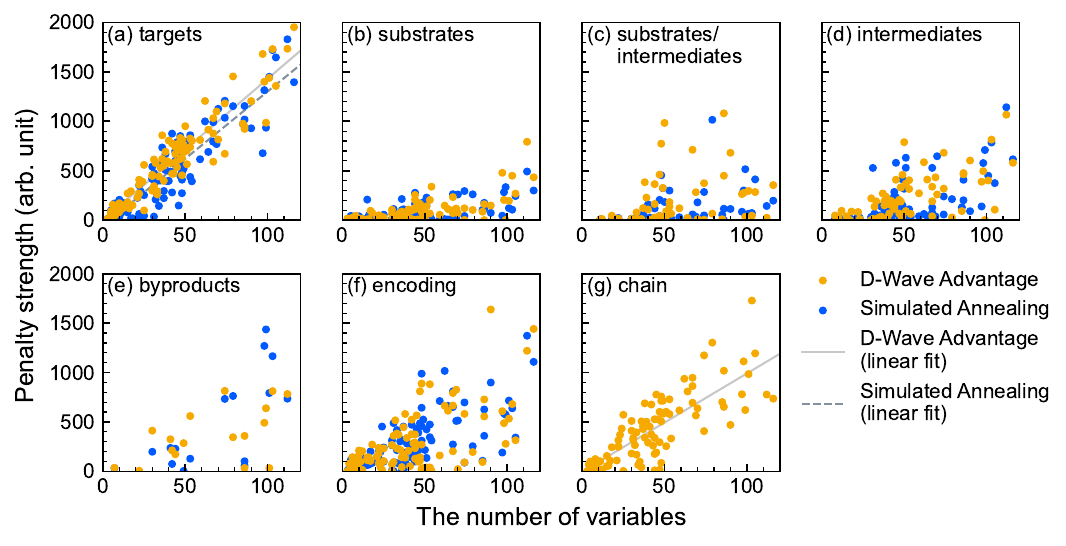}
  \caption{
    Scaling of the tuned penalty and chain strength parameter values
    with respect to the number of integer variables in the case of using
    the order encoding and category grouping methods.
    Panels (a)-(e) depict the penalty strength parameters for mass balance
    constraints of species in each group, panel (f) shows the penalty
    strength parameter for order encoding, and panel (g) shows the chain
    strength for embedding. Linear fits are plotted for datasets with
    a coefficient of determination ($R^2$) greater than 0.5.
    The $R^2$ values are 0.88 (panel (a) for D-Wave Advantage), 0.83
    (panel (a) for simulated annealing), and 0.65 (panel (g) for D-Wave
    Advantage), respectively.
  }
  \label{fig:penalty-strength-scaling}
\end{figure*}

Figure~\ref{fig:penalty-strength-scaling} depicts the penalty and chain strength
parameter values optimized through Bayesian optimization for the combination
of the category grouping and order encoding methods. We found that:
(1) the penalty strength for the mass balance constraints of the target species
is larger than that of the others and has a strong linear correlation with
the number of integer variables, $N_x$; (2) the chain strength parameter also
exhibits a linear correlation with $N_x$; and (3) the other penalty strength
parameters do not show any noticeable linear correlation with $N_x$.
These penalty strength scaling patterns (1)-(3) are common across all four
combinations of the unary/order encoding and depth/category grouping methods.
Refer to Appendix~\ref{appx:penalty-strength-scaling} for the results of
the other combinations not shown in Fig.~\ref{fig:penalty-strength-scaling}.

The observed penalty strength scaling patterns and the advantage of
the depth and category grouping methods can be explained as follows.
The minimum-cost infeasible solution, $\bm{x}=\bm{0}$ (i.e., the plan
of doing nothing), violates the mass balance constraints for the target
species. An Ising machine is likely to sample this solution with a high
probability unless the penalty for violating the mass balance constraints
of the target species is sufficiently larger than feasible solutions' cost
values. The steepest descent postprocessing may not improve this solution
as it relies on the same QUBO formulation as the Ising machine.
The inflow/outflow adjustment postprocessing also cannot reform the solution
$\bm{x}=\bm{0}$ because the target species are overconsumed in the pathway,
but the inflow reactions of the target species are unavailable by definition.
Therefore, the penalty strength for the mass balance constraints of
the target species must be large enough so that the Ising machine hardly
samples the solution $\bm{x}=\bm{0}$. Furthermore, as the order of
magnitude of feasible solutions' cost values increases almost linearly
with respect to $N_x$, the penalty strength for the mass balance
constraints of the target species must also increase linearly.
In contrast, the violation of the mass balance constraints for non-target
species can often be reconciled by adjusting the associated inflow
and outflow, although the effectiveness of the postprocessing may depend on
the problem structure. Thus, the penalty strength associated with non-target
species can be moderate. Keeping penalty strengths moderate may enhance
the exploration in the solution space during the annealing process,
potentially leading to high performance in finding solutions.
For the depth and category grouping methods, the penalty strength for
the mass balance constraints of the target species and the others can be
tuned separately; this is likely the reason for their advantage.

\subsubsection{Computation time}

We next examined the computation time scaling of the pathway-finding
algorithms using the D-Wave Advantage system and SA. Since the algorithms
output approximate solutions in a stochastic manner, the computation time
should be evaluated by the time taken to find a solution within a certain
cost tolerance $\rho$ with a specified tolerant failure probability $\epsilon$.
This computation time metric is known as time-to-solution (TTS)
\cite{Mohseni2022, McGeoch2022}, defined by
\begin{equation}
TTS(\rho, \epsilon, \tau) =
\tau_\mathrm{algo}(\tau)\left\lceil
\frac{\log\epsilon}{\log(1-p_\mathrm{s}(\rho, \tau))}\right\rceil.
\label{eq:tts}
\end{equation}
Here, $\tau$ is the annealing time; $\tau_\mathrm{algo}(\tau)$ is
the average execution time of a single run of the algorithm under
the annealing time $\tau$; $p_\mathrm{s}(\rho, \tau)$ is the success
probability for the algorithm under the annealing time $\tau$ to find
a solution whose cost is at most $\rho C_\mathrm{min}$, where $C_\mathrm{min}$
denotes the minimum cost of feasible solutions.

The computation time analysis was conducted on all 100 benchmark problems
and employed the order encoding and category grouping methods. We set
the cost tolerance $\rho$ to (a) 1, (b) 2, or (c) 3 and fixed the failure
tolerance $\epsilon$ to 0.01. We measured the TTS for different annealing time
$\tau$: 1, 2, 5, 10, 20, 50, 100, 200, and 500 \si{\micro s} for QA; 10, 20, 50,
100, 200, 500, 1000, 2000, 5000, and 10000 sweeps for SA. For each problem
and $\tau$, we ran the algorithm 1000 times with tuned penalty and chain
strengths as illustrated in Fig.~\ref{fig:penalty-strength-scaling}.
In the measurement, we took into account only the runtime of annealing
and postprocessing as $\tau_\mathrm{algo}$ and excluded the execution time
of embedding, translating, tuning, and network communication from
$\tau_\mathrm{algo}$. Lastly, we recorded the minimum value of TTS with
respect to $\tau$ as the computation time metric for each $\rho$.

We also computed the average runtime of the pathway-finding algorithm using
the Gurobi Optimizer for comparison. For each cost tolerance $\rho$,
the `MIPGap' parameter of the Gurobi Optimizer was set to $1-1/\rho$
so that the optimizer terminates when it is assured that the cost value
of the current best solution is less than or equal to $\rho C_\mathrm{min}$.
We recorded the average runtime of 1000 runs for each problem and $\rho$.

\begin{figure*}
  \centering
  \includegraphics{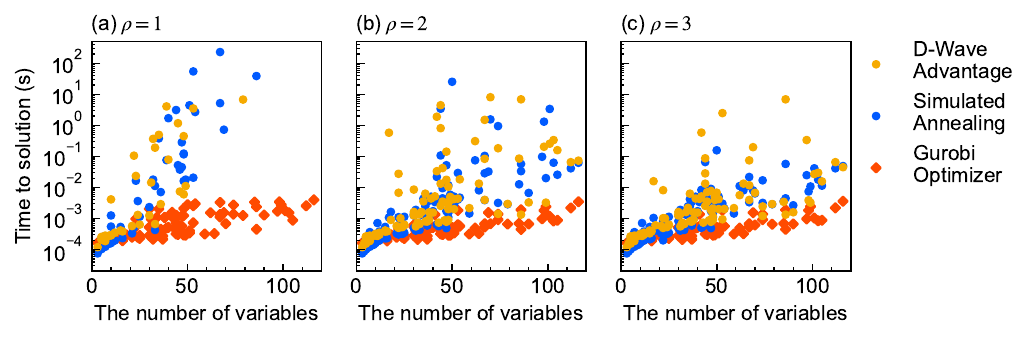}
  \caption{
    Computation time scaling for the pathway-finding algorithms using
    the D-Wave Advantage system, simulated annealing, and the Gurobi Optimizer.
    The computation times of the algorithms using the D-Wave Advantage system
    and simulated annealing are the time-to-solution defined by Eq.~\ref{eq:tts},
    the time taken to find a solution whose cost is at most
    $\rho\times(\text{the minimum cost})$ with a probability of at least 99\%.
    Here, the cost tolerance $\rho$ is (a) 1, (b) 2, and (c) 3, respectively.
    For the Gurobi Optimizer, the time-to-solution refers to the runtime
    it takes the solver to find a solution guaranteed to satisfy the cost
    tolerance requirement. Refer to the main text for additional details
    on the computing procedure.
  }
  \label{fig:tts}
\end{figure*}

Figure~\ref{fig:tts} shows the scaling of computation times for the pathway-finding
algorithms using the D-Wave Advantage system, SA, and the Gurobi Optimizer.
When $\rho=1$, the computation times for the D-Wave Advantage system and SA
increase drastically as the number of variables, $N_x$, increases, compared to
that for the Gurobi Optimizer. As $\rho$ increases, the rates of increase
in the computation times with respect to $N_x$ for the D-Wave Advantage system
and SA become slower. Consequently, these computation times approach
that of the Gurobi Optimizer. In addition, the pathway-finding algorithm
using the D-Wave Advantage system succeeded in finding optimal solutions
($\rho=1$) for 34 out of the 100 benchmark problems, while the SA-based
algorithm found optimal solutions for 43 problems. Approximate solutions
with $\rho=2$ were found for 96 problems when using the D-Wave Advantage
system and all problems when using SA. Both methods found approximate
solutions with $\rho=3$ for all problems.

\begin{figure}
  \centering
  \includegraphics[width=0.7\linewidth]{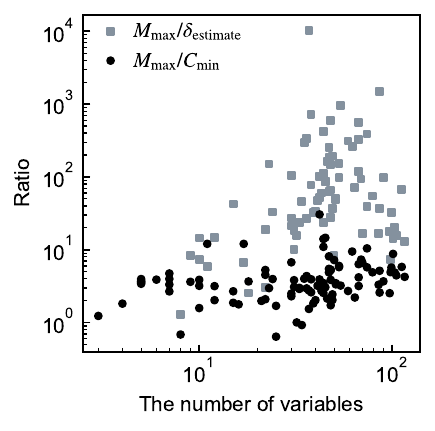}
  \caption{
    The ratio of the maximum penalty strength $M_\mathrm{max}$ relative to
    (1) the cost difference $\delta$ between the optimal and the second optimal
    solutions and (2) the minimum cost $C_\mathrm{min}$ of feasible solutions.
    The maximum penalty strength $M_\mathrm{max}$ was computed for the optimized
    penalty and chain strengths for simulated annealing (SA) shown in
    Fig.~\ref{fig:penalty-strength-scaling}. The minimum cost $C_\mathrm{min}$ was
    computed by the Gurobi Optimizer. The cost difference $\delta$ was estimated
    as follows: first, we gathered non-optimal feasible solutions from
    the samples used to compute the time-to-solution for SA shown in
    Fig.~\ref{fig:tts}; then, among these non-optimal feasible solutions,
    we identified the minimum cost, denoted by $C^\prime$; finally, we estimated
    $\delta$ as $\delta_\mathrm{estimate} = C^\prime - C_\mathrm{min}$.
    This estimate provides an upper bound of $\delta$ as $C^\prime$ is always
    greater than or equal to the true second minimum cost. Note that for some
    problems (especially with small sizes), all feasible solutions sampled by
    the SA-based algorithm are the optimal solution,
    thus $M_\mathrm{max}/\delta_\mathrm{estimate}$ is not plotted.
  }
  \label{fig:pcr}
\end{figure}

In the case of $\rho=1$, the fast increasing rate of the TTS of SA
with respect to $N_x$ is likely attributed to the increase in the maximum
penalty strength, $M_\mathrm{max}$. The reasoning behind this statement
is as follows. As discussed in Appendix~\ref{appx:sa}, the initial temperature
in SA should be proportional to the largest absolute energy difference between
any two states capable of direct transition, denoted by $\Delta$.
Such a high initial temperature enhances exploration in the solution space.
The final temperature, in contrast, should be sufficiently low relative to
the energy gap between the optimal and the second optimal solutions,
denoted by $\delta$. This ensures that the Boltzmann factor, and consequently
the sampling probability, for the optimal solution(s) are sufficiently larger
than those for the others. A larger difference between the initial and final
temperatures, or equivalently between $\Delta$ and $\delta$, leads to longer
annealing time. In fact, according to an estimate for the computational
complexity of SA (Eq.~\ref{eq:computational-complexity-sa} in Appendix~\ref{appx:sa}),
the necessary annealing time exponentially increases with respect to
$\Delta/\delta$. In the present Ising model, $\Delta$ is $O(M_\mathrm{max})$
as the penalty terms dominate energy barrier heights. Thus, the computational
complexity is expected to depend on $M_\mathrm{max}/\delta$. This ratio
increases as $N_x$ increases, as shown in Fig.~\ref{fig:pcr}. This occurs because
$M_\mathrm{max}$ is $O(N_x)$ as already shown in
Fig.~\ref{fig:penalty-strength-scaling}, whereas the cost difference $\delta$
may not necessarily depend on the problem size. Therefore, the increase
of $M_\mathrm{max}$ is likely to cause the increase of $\Delta/\delta$,
leading to the rapid increase in the TTS of SA for $\rho=1$.

\begin{figure}
  \centering
  \includegraphics[width=1\linewidth]{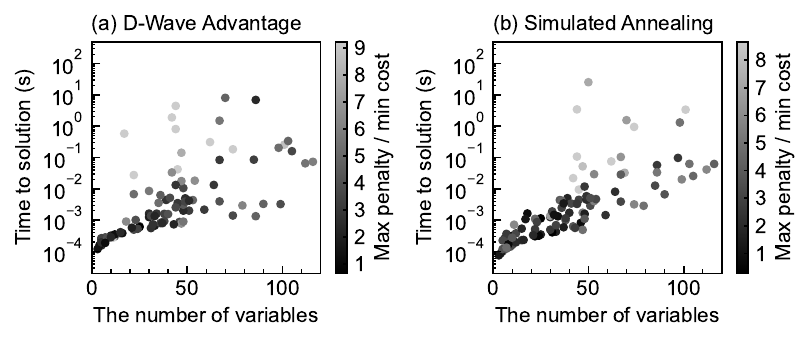}
  \caption{
    The relationship between the time-to-solution and the ratio of the maximum
    penalty strength to the minimum cost. Panels depict the time-to-solution
    with cost tolerance $\rho=2$ for the algorithms using (a) the D-Wave
    Advantage system and (b) simulated annealing, respectively.
    The time-to-solution data correspond to that in panel (b) in
    Fig.~\ref{fig:tts}. Each data point is colored according to the value
    of the ratio of the maximum penalty strength to the minimum cost.
    For clarity, outlier values (colors) of the penalty-cost ratio were
    treated using the following standard method based on the inter-quartile range:
    values outside the interval $[Q_1-1.5IQR, Q_3+1.5IQR]$ were truncated,
    where $Q_1$ and $Q_3$ represent the first and third quartiles, respectively,
    and $IQR(\coloneqq Q_3-Q_1)$ is the inter-quartile range; smaller outliers
    were set to the lower bound of the interval, while larger outliers were set
    to the upper bound.
  }
  \label{fig:tts-pcr}
\end{figure}

In the cases of $\rho=2$ and $3$, the relatively slow increasing rate of
the TTS of SA with respect to $N_x$ can be explained by a theoretical
expectation that the computational complexity of SA depends on
$(\rho-1)M_\mathrm{max}/C_\mathrm{min}$ instead of $M_\mathrm{max}/\delta$,
contrasting the case of $\rho=1$. This expectation is based on the following
arguments. The sampling probability of the optimal solution(s) is desired
to be sufficiently higher than those of non-allowable solutions whose costs
are greater than $\rho C_\mathrm{min}$ but not necessarily than those of
solutions within the cost tolerance range. This insight suggests that
the final temperature should be sufficiently low compared to the cost
difference between the optimal and the least-cost non-allowable solutions,
$(\rho-1)C_\mathrm{min}$, instead of $\delta$ in the case of $\rho=1$.
Therefore, based on a similar discussion presented in Appendix~\ref{appx:sa},
the necessary annealing time is expected to depend on
$(\rho-1)M_\mathrm{max}/C_\mathrm{min}$. In fact, as demonstrated in
Fig.~\ref{fig:tts-pcr}, problems with larger $M_\mathrm{max}/C_\mathrm{min}$
tend to have longer TTS for $\rho=2$. Furthermore, $M_\mathrm{max}/C_\mathrm{min}$
does not exhibit a clear increasing trend with respect to $N_x$,
as shown in Fig.~\ref{fig:pcr}. This observation is explained by the fact that
the minimum cost $C_\mathrm{min}$ also tends to increase almost linearly
as the problem size increases. Therefore, the reason for the moderate
increase in the TTS of SA for $\rho=2$ and $3$ is likely that
$M_\mathrm{max}/C_\mathrm{min}$ does not necessarily increase
with respect to $N_x$.

For the case of using the D-Wave Advantage system, the increasing trend of
the TTS with respect to $N_x$ can be partially explained in terms of
$M_\mathrm{max}/\delta$ and $M_\mathrm{max}/C_\mathrm{min}$, in a way
similar to the case of SA. According to an estimate for the computational
complexity of QA (Eq.~\ref{eq:computational-complexity-qa} in
Appendix~\ref{appx:qa}), the necessary annealing time required to find
the optimal solution(s), i.e., for the case of $\rho=1$, increases exponentially
as $\log(1/\delta^\prime)$ increases, where $\delta^\prime$ denotes the energy
gap between the ground and the first excited states of the physical Ising model.
When using the D-Wave Advantage system, the Ising Hamiltonian must be rescaled
by the maximum of $|J_{ij}|$ and $|h_i|$ due to their finite ranges.
This rescaling factor scales as $O(M_\mathrm{max})$. Therefore,
the factor $1/\delta^\prime$ in the necessary annealing time is
$O(M_\mathrm{max}/\delta)$ for $\rho=1$. Similar to the above arguments
for SA with $\rho>1$, the factor $1/\delta^\prime$ can be replaced by
$O(M_\mathrm{max}/C_\mathrm{min})$ for $\rho>1$. Thus, $M_\mathrm{max}/\delta$
and $M_\mathrm{max}/C_\mathrm{min}$ are likely important factors determining
the computational time scaling for QA as well as SA. In fact, Fig.~\ref{fig:tts-pcr}
demonstrates that problems with longer TTS tend to have larger
$M_\mathrm{max}/C_\mathrm{min}$ value, for problems with fewer than 50 variables.

However, despite the theoretical prediction that QA exhibits an advantage
over SA in terms of the computational complexity scaling with respect to
$M_\mathrm{max}/\delta$ or $M_\mathrm{max}/C_\mathrm{min}$
(see also Appendix~\ref{appx:qa}), the results presented in Fig.~\ref{fig:tts}
do not demonstrate an apparent quantum advantage. This gap between theory and
reality can be due to various factors, including device-specific
characteristics of the D-Wave Advantage system, which are further
discussed in the following paragraph.

Device-specific characteristics, such as the finite setting precision of
$J_{ij}$ and $h_i$ and embedding overheads, must impact the TTS for
the D-Wave Advantage system. First, the precision in setting $J_{ij}$ and $h_i$
is limited. Since the energy gap between the ground and the first excited
states of the rescaled Ising Hamiltonian is $O(\delta/M_\mathrm{max})$,
the energy gap can be smaller than the setting errors of $J_{ij}$ and $h_i$
for large-size problems, leading to poor performance of sampling the optimal
solution(s). Second, minor embedding necessitates more physical qubits
compared with the original logical Ising/QUBO formulation. As presented in
Appendix~\ref{appx:embedding-performance}, the necessary number of physical
qubits to represent a pathway-finding problem on the device increases almost
quadratically as the number of logical binary variables increases.
This embedding overhead contributes to the increased computation time of QA.
Lastly, other hardware characteristics, such as thermal noises and readout
errors, may also affect performance. These limitations specific to QA
hardware devices may nullify the theoretical quantum advantage in
the computational complexity scaling with respect to $M_\mathrm{max}/\delta$
or $M_\mathrm{max}/C_\mathrm{min}$.

\subsubsection{Computed synthesis pathways}

\begin{figure*}
    \centering
    \begin{minipage}[b]{0.49\linewidth}
        \centering
        \includegraphics[keepaspectratio, width=\linewidth]{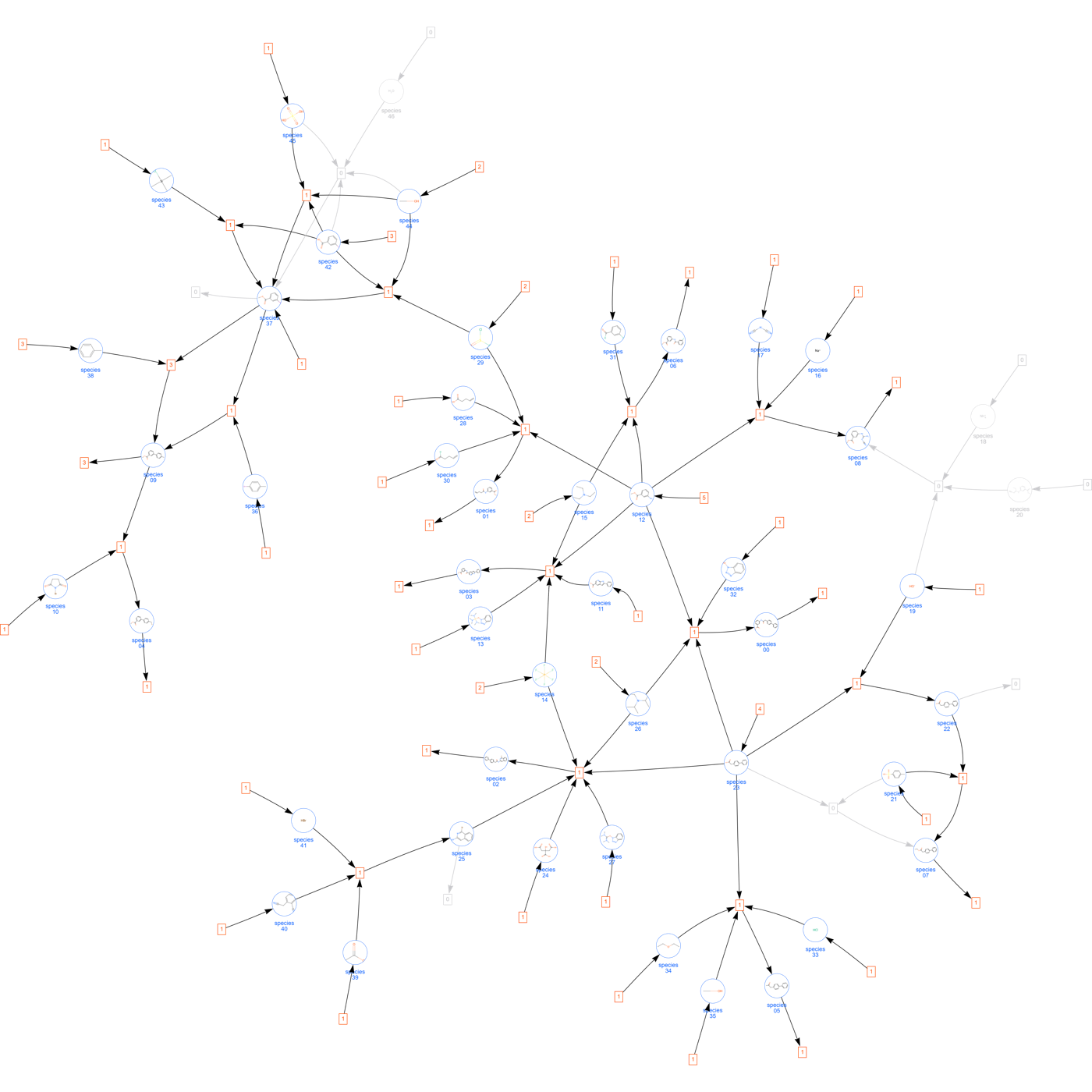}
        \subcaption{D-Wave Advantage}
    \end{minipage}
    \begin{minipage}[b]{0.49\linewidth}
        \centering
        \includegraphics[keepaspectratio, width=\linewidth]{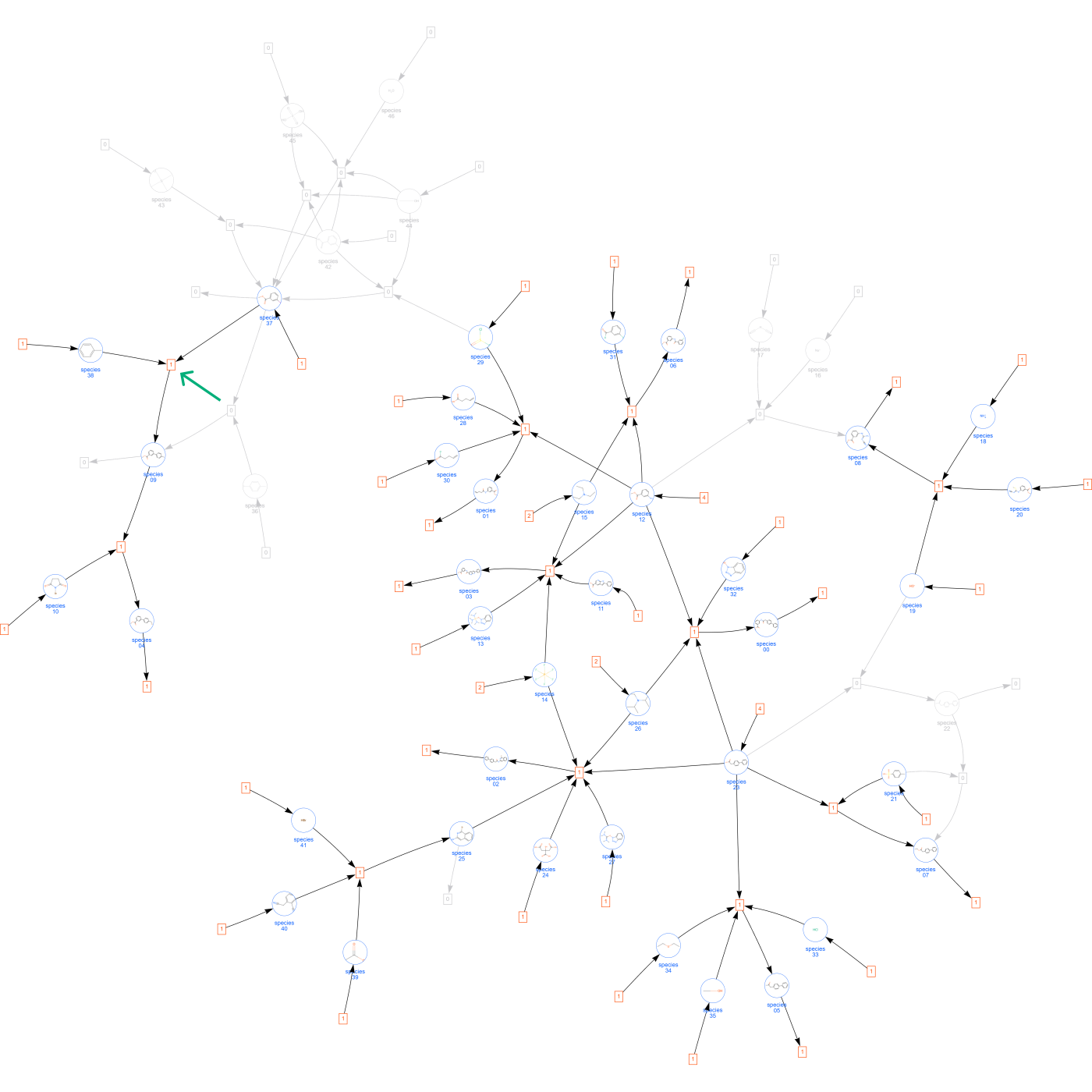}
        \subcaption{Simulated Annealing}
    \end{minipage}\\
    \begin{minipage}[b]{0.49\linewidth}
        \centering
        \includegraphics[keepaspectratio, width=\linewidth]{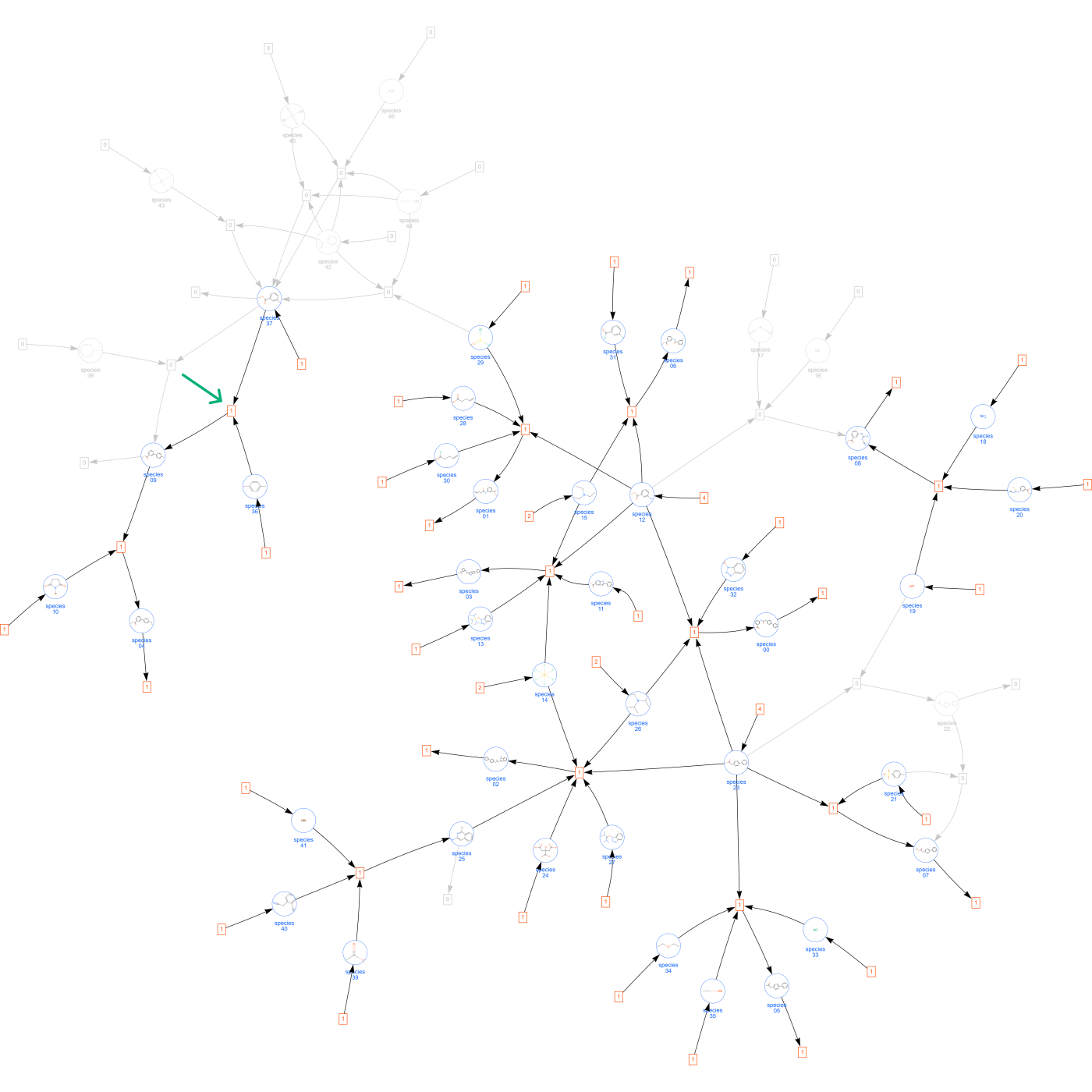}
        \subcaption{Gurobi Optimizer}
    \end{minipage}
    \caption{
        Synthesis pathways computed by (a) the D-Wave Advantage system,
        (b) simulated annealing, and (c) the Gurobi Optimizer.
        The problem solved is the 80th problem in ascending order of
        the number of integer variables (the same problem shown in panel (d)
        of Figs.~\ref{fig:encoding-grouping-performance-qa}
        and \ref{fig:encoding-grouping-performance-sa}).
        Circles and squares represent chemical species and reactions, respectively.
        Each reaction's multiplicity is indicated inside the reaction node.
        Colored parts correspond to computed synthesis pathways,
        whereas light gray parts are not included in these pathways;
        in other words, the multiplicity of each light gray reaction is zero.
        The green arrows in the upper left area of panels (b) and (c) indicate
        different choices of alternative reactions in these two pathways.
        The costs of the three synthesis pathways are (a) 372.7, (b) 251.4,
        and (c) 250.0, respectively.
    }
    \label{fig:pathway-80th}
\end{figure*}

\begin{figure*}
    \centering
    \begin{minipage}[b]{0.49\linewidth}
        \centering
        \includegraphics[keepaspectratio, width=\linewidth]{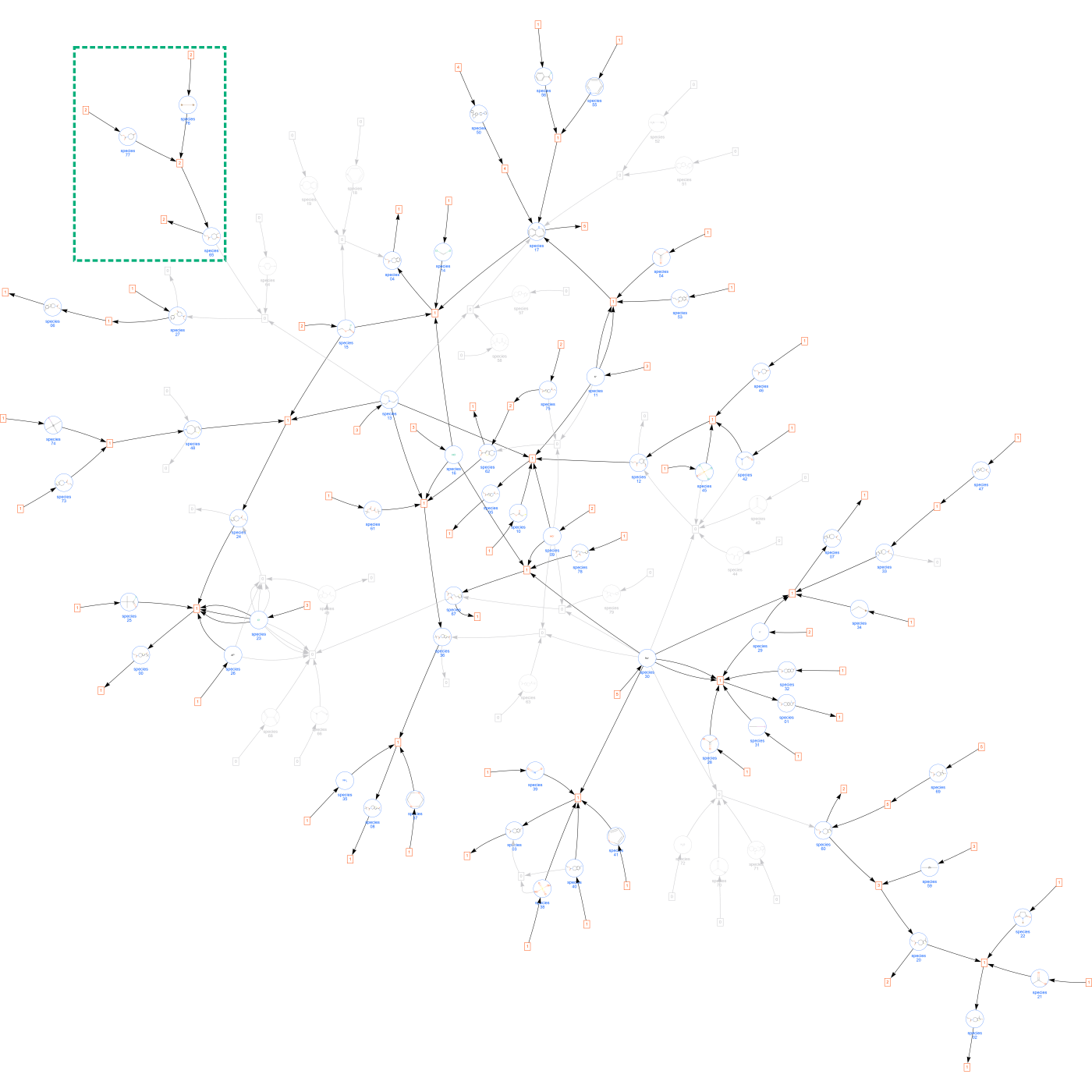}
        \subcaption{D-Wave Advantage}
    \end{minipage}
    \begin{minipage}[b]{0.49\linewidth}
        \centering
        \includegraphics[keepaspectratio, width=\linewidth]{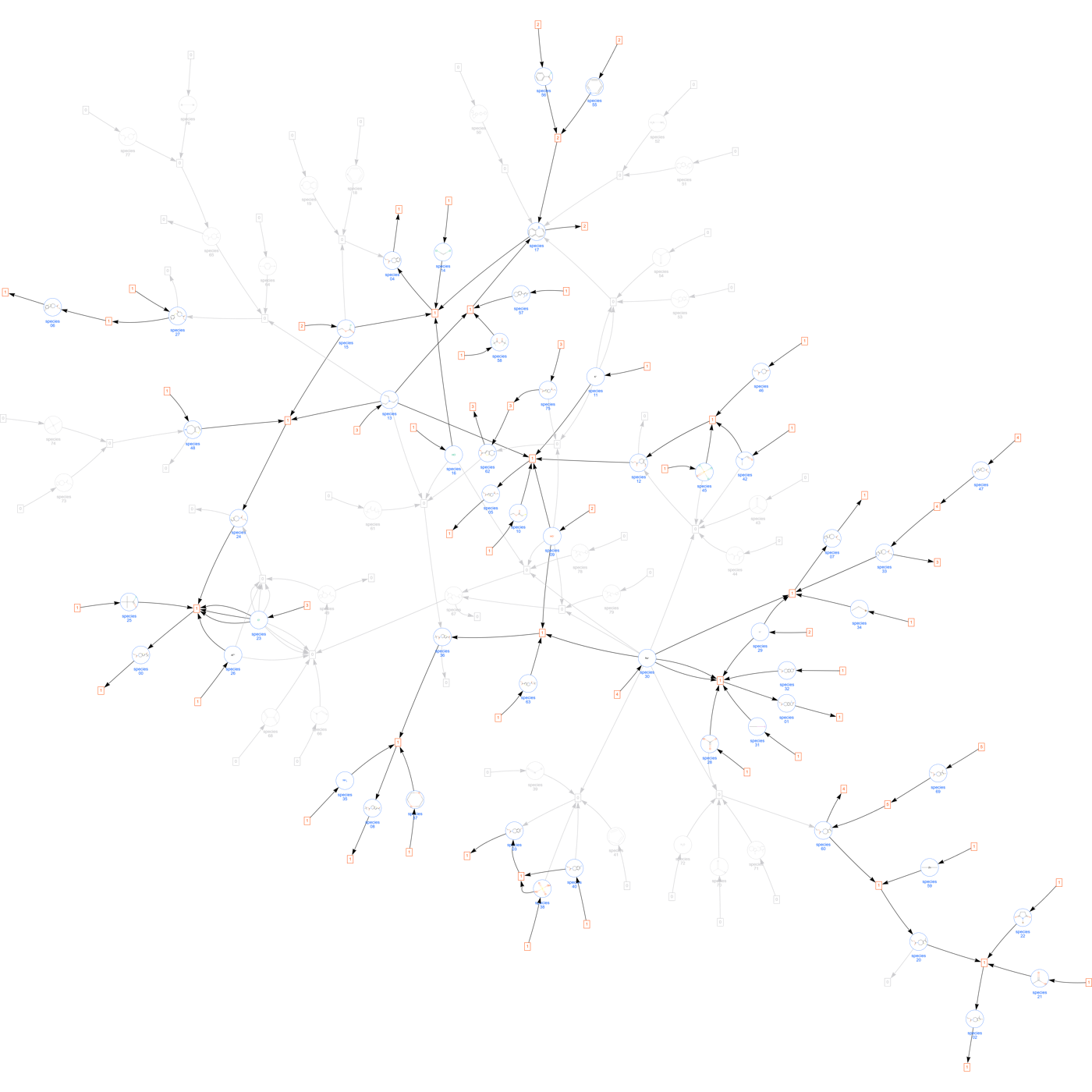}
        \subcaption{Simulated Annealing}
    \end{minipage}\\
    \begin{minipage}[b]{0.49\linewidth}
        \centering
        \includegraphics[keepaspectratio, width=\linewidth]{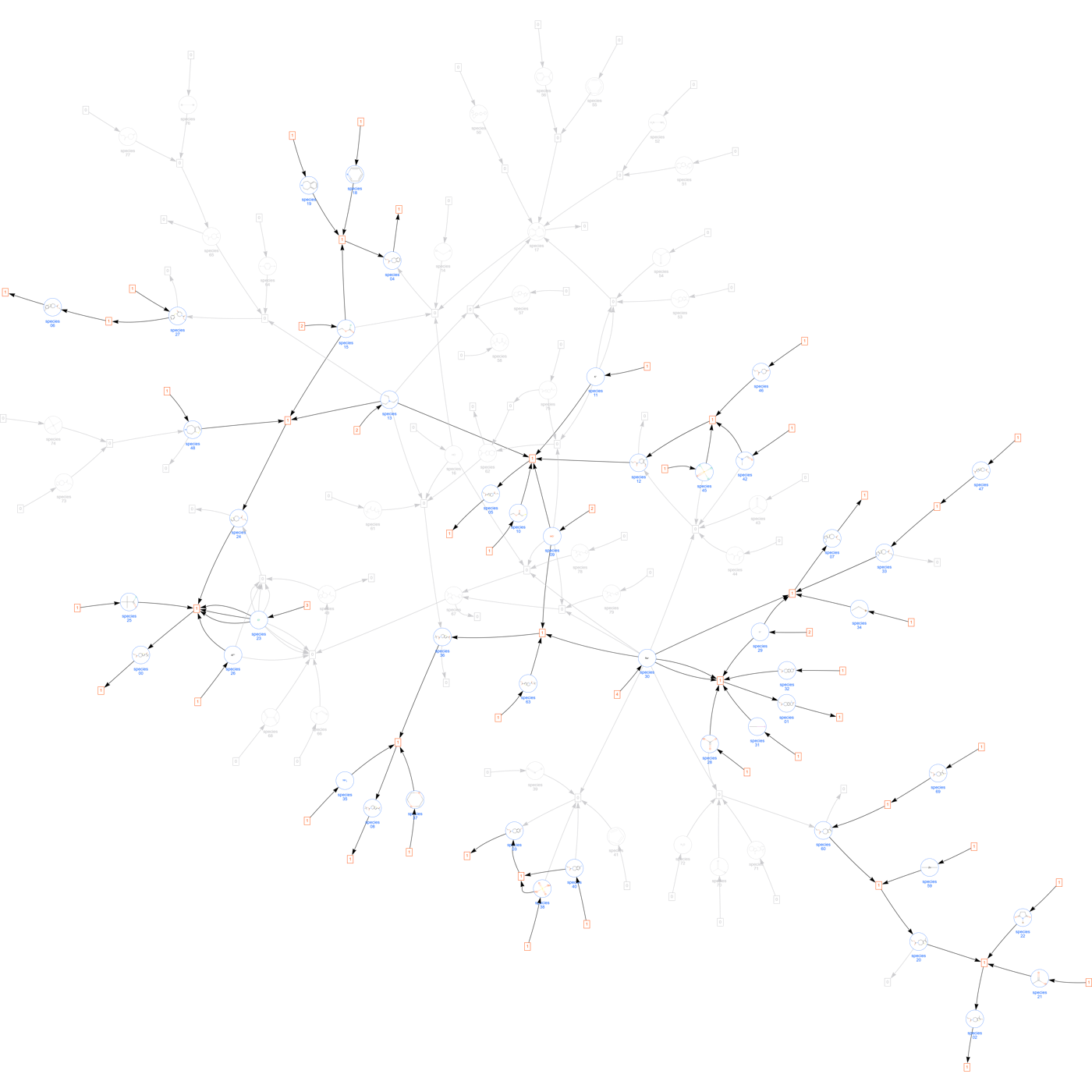}
        \subcaption{Gurobi Optimizer}
    \end{minipage}
    \caption{
        Synthesis pathways computed by (a) the D-Wave Advantage system,
        (b) simulated annealing, and (c) the Gurobi Optimizer.
        The problem solved is the 100th problem in ascending order of
        the number of integer variables (the same problem shown in panel (e)
        of Figs.~\ref{fig:encoding-grouping-performance-qa}
        and \ref{fig:encoding-grouping-performance-sa}).
        The green frame in the upper left area of panel (a) encloses
        a sub-pathway that produces an unnecessary precursor.
        See the caption of Fig.~\ref{fig:pathway-80th} for the details of
        other symbols' meaning. The costs of the three synthesis pathways
        are (a) 552.1, (b) 471.4, and (c) 328.3, respectively.
    }
    \label{fig:pathway-100th}
\end{figure*}

Finally, we compare synthesis pathways computed by the D-Wave Advantage system,
SA, and the Gurobi Optimizer. Figs.~\ref{fig:pathway-80th} and \ref{fig:pathway-100th}
depict synthesis pathways computed for two representative problems shown in
panels (d) and (e) of Figs.~\ref{fig:encoding-grouping-performance-qa} and
\ref{fig:encoding-grouping-performance-sa}, that is, the 80th and 100th problems
in ascending order of the number of integer variables, respectively.
In this computation, we employed the order encoding method and the penalty
and chain strength parameters optimized using the category grouping method.
The annealing time for the D-Wave Advantage system and the number of sweeps
for SA were set to their default values, i.e., 20 \si{\micro s} and 1000,
respectively. We ran the proposed Ising-computing algorithm 1000 times for
each problem and each Ising machine. The feasible synthesis pathways
at the minimum cost among the 1000 solutions for each problem and
each Ising machine are shown in panels (a) and (b) of
Figs.~\ref{fig:pathway-80th} and \ref{fig:pathway-100th}.
The synthesis pathways computed by the Gurobi Optimizer shown in panel (c)
of these figures are the exact optimal synthesis pathways for each problem.

The synthesis pathways computed by the D-Wave Advantage system and SA differ
from the optimal synthesis pathways computed by the Gurobi Optimizer
in the following two aspects. First, it has been observed that synthesis
pathways computed by the D-Wave Advantage system and SA make sub-optimal
choices when alternative reactions are available, in certain cases.
Pathways in panels (b) and (c) of Fig.~\ref{fig:pathway-80th} exemplify
such a situation; these two pathways differ only in the choice of reactions
indicated by the green arrows in the upper left area. Second, there are
instances where synthesis pathways computed by the D-Wave Advantage system
and SA include synthesis sub-pathways which produce precursors not used for
synthesizing any target. Panel (a) of Fig.~\ref{fig:pathway-100th} clearly
exemplifies such a situation; the sub-pathway enclosed by the green frame
in the upper left area is disconnected with the other part and is not included
in the optimal pathway; this sub-pathway produces a precursor not used for
synthesizing any target. In general, such unnecessary sub-pathways are not
always separated from main pathways and often overlap with them.

These differences between the optimal pathways and those computed by the D-Wave
Advantage system and SA may be attributed to the large penalty strength issue
and device-specific noise and errors mentioned above. When the penalty strength
is high, constraint satisfaction is more prioritized than cost minimization,
often leading to feasible but sub-optimal solutions. If the multiplicity of
an unnecessary reaction in a chemical reaction network becomes positive
due to device-specific noise or errors, the postprocessing methods may
add additional reactions to make the returned solution feasible,
which may form an unnecessary sub-pathway.

It may be possible to address the issue of unnecessary sub-pathways through
the following postprocessing steps: (1) detecting disposed precursors;
(2) finding sub-pathways that produces the excess precursors at maximum cost
in the original pathway, using the proposed pathway-finding algorithm;
(3) subtracting the sub-pathways---which are likely unnecessary---from
the original pathway. Unnecessary sub-pathways may also be suppressed by
imposing additional penalties for precursors disposal in the QUBO formulation.
Introducing these treatments, however, necessitates careful consideration
as precursor disposal does not necessarily indicate wasteful pathways and
economical reactions may produce precursors as byproducts in some cases.
Methods to remove or suppress unnecessary sub-pathways remain open
for further investigation.

\subsubsection{Other examinations}

Additionally, we confirmed that the BO-based parameter tuning used in
the present study contributes to performance improvement,
outperforming random search (see Appendix~\ref{appx:tuning-performance}).
An examination into the postprocessing methods showed that
the postprocessing methods indeed enhance the algorithm's performance
(see Appendix~\ref{appx:postprocessing-performance}).

\section{Conclusions} \label{sec:conclusion}

We have developed an algorithm for finding optimal pathways in chemical
reaction networks using Ising machines. A pathway-finding problem
in a chemical reaction network is formulated as a combinatorial optimization
problem that involves integer variables representing reaction multiplicity
and mass balance constraints corresponding to the law of conservation of mass.
The proposed algorithm translates a chemical pathway-finding problem into
the ground-state finding problem of an Ising model and solves the translated
problem using an Ising machine. In addition, the proposed algorithm applies
two postprocessing methods to enhance the feasibility and optimality of
solutions returned from the Ising machine.

We utilized Bayesian optimization to tune parameters determining penalty
strengths for constraint violations. To enhance tuning performance,
we have devised a dimensionality reduction technique for the parameter
search space. In this technique, parameters are grouped together
according to the associated constraint type, and all parameters in each
group are constrained to take the same value. We systematically elucidated
the effects of several different parameter grouping methods on tuning
performance. We found that it is crucial to design parameter groups
tailored according to the specific structure of problems for performance
improvement. Most Ising/QUBO formulations do not consider such elaborate
parameter grouping; instead, all penalty strengths of the same type are
often consolidated into one group, like our `unified' method (for instance,
see \cite{Lucas2014} and applications reviewed in \cite{Yarkoni2022}).
Our findings may inform the future development of automatic parameter
tuning methods for complex constrained problems in Ising computing.

Our performance evaluation and analysis of the proposed algorithm using
the D-Wave Advantage system and simulated annealing indicate that:
(1) the computation time required to find optimal pathways rapidly increases
as the problem size increases, likely due to the growth of penalty strengths
with respect to the problem size and hardware limitations of quantum annealing
devices; (2) the scaling of the computation time can be moderated if a relative
error in a cost value is allowable and if the maximum penalty strength to
the minimum cost hardly increases with respect to the problem size.
These findings provide insights into future directions of the algorithm
development for chemical pathway-finding problems. 

In the development of chemical pathway-finding algorithms utilizing
the penalty-based Ising/QUBO formulation, it might be advisable for
the algorithms to aim at finding approximate solutions within a relative
cost error tolerance. In addition, appropriate applications may be problems
with a moderate ratio of the maximum penalty strength to the minimum cost.
For further performance improvement, the advancement of parameter tuning
methods is needed. For instance, annealing schedule optimization
\cite{Herr2017, Chen2022} likely reduces the required annealing time,
and transfer learning techniques \cite{Chen2022} may reduce the overhead
runtime for parameter tuning by leveraging tuning results for other
problem instances. Hardware improvements, especially reducing embedding
overheads, are also necessary. Furthermore, a method presented in \cite{Ohzeki2020}
may mitigate the challenges in the penalty-based formulation by transforming
quadratic penalty terms into linear terms using the Hubbard--Stratonovich
transformation.

An alternative approach is to apply penalty-free quantum optimization
algorithms, such as constrained quantum annealing \cite{Hen2016} and
quantum alternating operator ansatz (QAOA) \cite{Hadfield2019}, which may help
bypass the penalty-related issues. However, these algorithms introduce their
own challenges, such as the need for a quantum processing unit capable of
handling complex qubit interactions like XX and YY interactions and
the difficulty of designing an effective driver Hamiltonian \cite{Leipold2021}.
Despite these technical difficulties, advancements in algorithms along
this direction hold promise for resolving the challenges identified
in the present study when solving chemical pathway-finding problems
using Ising machines.

Furthermore, extending the scope of the chemical pathway-finding problems
targeted by the Ising computing framework is a future challenge. For example,
a chemical pathway-finding problem may involve continuous variables
representing real-valued reaction multiplicity in moles. Real-valued reaction
multiplicity is necessary to be considered in synthesis planning when taking
into account reaction yields because the optimal multiplicity may differ from
a theoretical integer value based on stoichiometry. In such a case, the problem
can be generally formulated as mixed integer programming. For another instance,
a chemical pathway-finding problem can be formulated as multi-objective
optimization involving multiple objectives such as low monetary costs and
low harmfulness \cite{Szymkuc2016}. Moreover, chemists are sometimes interested in
enumerating near-optimal pathways or even all feasible pathways
\cite{Andersen2017, Shibukawa2020}.

Several Ising/quantum-computing algorithms aim at solving mixed integer
programming \cite{Zhao2022, Ajagekar2022}, multi-objective optimization
\cite{Baran2016}, and exhaustive enumeration \cite{Kumar2020, Mizuno2021}.
These algorithms may unlock further potential in chemical reaction network
analysis using Ising machines. Note also that all chemical pathway-finding
problems inherently involve mass balance constraints due to the law of
conservation of mass; hence, our methods and findings related to penalty
strength for mass balance constraints are likely broadly applicable
across these problems.

We hope this study will serve as a starting point for future interdisciplinary
technological advancements in chemical reaction network analysis using
next-generation computers such as Ising machines and quantum computers.

\begin{acknowledgments}
  This work was supported by JST, PRESTO Grant Number JPMJPR2018, Japan.
\end{acknowledgments}

\appendix

\section{Chemical Reaction Network Underlying a Synthesis-Planning Problem}
\label{appx:synthesis-relevant-network}

\begin{figure*}
  \centering
  \includegraphics[width=\linewidth]{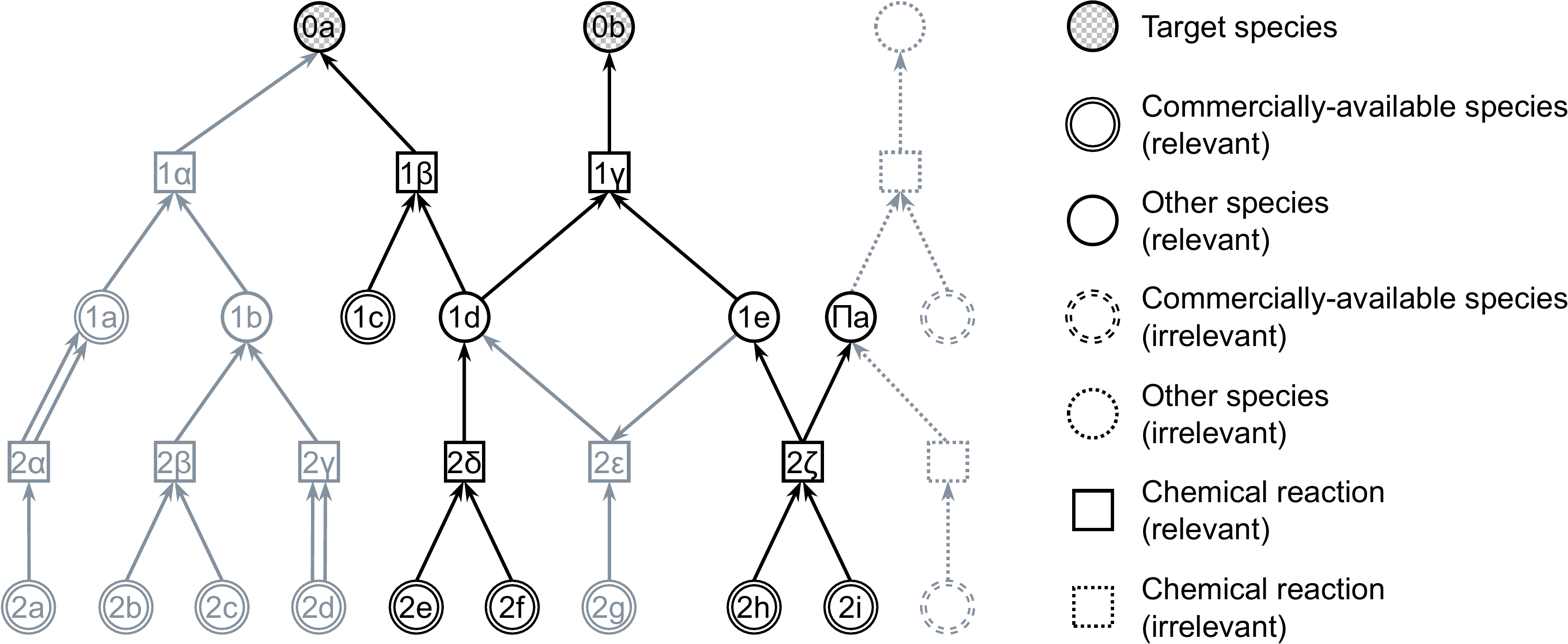}
  \caption{
    Chemical reactions and species relevant and irrelevant to the synthesis
    planning of specified targets. Relevant chemical reactions (solid squares)
    are candidates for synthesis steps because they yield one of the potential
    precursors for the targets. On the other hand, irrelevant chemical reactions
    (dashed square) are never used for synthesizing the targets. Relevant
    chemical species (solid circles) are defined as those participating in
    relevant chemical reactions. Note that not all relevant chemical reactions
    are always used in chemical synthesis. For instance, four chemical reactions,
    namely 1$\beta$, 1$\gamma$, 2$\delta$, and 2$\zeta$, constitute a feasible
    synthesis pathway, indicated in dark black in the figure. Determining the optimal
    combination of the candidate reactions is central to the synthesis-planning problem.
  }
  \label{fig:synthesis-relevant-network}
\end{figure*}

A synthesis-planning problem to find the optimal synthesis pathway
toward specified targets does not require consideration of all chemical
reactions in chemical databases such as Reaxys. As illustrated in
Fig.~\ref{fig:synthesis-relevant-network}, chemical reactions can be classified
into two types: (1) chemical reactions \textit{relevant} to synthesizing the targets,
which should be taken into account in the synthesis planning, and (2) other
\textit{irrelevant} chemical reactions. Relevant chemical reactions are recursively
defined as follows:

\begin{enumerate}
  \item All chemical reactions that directly produce target species are relevant
        to the synthesis planning.
  \item If a chemical reaction produces any potential precursor for the target
        species, that is, a reactant of another relevant reaction, the chemical
        reaction is relevant to the synthesis planning.
\end{enumerate}

We also define \textit{relevant} chemical species as chemical species participating
in the relevant chemical reactions. Such species can be either targets,
precursors, or byproducts. In Fig.~\ref{fig:synthesis-relevant-network},
species 0a and 0b are targets, species 1a–1e and 2a–2i are precursors,
and species $\Pi$a is a byproduct. Here, the number in the label of a target
or precursor node indicates the \textit{depth} of the species node. The node depth
is defined as the minimum number of synthesis steps from the species node to
any target node. The depth of a byproduct node is undefined.

\begin{figure}
  \centering
  \includegraphics[width=\linewidth]{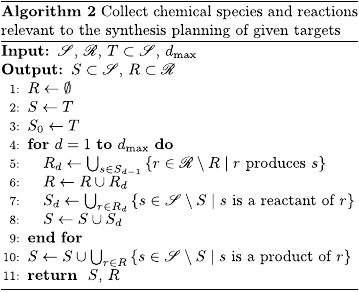}
  \caption{
    An algorithm for collecting relevant chemical species and reactions.
    The input arguments are a set of chemical species, $\mathscr{S}$,
    a set of chemical reactions, $\mathscr{R}$, a set of targets, $T$,
    and the maximum search depth, $d_\mathrm{max}$. This algorithm
    iteratively enumerates relevant chemical species and reactions
    up to depth $d_\mathrm{max}$, then returns the sets $S$ and $R$
    of the collected species and reactions. Note that the maximum
    recursive depth, $d_\mathrm{max}$, does not exceed the number of
    all chemical reactions, $|\mathscr{R}|$. For example,
    in Fig.~\ref{fig:synthesis-relevant-network}, a square node with
    a label starting with integer $d$ represents a relevant reaction
    in $R_d$ at line 5 of the code, and a circle node with a label
    starting with integer $d$ signifies a relevant species in $S_d$
    at lines 3 and 7. In addition, the byproduct species `$\Pi$a' is
    added to $S$ at line 10.
  }
  \label{fig:relevant-network-search-algorithm}
\end{figure}

The above recursive definition establishes an algorithm for collecting
chemical species and reactions relevant to the synthesis planning
(Fig.~\ref{fig:relevant-network-search-algorithm}). This algorithm
significantly reduces the number of chemical reactions to be considered
compared to that of the whole dataset.

In addition to the relevant chemical reactions enumerated by Algorithm 2
 (Fig.~\ref{fig:relevant-network-search-algorithm}), the inflow and outflow
dummy reactions associated with the relevant species are also relevant
to the synthesis planning. In synthesis planning, only commercially available
species have their inflow reactions. All target species must have outflow
reactions; others may also have outflow reactions representing disposal.
Note that inflow reactions of purchasable relevant byproducts and outflow
reactions of relevant precursors never produced by chemical reactions
in the database (e.g., species 1c in Fig.~\ref{fig:synthesis-relevant-network})
can be omitted because the purchase of byproducts and disposal of purchased
substrates are not economical choices.

\section{Ising Machines and Minor Embedding} \label{appx:Ising-machine}

Ising machines are dedicated computing devices for finding the ground state(s)
of Ising models \cite{Mohseni2022}. Major algorithms underlying Ising machines are
simulated annealing \cite{Kirkpatrick1983} and quantum annealing \cite{Kadowaki1998}.
This appendix surveys the simulated and quantum annealing algorithms.
Moreover, we also explain the minor embedding required for using Ising machines
with limited spin connectivity.

\subsection{Simulated annealing} \label{appx:sa}

Simulated annealing (SA) is a physics-inspired heuristic algorithm for
optimization, proposed by Kirkpatrick et al. in 1983 \cite{Kirkpatrick1983}.
SA simulates a thermal annealing process in a statistical mechanics system
by a Markov chain Monte Carlo method. The system is designed such that
its lowest energy state(s) correspond to the optimal solution(s) of
an optimization problem to be solved. In the annealing process,
the temperature of the system gradually decreases. If the cooling process
is slow enough, the system is expected to remain in thermal equilibrium.
Since the Gibbs distribution assigns a high probability to the lowest
energy state(s) at sufficiently low temperatures, the system is likely
to reach the lowest energy state(s) after annealing.

The annealing schedule, which dictates the rate of temperature decrease,
determines the success probability of sampling the lowest energy state(s).
Geman and Geman \cite{Geman1984} proved that the sampling-probability distribution
in SA converges to the uniform distribution on the lowest energy state(s)
as $t \to \infty$ if the temperature decreases in time as
\begin{equation}
T(t) = \frac{N_\sigma\Delta}{\log t} \quad (t>t_0).
\label{eq:annealing-schedule-sa}
\end{equation}
Here, $t$ denotes time (Monte Carlo step), $t_0 (>1)$ is a constant,
$T(t)$ is the temperature at time $t$, $N_\sigma$ is the number of spin
variables, and $\Delta$ is the largest absolute energy difference between
any two states such that the system can directly transition between them
in a single step of the Markov chain.

The computational complexity of SA with the inverse-logarithmic annealing
schedule (Eq.~\ref{eq:annealing-schedule-sa}) is estimated as follows. Assume that
the probability distribution stays close to the Gibbs distribution during
the annealing process. Let $E_0$ and $E_1$ denote the lowest and the second
lowest energies, respectively. The temperature at final time $\tau$ needs
to be low enough compared with the energy gap $\delta\ (\coloneqq E_1-E_0)$;
this is because the lowest energy state(s) should have a sufficiently larger
Boltzmann factor than the second lowest energy state(s):
\begin{equation}
\begin{alignedat}{2}
&\frac{\mathrm{e}^{-\frac{E_0}{T(\tau)}}}{\mathrm{e}^{-\frac{E_1}{T(\tau)}}}
&&= \mathrm{e}^{\frac{\delta}{T(\tau)}} \gg 1 \\
\Rightarrow \quad &T(\tau) &&= \frac{N_\sigma\Delta}{\log \tau} \ll \delta.
\end{alignedat}
\end{equation}
This suggests that the necessary annealing time in SA scales as
\begin{equation}
\tau \sim \exp\left(c\frac{N_\sigma\Delta}{\delta}\right),
\label{eq:computational-complexity-sa}
\end{equation}
where $c$ is a positive constant of $O(N_\sigma^0)$.
This computational complexity is consistent with that implied
by a theoretical result in \cite{Mitra1986}:
the total-variation distance between the sampling probability distribution
in SA with the annealing schedule given by Eq.~\ref{eq:annealing-schedule-sa}
and the Gibbs distribution at absolute zero temperature is upper bounded by
$O((t/N_\sigma)^{-\delta/N_\sigma\Delta})$.

Note that most SA algorithms do not employ the inverse-logarithmic schedule
due to impractical slowness in real-world applications. Practical implementations
of SA utilize other fast cooling schedules, such as the geometric schedule
$T(t) = T(0) r_T^t \ (0< r_T < 1)$, to find approximate optimal solutions.
For instance, the SA algorithm implemented in the D-Wave Ocean Software
\cite{DWaveOcean} employs the geometric annealing schedule as default,
which was also used in our benchmarking in Sec.~\ref{sec:evaluation}.
Thus, we reference the theoretical computational complexity given by
Eq.~\ref{eq:computational-complexity-sa} merely as a coarse measure to help
interpret the observed computation time scaling of SA with respect to
the problem size in Sec.~\ref{sec:evaluation}.

\subsection{Quantum annealing} \label{appx:qa}

Quantum annealing (QA) is a heuristic quantum optimization algorithm inspired
by SA. QA for Ising models was first proposed by Kadowaki and Nishimori
in 1998 \cite{Kadowaki1998}. The QA algorithm utilizes the time evolution of
a quantum system with the time-dependent Hamiltonian
\begin{equation}
\begin{alignedat}{1}
\hat{H}_\mathrm{QA}(t)
= &-A(t) \sum_{i=1}^{N_\sigma} \hat{\sigma}^x_i \\
  &-B(t) \left[\sum_{i=1}^{N_\sigma-1} \sum_{j=i+1}^{N_\sigma} J_{ij}
              \hat{\sigma}^z_i \hat{\sigma}^z_j +
              \sum_{i=1}^{N_\sigma} h_i \hat{\sigma}^z_i \right],
\end{alignedat}
\label{eq:ising-model-qa}
\end{equation}
from time $0$ to $\tau$. Here, $A(t)$ and $B(t)$ are monotonic functions
such that $A(0)=1, B(0)=0$ and $A(\tau)=0, B(\tau)=1$, and $\hat{\sigma}^x_i$
and $\hat{\sigma}^z_i$ are the Pauli $x$ and $z$ operators acting on the $i$-th
qubit, respectively. The system Hamiltonian varies with time from
the first term (the transverse field Hamiltonian) to the second term
(the quantum Ising Hamiltonian). In the computation, the quantum system
is initialized to the ground state of the initial Hamiltonian,
$\left|+\right\rangle^{\otimes N_\sigma}$, where
$\left|+\right\rangle = \left(
  \left|\uparrow\right\rangle + \left|\downarrow\right\rangle
\right) / \sqrt{2}$, and $\left|\uparrow\right\rangle$ and
$\left|\downarrow\right\rangle$ are the eigenstates of the Pauli $z$ operator
with eigenvalues $1$ and $-1$, respectively. According to the quantum
adiabatic theorem, if the system Hamiltonian varies slowly enough,
the quantum state is expected to stay close to the instantaneous ground state.
Therefore, after the adiabatic time evolution of the quantum system,
it reaches a quantum state close to the ground state of the Ising Hamiltonian
at time $\tau$, allowing us to observe the lowest-energy spin configuration(s)
with a high probability.

A convergence condition of QA is known for the following transverse-field
Ising model:
\begin{equation}
\hat{H^\prime}_\mathrm{QA}(t) =
-\Gamma(t)\sum_{i=1}^{N_\sigma} \hat{\sigma}^x_i
-\sum_{i=1}^{N_\sigma-1} \sum_{j=i+1}^{N_\sigma}
 J_{ij} \hat{\sigma}^z_i \hat{\sigma}^z_j
-\sum_{i=1}^{N_\sigma} h_i \hat{\sigma}^z_i.
\end{equation}
This Hamiltonian is related to the Hamiltonian in Eq.~\ref{eq:ising-model-qa}
as $\hat{H^\prime}_\mathrm{QA}(t) = \hat{H}_\mathrm{QA}(t)/B(t)$.
The function $\Gamma(t) \ (= A(t)/B(t))$ controls the magnitude of
quantum fluctuation induced by the transverse field. Morita and Nishimori
\cite{Morita2007, Morita2008} proved that the excitation probability is
bounded by an arbitrarily small constant $\epsilon$ at each time
if $\Gamma(t)$ decreases in time as
\begin{equation}
\Gamma(t) = a(\sqrt{\epsilon} t + b)^{-\frac{1}{2N_\sigma-1}} \quad (t>t_0),
\label{eq:annealing-schedule-qa}
\end{equation}
where $a$ and $b$ are constants of $O(N_\sigma^0)$
and $t_0$ is a positive constant.

The computational complexity of QA with the power-law annealing schedule
(Eq.~\ref{eq:annealing-schedule-qa}) is estimated as follows. At the final
time $\tau$, the perturbative transverse field needs to be small enough
compared with the energy gap $\delta$ between the ground and the first
excited states of the non-perturbed Ising model. This is because
the contributions of excited states of the non-perturbed Ising model
to the perturbed ground state should be sufficiently small:
\begin{equation}
\begin{alignedat}{1}
\left|
\frac{\langle k| \Gamma(\tau) \sum_i \hat{\sigma}^x_i | 0 \rangle}{E_k-E_0}
\right|
&\propto \frac{(\sqrt{\epsilon}\tau + b)^{-\frac{1}{2N_\sigma-1}}}{E_k-E_0}
 \quad (k \ne 0)\\
&\le \frac{(\sqrt{\epsilon}\tau + b)^{-\frac{1}{2N_\sigma-1}}}{\delta} \ll 1,
\end{alignedat}
\end{equation}
where $|k\rangle$ denotes the $(k+1)$-th lowest energy eigenstate of
the non-perturbed Ising model with energy $E_k$. This suggests
that the necessary annealing time in QA scales as
\begin{equation}
\tau \sim
\exp\left[(2N_\sigma-1)\log\left(\frac{c^\prime}{\delta}\right)\right],
\label{eq:computational-complexity-qa}
\end{equation}
where $c^\prime$ is a positive constant. Therefore, QA exhibits
an advantage over SA in terms of the computational complexity scaling
with respect to $\delta$.

Note that D-Wave quantum annealing devices are based on the Hamiltonian
defined by Eq.~\ref{eq:ising-model-qa} and do not employ the power-law
annealing schedule given by Eq.~\ref{eq:annealing-schedule-qa}.
Thus, the theoretical computational complexity given by
Eq.~\ref{eq:computational-complexity-qa} is merely a coarse measure to
help interpret the observed computation time scaling of QA with respect to
the problem size. In addition, due to minor embedding, the necessary number
of qubits, $N_\sigma$, is often greater than the number of spins of
an Ising model that one wants to solve. We explain the minor embedding
in the next subsection.

\subsection{Minor embedding} \label{appx:embedding}

Due to physical hardware topology, some Ising machines, including D-Wave
quantum annealing devices, can only solve Ising models with limited spin
interactions. Therefore, one needs preprocessing to represent
the \textit{logical} Ising model one wants to solve by the \textit{physical}
Ising model implemented in hardware.

The mapping from the logical model to the physical model is called minor
embedding \cite{Cai2014}. In the embedding, a collection of multiple physical
spins, termed a \textit{chain}, represents a single logical spin. To ensure that
the physical spins in a chain collectively behave as a single logical
variable, a constraint that all spins in the chain take the same value
is imposed. This constraint can be physically implemented by strong
ferromagnetic interactions ($J_{ij} \gg 1$) between adjacent spins
in the chain. The strength of the ferromagnetic interactions is called
\textit{chain strength}. In addition, the embedding ensures that if logical spin
variables $\sigma$ and $\sigma^\prime$ interact, at least one physical spin
in the chain of $\sigma$ and one spin in the chain of $\sigma^\prime$ interact.

Finding minor embedding is an NP-hard problem. However, several heuristic
algorithms exist for finding embeddings \cite{Cai2014, Sugie2021}.
The algorithm proposed in \cite{Cai2014} is available in the \texttt{minorminer}
package of the D-Wave Ocean Software \cite{DWaveOcean}.

Physical spins in a chain often take different values. In such a case,
postprocessing is performed on a classical computer, assigning the most common
physical spin value in the chain to the corresponding logical spin variable.
The probability of such chain breaks mainly depends on the chain strength.
Therefore, careful tuning of the chain strength is necessary to obtain
feasible, low-cost solutions. Sec.~\ref{sec:tuning} describes chain strength
tuning in detail.

\section{Benchmark Problem Generation} \label{appx:benchmark-problems}

The benchmark problems used in Sec.~\ref{sec:evaluation} were generated as follows.

First, we constructed a chemical reaction network from the USPTO dataset,
a dataset of chemical reactions extracted from the United States patents
published between 1976 and September 2016 \cite{Lowe2017}. The dataset
contains invalid entries, such as species with invalid chemical structures and
chemical reactions with no reactants/products; we removed these invalid
reactions from the network. In addition, we ignored catalysis information.
The numbers of chemical reactions and species in the whole network of
the USPTO dataset are around 1.1 and 1.5 million, respectively.

Second, we defined a set of commercially available substrates
and a set of target candidates. Instead of referring to actual
chemical makers' catalogs, we generated a set of commercially available
substrates and a set of target candidates based solely on the USPTO network
structure as follows:
(1) We identified species with an in-degree or out-degree of 20 or more
as \textit{hub}. We assumed that a hub species has an established standard way
of purchasing or synthesizing it because it can be regarded as popular
in chemistry. In other words, we assumed that the hub species are virtually
commercially available and that finding synthesis pathways to them is
unnecessary. Thus, we omitted deeper searches from hub species
in Algorithm 2 (Fig.~\ref{fig:relevant-network-search-algorithm}).
This treatment prevents a synthesis planning problem from becoming too large
and complex to solve by the D-Wave Advantage machine.
(2) We classified non-hub species into \textit{source} (with zero in-degree),
\textit{sink} (with zero out-degree), and \textit{non-terminal} (others).
(3) We assumed that all source species are commercially available
because source species are always used as starting materials in the dataset.
(4) Non-terminal species may be either commercially available or not;
we randomly assigned 25\% of non-terminal species as commercially available
substrates.
(5) We designated all commercially unavailable species to target candidates.
These target candidates comprise the sink species and the commercially
unavailable non-terminal species.

Third, we randomly determined a unit purchase price of each commercially
available species and a fixed execution cost for each chemical reaction.
In our experiment, we used uniform distribution ranging from 1 to 10
for the random cost assignment.

Fourth, we randomly selected multiple targets for each synthesis planning:
The number of targets was determined at uniformly random between 1 to 10;
we then randomly selected a set of multiple species with high Tanimoto
similarity \cite{Gasteiger2003} based on Morgan fingerprints
\cite{Rogers2010, RDKit-Morgan} from the set of target candidates.
Synthesis plans for multiple similar species can save monetary costs
by sharing precursors and reactions \cite{Kowalik2012}.

Fifth, we extracted a subnetwork underlying each synthesis planning
from the USPTO chemical reaction network. Algorithm 2
(Fig.~\ref{fig:relevant-network-search-algorithm}) identified the relevant
chemical species and reactions for each synthesis. Here, the maximum search
depth parameter, $d_\mathrm{max}$, was set to $|\mathscr{R}|$, the upper bound
of synthesis steps in a chemical reaction network with $|\mathscr{R}|$ reactions.
Then, inflow dummy reactions of commercially available precursors were added
to represent purchasing starting materials; outflow dummy reactions of species
other than the source species (with zero in-degree) were added to represent
shipments of target chemicals or disposals of synthesized byproducts.

Finally, we created optimization problems of synthesis planning
based on the random costs and the underlying chemical reaction networks.
In all problems, the multiplicity of the outflow reaction of each target
was fixed to 1 by setting the lower and upper bounds to 1, and the lower
and upper bounds of the multiplicity of other reactions were set to 0 and 5,
respectively.

Note that problems with disconnected relevant chemical reaction networks
were not adopted as the benchmark problems because such problems can easily be
decomposed into smaller problems. In addition, problems that are infeasible
or unable to be embedded into the D-Wave Advantage machine were also not adopted
as the benchmark problems.

\section{Supplemental Performance Evaluation} \label{appx:supp-perf-eval}

\subsection{Embedding performance} \label{appx:embedding-performance}

We calculated embeddings of the benchmark QUBO problems into the D-Wave
Advantage Pegasus topology QPU using the \texttt{minorminer.find\_embedding}
function from the D-Wave Ocean Software \cite{DWaveOcean}. For each problem,
we performed the calculation ten times and selected the embedding with
the shortest maximum and mean chain length from those obtained.

\begin{figure*}
  \centering
  \includegraphics{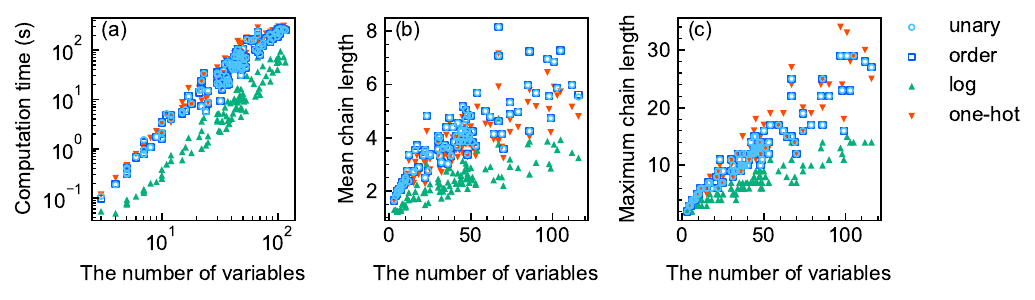}
  \caption{
    Scaling of (a) the computation time required to find an embedding
    using the \texttt{minorminer.find\_embedding} function, (b) the mean
    chain length, and (c) the maximum chain length, with respect to
    the number of integer variables.
  }
  \label{fig:embedding-performance}
\end{figure*}

Figure~\ref{fig:embedding-performance} illustrates the scaling of
(a) the computation time required to find an embedding,
(b) the mean chain length, and (c) the maximum chain length,
with respect to the number of integer variables. The computation time
of the \texttt{minorminer.find\_embedding} function tends to increase
nearly quadratically as the problem size increases. The mean and maximum chain
lengths exhibit roughly linear scaling, leading to quadratic scaling of
the number of physical qubits with respect to the number of logical qubits.
Since the log encoding method requires fewer binary variables than the other
encoding methods, both the computation time and chain length for the log
encoding method are less than those for the others.

\subsection{Tuning performance} \label{appx:tuning-performance}

\begin{figure*}[tb]
  \centering
  \includegraphics{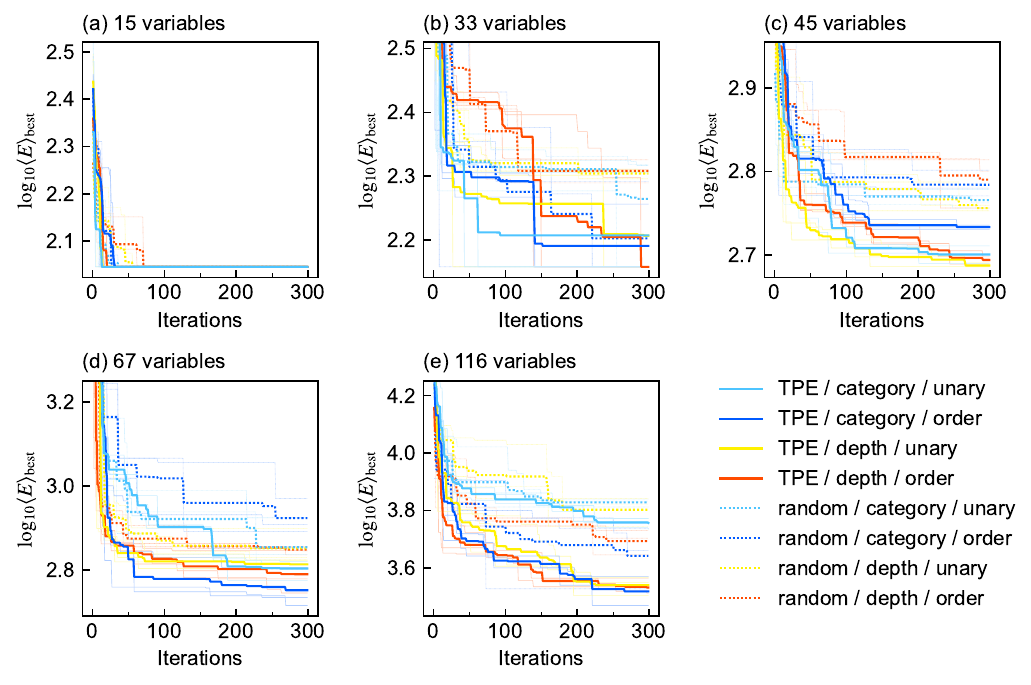}
  \caption{
    Convergence plots of parameter tuning for the D-Wave Advantage case.
    The solid lines illustrate the convergence trends for Bayesian optimization
    with the multivariate tree-structured Parzen estimator (TPE), and the dotted
    lines illustrate those for random search. We performed parameter tuning
    three times for each problem, grouping, and encoding. The thick, dark lines
    represent the mean of the three tuning processes, while the thin, light lines
    depict each individual process to highlight variations in the tuning
    processes. The problems selected are the same as those in
    Fig.~\ref{fig:encoding-grouping-performance-qa}.
  }
  \label{fig:tuning-convergence-qa}
\end{figure*}

\begin{figure*}[tb]
  \centering
  \includegraphics{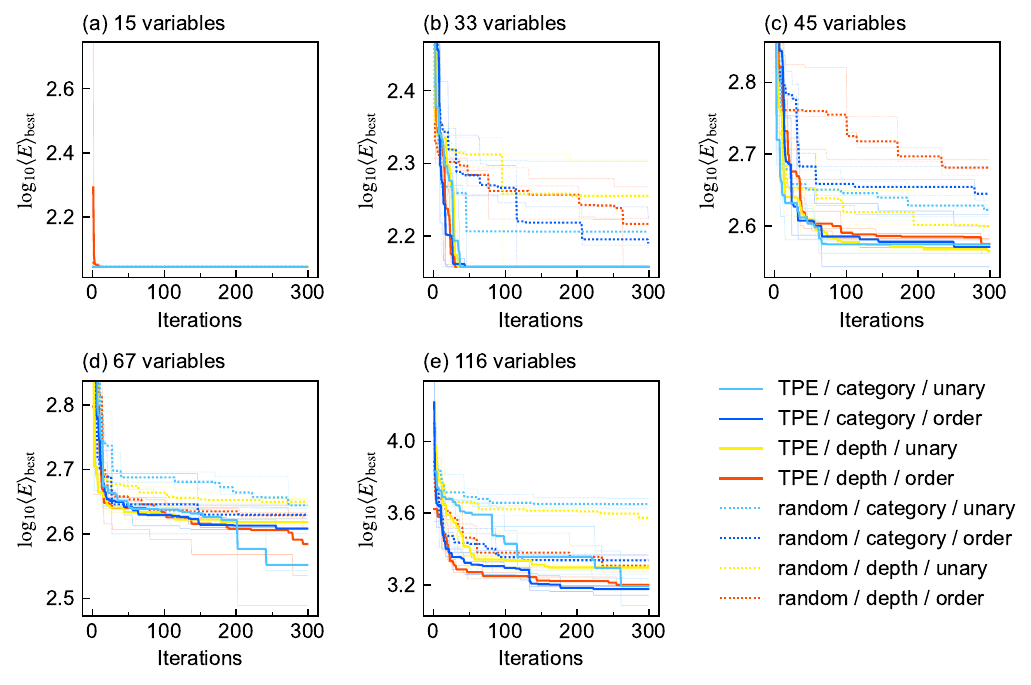}
  \caption{
    Convergence plots of parameter tuning for the simulated annealing case.
    The notation is the same as that in Fig.~\ref{fig:tuning-convergence-qa}.
  }
  \label{fig:tuning-convergence-sa}
\end{figure*}

To evaluate the effectiveness of Bayesian optimization in penalty strength
tuning, we compared its results with those of random search.
The results for the D-Wave Advantage system and simulated annealing are
depicted in Figs.~\ref{fig:tuning-convergence-qa} and \ref{fig:tuning-convergence-sa},
respectively. For the smallest size problem, both Bayesian optimization
and random search converged rapidly to the same value. However, for larger
problems, Bayesian optimization converged to a better value than random search
when using the same encoding and grouping methods.

\subsection{Postprocessing performance} \label{appx:postprocessing-performance}

\begin{figure*}
  \centering
  \includegraphics{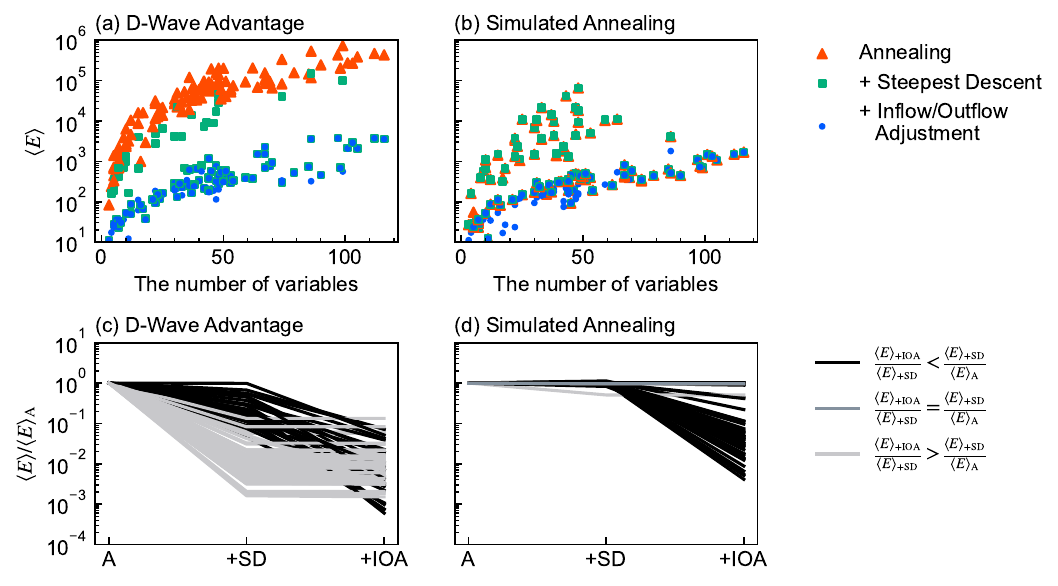}
  \caption{
    Effectiveness of the postprocessing methods.
    Panel (a) and (b) depict the tuning score $\langle E \rangle$, defined by
    Eq.~\ref{eq:mean-standard-energy}, for solutions at each step illustrated in
    Fig.~\ref{fig:algorithm-structure}, i.e., QUBO solutions, QUBO solutions (improved),
    and pathway solutions (improved). Red triangles represent the score
    of QUBO solutions sampled by an Ising machine, $\langle E \rangle_\mathrm{A}$.
    Green squares denote the score of improved QUBO solutions processed by
    steepest descent, $\langle E \rangle_\mathrm{+SD}$. Blue circles indicate
    the score of improved pathway solutions processed by both steepest descent
    and inflow/outflow adjustment, $\langle E \rangle_\mathrm{+IOA}$.
    Panel (c) and (d) illustrate changes in the score relative to the score of
    raw solutions, $\langle E \rangle_\mathrm{A}$. Points at position $X$
    ($X=\mathrm{A}, \mathrm{+SD}, \mathrm{+IOA}$) represent the relative score
    $\langle E \rangle_X/\langle E \rangle_\mathrm{A}$. Each line corresponds
    to one problem and is colored according to the criteria shown in the lower
    legend to highlight the type of improvement process.
  }
  \label{fig:postprocessing-performance}
\end{figure*}

To evaluate the effectiveness of the two postprocessing methods, we analyzed
changes in the tuning score function $\langle E \rangle$, defined by
Eq.~\ref{eq:mean-standard-energy}, during postprocessing. The results are shown
in Fig.~\ref{fig:postprocessing-performance}. In the D-Wave Advantage case,
the postprocessing enhances the $\langle E \rangle$ score by several
orders of magnitude. The score improvement was primarily due to
steepest descent (SD) or inflow/outflow adjustment (IOA).
On the other hand, in the simulated annealing case, SD does not
contribute to score improvement, whereas IOA does enhance solution
quality for some problems.

\subsection{Penalty strength scaling (supplemental)} \label{appx:penalty-strength-scaling}

\begin{figure*}
  \centering
  \includegraphics{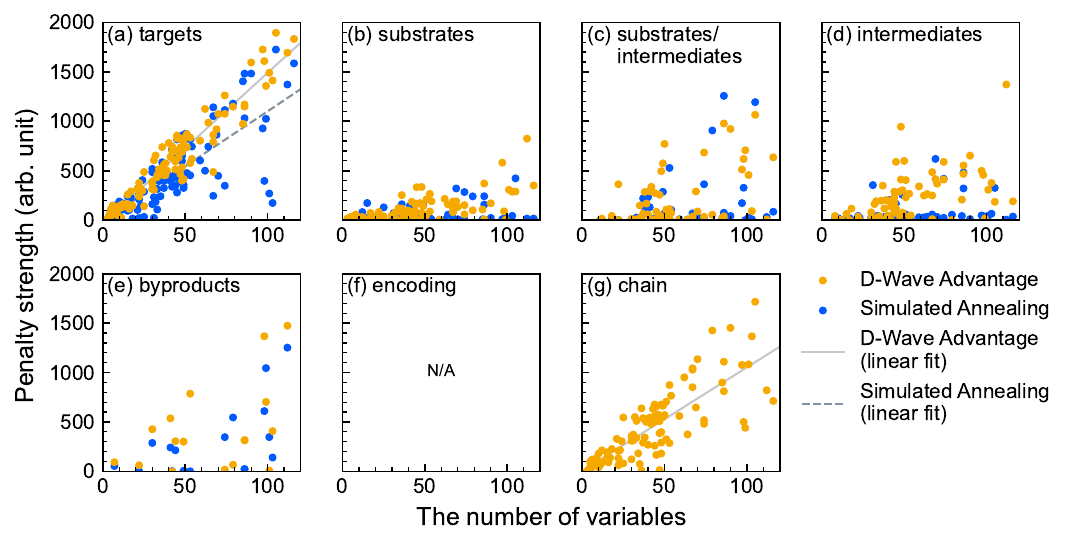}
  \caption{
    Scaling of the tuned penalty and chain strength parameter values
    with respect to the number of integer variables in the case of using
    the unary encoding and category grouping methods. The notation is
    the same as that in Fig.~\ref{fig:penalty-strength-scaling} in
    Sec.~\ref{sec:results-discussion}. Note that panel (f) is blank because
    the unary encoding method does not require any penalty for encoding.
    The $R^2$ values are 0.92 (panel (a) for D-Wave Advantage),
    0.63 (panel (a) for simulated annealing), and
    0.65 (panel (g) for D-Wave Advantage), respectively.
  }
  \label{fig:penalty-strength-scaling-unary-category}
\end{figure*}

\begin{figure*}
  \centering
  \includegraphics{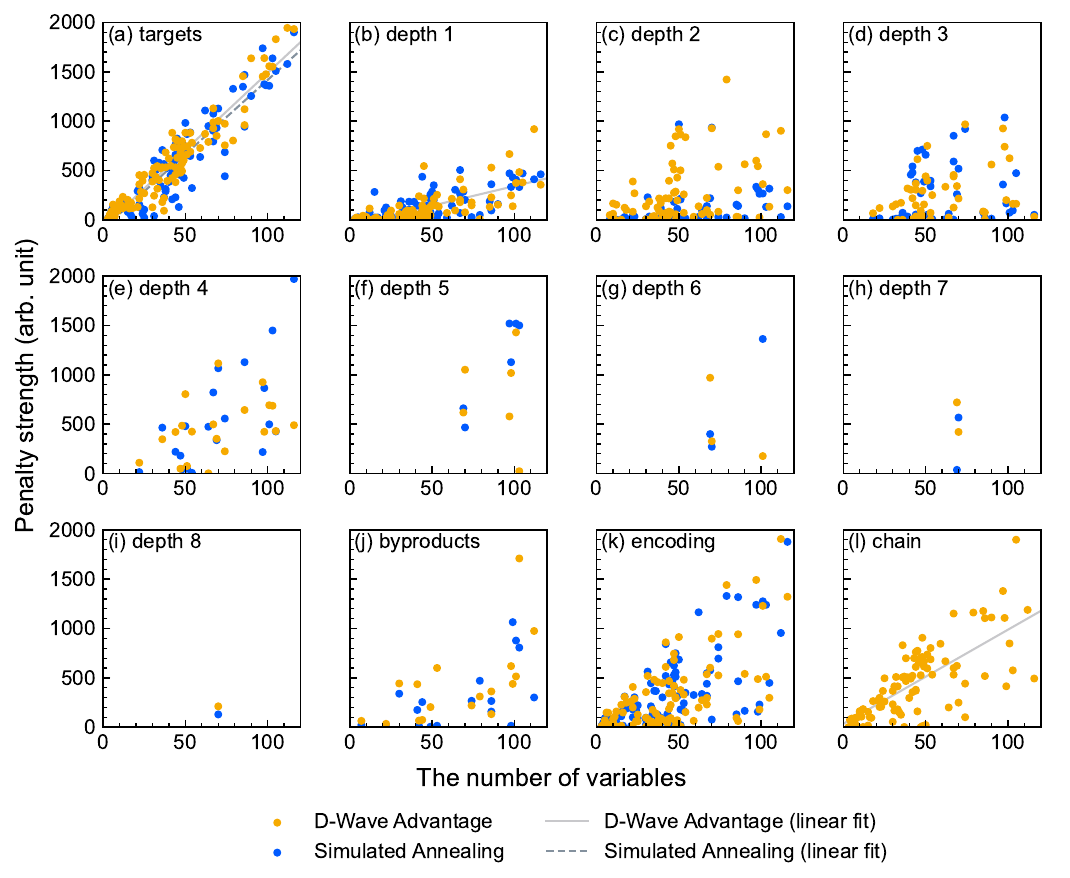}
  \caption{
    Scaling of the tuned penalty and chain strength parameter values
    with respect to the number of integer variables in the case of using
    the order encoding and depth grouping methods.
    Panels (a)-(j) depict the penalty strength parameters for mass balance
    constraints of species in each group, panel (k) shows the penalty
    strength parameter for order encoding, and panel (l) shows the chain
    strength for embedding. Linear fits are plotted for datasets with
    a coefficient of determination ($R^2$) greater than 0.5, excluding those
    with fewer than ten data points.
    The $R^2$ values are 0.91 (panel (a) for D-Wave Advantage),
    0.86 (panel (a) for simulated annealing),
    0.52 (panel (b) for D-Wave Advantage),
    and 0.54 (panel (l) for D-Wave Advantage), respectively.
  }
  \label{fig:penalty-strength-scaling-order-depth}
\end{figure*}

\begin{figure*}
  \centering
  \includegraphics{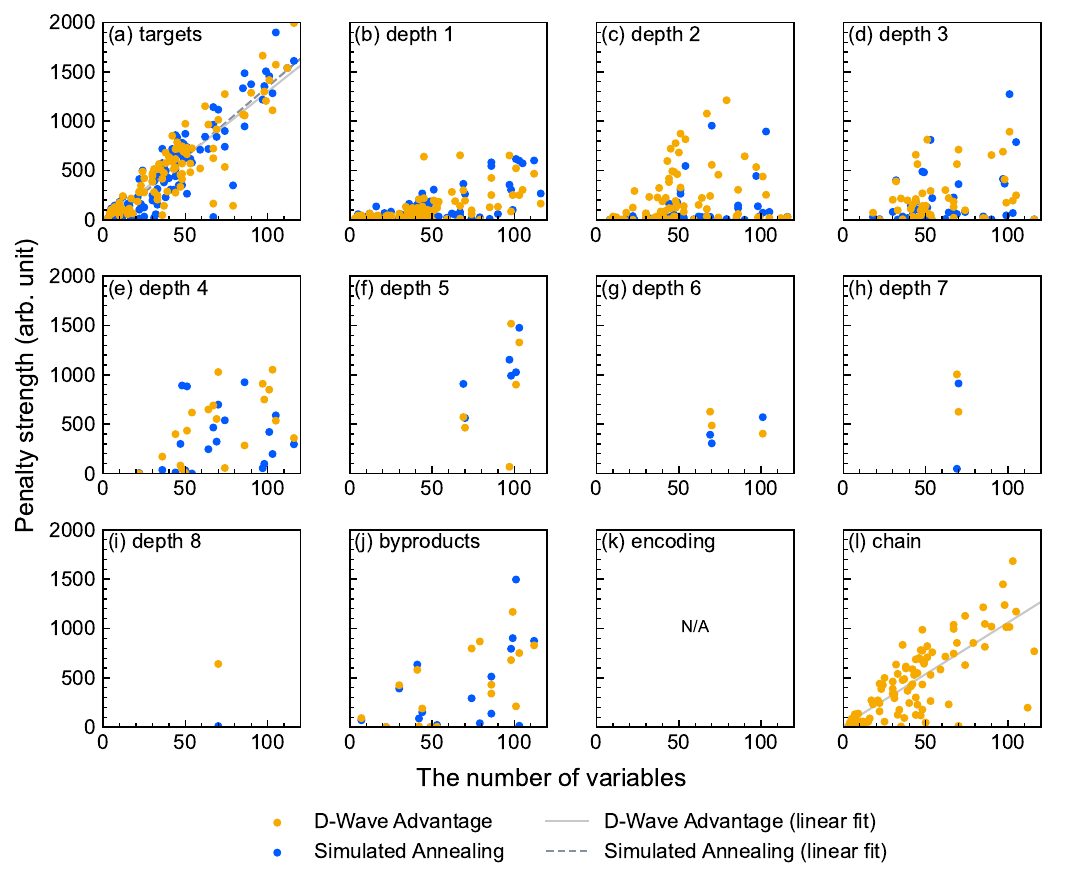}
  \caption{
    Scaling of the tuned penalty and chain strength parameter values
    with respect to the number of integer variables in the case of using
    the unary encoding and depth grouping methods. The notation is the same
    as that in Fig.~\ref{fig:penalty-strength-scaling-order-depth}. Note that
    panel (k) is blank because the unary encoding method does not require
    any penalty for encoding.
    The $R^2$ values are 0.79 (panel (a) for D-Wave Advantage),
    0.82 (panel (a) for simulated annealing),
    and 0.61 (panel (l) for D-Wave Advantage), respectively.
  }
  \label{fig:penalty-strength-scaling-unary-depth}
\end{figure*}

Figure~\ref{fig:penalty-strength-scaling} in Sec.~\ref{sec:results-discussion} and
Figs.~\ref{fig:penalty-strength-scaling-unary-category}--\ref{fig:penalty-strength-scaling-unary-depth}
in this appendix demonstrate that the optimized penalty strength
for the mass balance constraints of the target species tends to be larger
than that of the others and exhibits a strong linear correlation with
the number of integer variables when using the unary/order encoding and
depth/category grouping methods.

\bibliography{main}

\begin{thebibliography}{63}%
\makeatletter
\providecommand \@ifxundefined [1]{%
 \@ifx{#1\undefined}
}%
\providecommand \@ifnum [1]{%
 \ifnum #1\expandafter \@firstoftwo
 \else \expandafter \@secondoftwo
 \fi
}%
\providecommand \@ifx [1]{%
 \ifx #1\expandafter \@firstoftwo
 \else \expandafter \@secondoftwo
 \fi
}%
\providecommand \natexlab [1]{#1}%
\providecommand \enquote  [1]{``#1''}%
\providecommand \bibnamefont  [1]{#1}%
\providecommand \bibfnamefont [1]{#1}%
\providecommand \citenamefont [1]{#1}%
\providecommand \href@noop [0]{\@secondoftwo}%
\providecommand \href [0]{\begingroup \@sanitize@url \@href}%
\providecommand \@href[1]{\@@startlink{#1}\@@href}%
\providecommand \@@href[1]{\endgroup#1\@@endlink}%
\providecommand \@sanitize@url [0]{\catcode `\\12\catcode `\$12\catcode
  `\&12\catcode `\#12\catcode `\^12\catcode `\_12\catcode `\%12\relax}%
\providecommand \@@startlink[1]{}%
\providecommand \@@endlink[0]{}%
\providecommand \url  [0]{\begingroup\@sanitize@url \@url }%
\providecommand \@url [1]{\endgroup\@href {#1}{\urlprefix }}%
\providecommand \urlprefix  [0]{URL }%
\providecommand \Eprint [0]{\href }%
\providecommand \doibase [0]{https://doi.org/}%
\providecommand \selectlanguage [0]{\@gobble}%
\providecommand \bibinfo  [0]{\@secondoftwo}%
\providecommand \bibfield  [0]{\@secondoftwo}%
\providecommand \translation [1]{[#1]}%
\providecommand \BibitemOpen [0]{}%
\providecommand \bibitemStop [0]{}%
\providecommand \bibitemNoStop [0]{.\EOS\space}%
\providecommand \EOS [0]{\spacefactor3000\relax}%
\providecommand \BibitemShut  [1]{\csname bibitem#1\endcsname}%
\let\auto@bib@innerbib\@empty
\bibitem [{\citenamefont {{Elsevier}}(2023)}]{Reaxys}%
  \BibitemOpen
  \bibfield  {author} {\bibinfo {author} {\bibnamefont {{Elsevier}}},\ }\href
  {https://www.reaxys.com/} {\bibinfo {title} {{Reaxys}}} (\bibinfo {year}
  {2023}),\ \bibinfo {note} {{Accessed: 2023-5-1}}\BibitemShut {NoStop}%
\bibitem [{\citenamefont {{American Chemical Society}}(2022)}]{SciFinder}%
  \BibitemOpen
  \bibfield  {author} {\bibinfo {author} {\bibnamefont {{American Chemical
  Society}}},\ }\href
  {https://www.cas.org/solutions/cas-scifinder-discovery-platform/cas-scifinder}
  {\bibinfo {title} {{CAS SciFinder$^\text{n}$}}} (\bibinfo {year} {2022}),\
  \bibinfo {note} {{Accessed: 2022-12-12}}\BibitemShut {NoStop}%
\bibitem [{\citenamefont {Jacob}\ \emph {et~al.}(2017)\citenamefont {Jacob},
  \citenamefont {Lan}, \citenamefont {Goodman},\ and\ \citenamefont
  {Lapkin}}]{Jacob2017}%
  \BibitemOpen
  \bibfield  {author} {\bibinfo {author} {\bibfnamefont {P.-M.}\ \bibnamefont
  {Jacob}}, \bibinfo {author} {\bibfnamefont {T.}~\bibnamefont {Lan}}, \bibinfo
  {author} {\bibfnamefont {J.~M.}\ \bibnamefont {Goodman}},\ and\ \bibinfo
  {author} {\bibfnamefont {A.~A.}\ \bibnamefont {Lapkin}},\ }\bibfield  {title}
  {\bibinfo {title} {A possible extension to the {RInChI} as a means of
  providing machine readable process data},\ }\href@noop {} {\bibfield
  {journal} {\bibinfo  {journal} {Journal of Cheminformatics}\ }\textbf
  {\bibinfo {volume} {9}},\ \bibinfo {pages} {1} (\bibinfo {year}
  {2017})}\BibitemShut {NoStop}%
\bibitem [{\citenamefont {Granda}\ \emph {et~al.}(2018)\citenamefont {Granda},
  \citenamefont {Donina}, \citenamefont {Dragone}, \citenamefont {Long},\ and\
  \citenamefont {Cronin}}]{Granda2018}%
  \BibitemOpen
  \bibfield  {author} {\bibinfo {author} {\bibfnamefont {J.~M.}\ \bibnamefont
  {Granda}}, \bibinfo {author} {\bibfnamefont {L.}~\bibnamefont {Donina}},
  \bibinfo {author} {\bibfnamefont {V.}~\bibnamefont {Dragone}}, \bibinfo
  {author} {\bibfnamefont {D.-L.}\ \bibnamefont {Long}},\ and\ \bibinfo
  {author} {\bibfnamefont {L.}~\bibnamefont {Cronin}},\ }\bibfield  {title}
  {\bibinfo {title} {Controlling an organic synthesis robot with machine
  learning to search for new reactivity},\ }\href@noop {} {\bibfield  {journal}
  {\bibinfo  {journal} {Nature}\ }\textbf {\bibinfo {volume} {559}},\ \bibinfo
  {pages} {377} (\bibinfo {year} {2018})}\BibitemShut {NoStop}%
\bibitem [{\citenamefont {Coley}\ \emph {et~al.}(2019)\citenamefont {Coley},
  \citenamefont {Thomas}, \citenamefont {Lummiss}, \citenamefont {Jaworski},
  \citenamefont {Breen}, \citenamefont {Schultz}, \citenamefont {Hart},
  \citenamefont {Fishman}, \citenamefont {Rogers}, \citenamefont {Gao},
  \citenamefont {Hicklin}, \citenamefont {Plehiers}, \citenamefont {Byington},
  \citenamefont {Piotti}, \citenamefont {Green}, \citenamefont {Hart},
  \citenamefont {Jamison},\ and\ \citenamefont {Jensen}}]{Coley2019}%
  \BibitemOpen
  \bibfield  {author} {\bibinfo {author} {\bibfnamefont {C.~W.}\ \bibnamefont
  {Coley}}, \bibinfo {author} {\bibfnamefont {D.~A.}\ \bibnamefont {Thomas}},
  \bibinfo {author} {\bibfnamefont {J.~A.~M.}\ \bibnamefont {Lummiss}},
  \bibinfo {author} {\bibfnamefont {J.~N.}\ \bibnamefont {Jaworski}}, \bibinfo
  {author} {\bibfnamefont {C.~P.}\ \bibnamefont {Breen}}, \bibinfo {author}
  {\bibfnamefont {V.}~\bibnamefont {Schultz}}, \bibinfo {author} {\bibfnamefont
  {T.}~\bibnamefont {Hart}}, \bibinfo {author} {\bibfnamefont {J.~S.}\
  \bibnamefont {Fishman}}, \bibinfo {author} {\bibfnamefont {L.}~\bibnamefont
  {Rogers}}, \bibinfo {author} {\bibfnamefont {H.}~\bibnamefont {Gao}},
  \bibinfo {author} {\bibfnamefont {R.~W.}\ \bibnamefont {Hicklin}}, \bibinfo
  {author} {\bibfnamefont {P.~P.}\ \bibnamefont {Plehiers}}, \bibinfo {author}
  {\bibfnamefont {J.}~\bibnamefont {Byington}}, \bibinfo {author}
  {\bibfnamefont {J.~S.}\ \bibnamefont {Piotti}}, \bibinfo {author}
  {\bibfnamefont {W.~H.}\ \bibnamefont {Green}}, \bibinfo {author}
  {\bibfnamefont {A.~J.}\ \bibnamefont {Hart}}, \bibinfo {author}
  {\bibfnamefont {T.~F.}\ \bibnamefont {Jamison}},\ and\ \bibinfo {author}
  {\bibfnamefont {K.~F.}\ \bibnamefont {Jensen}},\ }\bibfield  {title}
  {\bibinfo {title} {A robotic platform for flow synthesis of organic compounds
  informed by {AI} planning},\ }\href@noop {} {\bibfield  {journal} {\bibinfo
  {journal} {Science}\ }\textbf {\bibinfo {volume} {365}},\ \bibinfo {pages}
  {eaax1566} (\bibinfo {year} {2019})}\BibitemShut {NoStop}%
\bibitem [{\citenamefont {Segler}\ \emph {et~al.}(2018)\citenamefont {Segler},
  \citenamefont {Preuss},\ and\ \citenamefont {Waller}}]{Segler2018}%
  \BibitemOpen
  \bibfield  {author} {\bibinfo {author} {\bibfnamefont {M.~H.~S.}\
  \bibnamefont {Segler}}, \bibinfo {author} {\bibfnamefont {M.}~\bibnamefont
  {Preuss}},\ and\ \bibinfo {author} {\bibfnamefont {M.~P.}\ \bibnamefont
  {Waller}},\ }\bibfield  {title} {\bibinfo {title} {Planning chemical
  syntheses with deep neural networks and symbolic {AI}},\ }\href@noop {}
  {\bibfield  {journal} {\bibinfo  {journal} {Nature}\ }\textbf {\bibinfo
  {volume} {555}},\ \bibinfo {pages} {604} (\bibinfo {year}
  {2018})}\BibitemShut {NoStop}%
\bibitem [{\citenamefont {Tsuji}\ \emph {et~al.}(2023)\citenamefont {Tsuji},
  \citenamefont {Sidorov}, \citenamefont {Zhu}, \citenamefont {Nagata},
  \citenamefont {Gimadiev}, \citenamefont {Varnek},\ and\ \citenamefont
  {List}}]{Tsuji2023}%
  \BibitemOpen
  \bibfield  {author} {\bibinfo {author} {\bibfnamefont {N.}~\bibnamefont
  {Tsuji}}, \bibinfo {author} {\bibfnamefont {P.}~\bibnamefont {Sidorov}},
  \bibinfo {author} {\bibfnamefont {C.}~\bibnamefont {Zhu}}, \bibinfo {author}
  {\bibfnamefont {Y.}~\bibnamefont {Nagata}}, \bibinfo {author} {\bibfnamefont
  {T.}~\bibnamefont {Gimadiev}}, \bibinfo {author} {\bibfnamefont
  {A.}~\bibnamefont {Varnek}},\ and\ \bibinfo {author} {\bibfnamefont
  {B.}~\bibnamefont {List}},\ }\bibfield  {title} {\bibinfo {title} {Predicting
  highly enantioselective catalysts using tunable fragment descriptors},\
  }\href@noop {} {\bibfield  {journal} {\bibinfo  {journal} {Angewandte Chemie
  International Edition}\ }\textbf {\bibinfo {volume} {62}},\ \bibinfo {pages}
  {e202218659} (\bibinfo {year} {2023})}\BibitemShut {NoStop}%
\bibitem [{\citenamefont {Maeda}\ and\ \citenamefont
  {Harabuchi}(2021)}]{Maeda2021}%
  \BibitemOpen
  \bibfield  {author} {\bibinfo {author} {\bibfnamefont {S.}~\bibnamefont
  {Maeda}}\ and\ \bibinfo {author} {\bibfnamefont {Y.}~\bibnamefont
  {Harabuchi}},\ }\bibfield  {title} {\bibinfo {title} {Exploring paths of
  chemical transformations in molecular and periodic systems: An approach
  utilizing force},\ }\href@noop {} {\bibfield  {journal} {\bibinfo  {journal}
  {{WIREs} Computational Molecular Science}\ }\textbf {\bibinfo {volume}
  {11}},\ \bibinfo {pages} {e1538} (\bibinfo {year} {2021})}\BibitemShut
  {NoStop}%
\bibitem [{\citenamefont {Temkin}\ \emph {et~al.}(1996)\citenamefont {Temkin},
  \citenamefont {Zeigarnik},\ and\ \citenamefont {Bonchev}}]{Temkin1996}%
  \BibitemOpen
  \bibfield  {author} {\bibinfo {author} {\bibfnamefont {O.~N.}\ \bibnamefont
  {Temkin}}, \bibinfo {author} {\bibfnamefont {A.~V.}\ \bibnamefont
  {Zeigarnik}},\ and\ \bibinfo {author} {\bibfnamefont {D.}~\bibnamefont
  {Bonchev}},\ }\href@noop {} {\emph {\bibinfo {title} {Chemical reaction
  networks: a graph-theoretical approach}}}\ (\bibinfo  {publisher} {{CRC
  Press}},\ \bibinfo {address} {Boca Raton, FL},\ \bibinfo {year}
  {1996})\BibitemShut {NoStop}%
\bibitem [{\citenamefont {Szymkuć}\ \emph {et~al.}(2016)\citenamefont
  {Szymkuć}, \citenamefont {Gajewska}, \citenamefont {Klucznik}, \citenamefont
  {Molga}, \citenamefont {Dittwald}, \citenamefont {Startek}, \citenamefont
  {Bajczyk},\ and\ \citenamefont {Grzybowski}}]{Szymkuc2016}%
  \BibitemOpen
  \bibfield  {author} {\bibinfo {author} {\bibfnamefont {S.}~\bibnamefont
  {Szymkuć}}, \bibinfo {author} {\bibfnamefont {E.~P.}\ \bibnamefont
  {Gajewska}}, \bibinfo {author} {\bibfnamefont {T.}~\bibnamefont {Klucznik}},
  \bibinfo {author} {\bibfnamefont {K.}~\bibnamefont {Molga}}, \bibinfo
  {author} {\bibfnamefont {P.}~\bibnamefont {Dittwald}}, \bibinfo {author}
  {\bibfnamefont {M.}~\bibnamefont {Startek}}, \bibinfo {author} {\bibfnamefont
  {M.}~\bibnamefont {Bajczyk}},\ and\ \bibinfo {author} {\bibfnamefont {B.~A.}\
  \bibnamefont {Grzybowski}},\ }\bibfield  {title} {\bibinfo {title}
  {Computer-assisted synthetic planning: The end of the beginning},\
  }\href@noop {} {\bibfield  {journal} {\bibinfo  {journal} {Angewandte Chemie
  International Edition}\ }\textbf {\bibinfo {volume} {55}},\ \bibinfo {pages}
  {5904} (\bibinfo {year} {2016})}\BibitemShut {NoStop}%
\bibitem [{\citenamefont {Kowalik}\ \emph {et~al.}(2012)\citenamefont
  {Kowalik}, \citenamefont {Gothard}, \citenamefont {Drews}, \citenamefont
  {Gothard}, \citenamefont {Weckiewicz}, \citenamefont {Fuller}, \citenamefont
  {Grzybowski},\ and\ \citenamefont {Bishop}}]{Kowalik2012}%
  \BibitemOpen
  \bibfield  {author} {\bibinfo {author} {\bibfnamefont {M.}~\bibnamefont
  {Kowalik}}, \bibinfo {author} {\bibfnamefont {C.~M.}\ \bibnamefont
  {Gothard}}, \bibinfo {author} {\bibfnamefont {A.~M.}\ \bibnamefont {Drews}},
  \bibinfo {author} {\bibfnamefont {N.~A.}\ \bibnamefont {Gothard}}, \bibinfo
  {author} {\bibfnamefont {A.}~\bibnamefont {Weckiewicz}}, \bibinfo {author}
  {\bibfnamefont {P.~E.}\ \bibnamefont {Fuller}}, \bibinfo {author}
  {\bibfnamefont {B.~A.}\ \bibnamefont {Grzybowski}},\ and\ \bibinfo {author}
  {\bibfnamefont {K.~J.~M.}\ \bibnamefont {Bishop}},\ }\bibfield  {title}
  {\bibinfo {title} {Parallel optimization of synthetic pathways within the
  network of organic chemistry},\ }\href@noop {} {\bibfield  {journal}
  {\bibinfo  {journal} {Angewandte Chemie International Edition}\ }\textbf
  {\bibinfo {volume} {51}},\ \bibinfo {pages} {7928} (\bibinfo {year}
  {2012})}\BibitemShut {NoStop}%
\bibitem [{\citenamefont {Shibukawa}\ \emph {et~al.}(2020)\citenamefont
  {Shibukawa}, \citenamefont {Ishida}, \citenamefont {Yoshizoe}, \citenamefont
  {Wasa}, \citenamefont {Takasu}, \citenamefont {Okuno}, \citenamefont
  {Terayama},\ and\ \citenamefont {Tsuda}}]{Shibukawa2020}%
  \BibitemOpen
  \bibfield  {author} {\bibinfo {author} {\bibfnamefont {R.}~\bibnamefont
  {Shibukawa}}, \bibinfo {author} {\bibfnamefont {S.}~\bibnamefont {Ishida}},
  \bibinfo {author} {\bibfnamefont {K.}~\bibnamefont {Yoshizoe}}, \bibinfo
  {author} {\bibfnamefont {K.}~\bibnamefont {Wasa}}, \bibinfo {author}
  {\bibfnamefont {K.}~\bibnamefont {Takasu}}, \bibinfo {author} {\bibfnamefont
  {Y.}~\bibnamefont {Okuno}}, \bibinfo {author} {\bibfnamefont
  {K.}~\bibnamefont {Terayama}},\ and\ \bibinfo {author} {\bibfnamefont
  {K.}~\bibnamefont {Tsuda}},\ }\bibfield  {title} {\bibinfo {title}
  {{CompRet}: a comprehensive recommendation framework for chemical synthesis
  planning with algorithmic enumeration},\ }\href@noop {} {\bibfield  {journal}
  {\bibinfo  {journal} {Journal of Cheminformatics}\ }\textbf {\bibinfo
  {volume} {12}},\ \bibinfo {pages} {1} (\bibinfo {year} {2020})}\BibitemShut
  {NoStop}%
\bibitem [{\citenamefont {Gao}\ \emph {et~al.}(2021)\citenamefont {Gao},
  \citenamefont {Pauphilet}, \citenamefont {Struble}, \citenamefont {Coley},\
  and\ \citenamefont {Jensen}}]{Gao2021}%
  \BibitemOpen
  \bibfield  {author} {\bibinfo {author} {\bibfnamefont {H.}~\bibnamefont
  {Gao}}, \bibinfo {author} {\bibfnamefont {J.}~\bibnamefont {Pauphilet}},
  \bibinfo {author} {\bibfnamefont {T.~J.}\ \bibnamefont {Struble}}, \bibinfo
  {author} {\bibfnamefont {C.~W.}\ \bibnamefont {Coley}},\ and\ \bibinfo
  {author} {\bibfnamefont {K.~F.}\ \bibnamefont {Jensen}},\ }\bibfield  {title}
  {\bibinfo {title} {Direct optimization across computer-generated reaction
  networks balances materials use and feasibility of synthesis plans for
  molecule libraries},\ }\href@noop {} {\bibfield  {journal} {\bibinfo
  {journal} {Journal of Chemical Information and Modeling}\ }\textbf {\bibinfo
  {volume} {61}},\ \bibinfo {pages} {493} (\bibinfo {year} {2021})}\BibitemShut
  {NoStop}%
\bibitem [{\citenamefont {Andersen}\ \emph {et~al.}(2012)\citenamefont
  {Andersen}, \citenamefont {Flamm}, \citenamefont {Merkle},\ and\
  \citenamefont {Stadler}}]{Andersen2012}%
  \BibitemOpen
  \bibfield  {author} {\bibinfo {author} {\bibfnamefont {J.~L.}\ \bibnamefont
  {Andersen}}, \bibinfo {author} {\bibfnamefont {C.}~\bibnamefont {Flamm}},
  \bibinfo {author} {\bibfnamefont {D.}~\bibnamefont {Merkle}},\ and\ \bibinfo
  {author} {\bibfnamefont {P.~F.}\ \bibnamefont {Stadler}},\ }\bibfield
  {title} {\bibinfo {title} {Maximizing output and recognizing autocatalysis in
  chemical reaction networks is {NP}-complete},\ }\href@noop {} {\bibfield
  {journal} {\bibinfo  {journal} {Journal of Systems Chemistry}\ }\textbf
  {\bibinfo {volume} {3}},\ \bibinfo {pages} {1} (\bibinfo {year}
  {2012})}\BibitemShut {NoStop}%
\bibitem [{\citenamefont {Andersen}\ \emph {et~al.}(2017)\citenamefont
  {Andersen}, \citenamefont {Flamm}, \citenamefont {Merkle},\ and\
  \citenamefont {Stadler}}]{Andersen2017}%
  \BibitemOpen
  \bibfield  {author} {\bibinfo {author} {\bibfnamefont {J.~L.}\ \bibnamefont
  {Andersen}}, \bibinfo {author} {\bibfnamefont {C.}~\bibnamefont {Flamm}},
  \bibinfo {author} {\bibfnamefont {D.}~\bibnamefont {Merkle}},\ and\ \bibinfo
  {author} {\bibfnamefont {P.~F.}\ \bibnamefont {Stadler}},\ }\bibfield
  {title} {\bibinfo {title} {Chemical transformation motifs---modelling
  pathways as integer hyperflows},\ }\href@noop {} {\bibfield  {journal}
  {\bibinfo  {journal} {{IEEE/ACM} transactions on computational biology and
  bioinformatics}\ }\textbf {\bibinfo {volume} {16}},\ \bibinfo {pages} {510}
  (\bibinfo {year} {2017})}\BibitemShut {NoStop}%
\bibitem [{\citenamefont {Mohseni}\ \emph {et~al.}(2022)\citenamefont
  {Mohseni}, \citenamefont {McMahon},\ and\ \citenamefont
  {Byrnes}}]{Mohseni2022}%
  \BibitemOpen
  \bibfield  {author} {\bibinfo {author} {\bibfnamefont {N.}~\bibnamefont
  {Mohseni}}, \bibinfo {author} {\bibfnamefont {P.~L.}\ \bibnamefont
  {McMahon}},\ and\ \bibinfo {author} {\bibfnamefont {T.}~\bibnamefont
  {Byrnes}},\ }\bibfield  {title} {\bibinfo {title} {{Ising} machines as
  hardware solvers of combinatorial optimization problems},\ }\href@noop {}
  {\bibfield  {journal} {\bibinfo  {journal} {Nature Reviews Physics}\ }\textbf
  {\bibinfo {volume} {4}},\ \bibinfo {pages} {363} (\bibinfo {year}
  {2022})}\BibitemShut {NoStop}%
\bibitem [{\citenamefont {Johnson}\ \emph {et~al.}(2011)\citenamefont
  {Johnson}, \citenamefont {Amin}, \citenamefont {Gildert}, \citenamefont
  {Lanting}, \citenamefont {Hamze}, \citenamefont {Dickson}, \citenamefont
  {Harris}, \citenamefont {Berkley}, \citenamefont {Johansson}, \citenamefont
  {Bunyk}, \citenamefont {Chapple}, \citenamefont {Enderud}, \citenamefont
  {Hilton}, \citenamefont {Karimi}, \citenamefont {Ladizinsky}, \citenamefont
  {Ladizinsky}, \citenamefont {Oh}, \citenamefont {Perminov}, \citenamefont
  {Rich}, \citenamefont {Thom}, \citenamefont {Tolkacheva}, \citenamefont
  {Truncik}, \citenamefont {Uchaikin}, \citenamefont {Wang}, \citenamefont
  {Wilson},\ and\ \citenamefont {Rose}}]{Johnson2011}%
  \BibitemOpen
  \bibfield  {author} {\bibinfo {author} {\bibfnamefont {M.~W.}\ \bibnamefont
  {Johnson}}, \bibinfo {author} {\bibfnamefont {M.~H.~S.}\ \bibnamefont
  {Amin}}, \bibinfo {author} {\bibfnamefont {S.}~\bibnamefont {Gildert}},
  \bibinfo {author} {\bibfnamefont {T.}~\bibnamefont {Lanting}}, \bibinfo
  {author} {\bibfnamefont {F.}~\bibnamefont {Hamze}}, \bibinfo {author}
  {\bibfnamefont {N.~G.}\ \bibnamefont {Dickson}}, \bibinfo {author}
  {\bibfnamefont {R.}~\bibnamefont {Harris}}, \bibinfo {author} {\bibfnamefont
  {A.~J.}\ \bibnamefont {Berkley}}, \bibinfo {author} {\bibfnamefont
  {J.}~\bibnamefont {Johansson}}, \bibinfo {author} {\bibfnamefont {P.~I.}\
  \bibnamefont {Bunyk}}, \bibinfo {author} {\bibfnamefont {E.~M.}\ \bibnamefont
  {Chapple}}, \bibinfo {author} {\bibfnamefont {C.}~\bibnamefont {Enderud}},
  \bibinfo {author} {\bibfnamefont {J.~P.}\ \bibnamefont {Hilton}}, \bibinfo
  {author} {\bibfnamefont {K.}~\bibnamefont {Karimi}}, \bibinfo {author}
  {\bibfnamefont {E.}~\bibnamefont {Ladizinsky}}, \bibinfo {author}
  {\bibfnamefont {N.}~\bibnamefont {Ladizinsky}}, \bibinfo {author}
  {\bibfnamefont {T.}~\bibnamefont {Oh}}, \bibinfo {author} {\bibfnamefont
  {I.~G.}\ \bibnamefont {Perminov}}, \bibinfo {author} {\bibfnamefont
  {C.}~\bibnamefont {Rich}}, \bibinfo {author} {\bibfnamefont {M.~C.}\
  \bibnamefont {Thom}}, \bibinfo {author} {\bibfnamefont {E.}~\bibnamefont
  {Tolkacheva}}, \bibinfo {author} {\bibfnamefont {C.~J.~S.}\ \bibnamefont
  {Truncik}}, \bibinfo {author} {\bibfnamefont {S.}~\bibnamefont {Uchaikin}},
  \bibinfo {author} {\bibfnamefont {J.}~\bibnamefont {Wang}}, \bibinfo {author}
  {\bibfnamefont {B.~A.}\ \bibnamefont {Wilson}},\ and\ \bibinfo {author}
  {\bibfnamefont {G.}~\bibnamefont {Rose}},\ }\bibfield  {title} {\bibinfo
  {title} {Quantum annealing with manufactured spins},\ }\href@noop {}
  {\bibfield  {journal} {\bibinfo  {journal} {Nature}\ }\textbf {\bibinfo
  {volume} {473}},\ \bibinfo {pages} {194} (\bibinfo {year}
  {2011})}\BibitemShut {NoStop}%
\bibitem [{\citenamefont {McGeoch}\ and\ \citenamefont
  {Farré}(2022)}]{McGeoch2022}%
  \BibitemOpen
  \bibfield  {author} {\bibinfo {author} {\bibfnamefont {C.}~\bibnamefont
  {McGeoch}}\ and\ \bibinfo {author} {\bibfnamefont {P.}~\bibnamefont
  {Farré}},\ }\href
  {https://www.dwavesys.com/media/3xvdipcn/14-1058a-a_advantage_processor_overview.pdf}
  {\emph {\bibinfo {title} {{Advantage} Processor Overview}}},\ \bibinfo {type}
  {Tech. Rep.}\ (\bibinfo  {institution} {{D-Wave Systems Inc.}},\ \bibinfo
  {year} {2022})\BibitemShut {NoStop}%
\bibitem [{\citenamefont {Yamaoka}\ \emph {et~al.}(2015)\citenamefont
  {Yamaoka}, \citenamefont {Yoshimura}, \citenamefont {Hayashi}, \citenamefont
  {Okuyama}, \citenamefont {Aoki},\ and\ \citenamefont {Mizuno}}]{Yamaoka2015}%
  \BibitemOpen
  \bibfield  {author} {\bibinfo {author} {\bibfnamefont {M.}~\bibnamefont
  {Yamaoka}}, \bibinfo {author} {\bibfnamefont {C.}~\bibnamefont {Yoshimura}},
  \bibinfo {author} {\bibfnamefont {M.}~\bibnamefont {Hayashi}}, \bibinfo
  {author} {\bibfnamefont {T.}~\bibnamefont {Okuyama}}, \bibinfo {author}
  {\bibfnamefont {H.}~\bibnamefont {Aoki}},\ and\ \bibinfo {author}
  {\bibfnamefont {H.}~\bibnamefont {Mizuno}},\ }\bibfield  {title} {\bibinfo
  {title} {20k-spin {Ising} chip for combinational optimization problem with
  {CMOS} annealing},\ }in\ \href@noop {} {\emph {\bibinfo {booktitle} {2015
  {IEEE} International Solid-State Circuits Conference ({ISSCC}) Digest of
  Technical Papers}}}\ (\bibinfo {organization} {{IEEE}},\ \bibinfo {year}
  {2015})\ pp.\ \bibinfo {pages} {1--3}\BibitemShut {NoStop}%
\bibitem [{\citenamefont {Okuyama}\ \emph {et~al.}(2019)\citenamefont
  {Okuyama}, \citenamefont {Sonobe}, \citenamefont {Kawarabayashi},\ and\
  \citenamefont {Yamaoka}}]{Okuyama2019}%
  \BibitemOpen
  \bibfield  {author} {\bibinfo {author} {\bibfnamefont {T.}~\bibnamefont
  {Okuyama}}, \bibinfo {author} {\bibfnamefont {T.}~\bibnamefont {Sonobe}},
  \bibinfo {author} {\bibfnamefont {K.}~\bibnamefont {Kawarabayashi}},\ and\
  \bibinfo {author} {\bibfnamefont {M.}~\bibnamefont {Yamaoka}},\ }\bibfield
  {title} {\bibinfo {title} {Binary optimization by momentum annealing},\
  }\href@noop {} {\bibfield  {journal} {\bibinfo  {journal} {Physical Review
  E}\ }\textbf {\bibinfo {volume} {100}},\ \bibinfo {pages} {012111} (\bibinfo
  {year} {2019})}\BibitemShut {NoStop}%
\bibitem [{\citenamefont {Matsubara}\ \emph {et~al.}(2020)\citenamefont
  {Matsubara}, \citenamefont {Takatsu}, \citenamefont {Miyazawa}, \citenamefont
  {Shibasaki}, \citenamefont {Watanabe}, \citenamefont {Takemoto},\ and\
  \citenamefont {Tamura}}]{Matsubara2020}%
  \BibitemOpen
  \bibfield  {author} {\bibinfo {author} {\bibfnamefont {S.}~\bibnamefont
  {Matsubara}}, \bibinfo {author} {\bibfnamefont {M.}~\bibnamefont {Takatsu}},
  \bibinfo {author} {\bibfnamefont {T.}~\bibnamefont {Miyazawa}}, \bibinfo
  {author} {\bibfnamefont {T.}~\bibnamefont {Shibasaki}}, \bibinfo {author}
  {\bibfnamefont {Y.}~\bibnamefont {Watanabe}}, \bibinfo {author}
  {\bibfnamefont {K.}~\bibnamefont {Takemoto}},\ and\ \bibinfo {author}
  {\bibfnamefont {H.}~\bibnamefont {Tamura}},\ }\bibfield  {title} {\bibinfo
  {title} {Digital annealer for high-speed solving of combinatorial
  optimization problems and its applications},\ }in\ \href@noop {} {\emph
  {\bibinfo {booktitle} {2020 25th {Asia} and South Pacific Design Automation
  Conference ({ASP-DAC})}}}\ (\bibinfo {organization} {{IEEE}},\ \bibinfo
  {year} {2020})\ pp.\ \bibinfo {pages} {667--672}\BibitemShut {NoStop}%
\bibitem [{\citenamefont {Yamamoto}\ \emph {et~al.}(2020)\citenamefont
  {Yamamoto}, \citenamefont {Ando}, \citenamefont {Mertig}, \citenamefont
  {Takemoto}, \citenamefont {Yamaoka}, \citenamefont {Teramoto}, \citenamefont
  {Sakai}, \citenamefont {Takamaeda-Yamazaki},\ and\ \citenamefont
  {Motomura}}]{Yamamoto2020}%
  \BibitemOpen
  \bibfield  {author} {\bibinfo {author} {\bibfnamefont {K.}~\bibnamefont
  {Yamamoto}}, \bibinfo {author} {\bibfnamefont {K.}~\bibnamefont {Ando}},
  \bibinfo {author} {\bibfnamefont {N.}~\bibnamefont {Mertig}}, \bibinfo
  {author} {\bibfnamefont {T.}~\bibnamefont {Takemoto}}, \bibinfo {author}
  {\bibfnamefont {M.}~\bibnamefont {Yamaoka}}, \bibinfo {author} {\bibfnamefont
  {H.}~\bibnamefont {Teramoto}}, \bibinfo {author} {\bibfnamefont
  {A.}~\bibnamefont {Sakai}}, \bibinfo {author} {\bibfnamefont
  {S.}~\bibnamefont {Takamaeda-Yamazaki}},\ and\ \bibinfo {author}
  {\bibfnamefont {M.}~\bibnamefont {Motomura}},\ }\bibfield  {title} {\bibinfo
  {title} {{STATICA}: A 512-spin 0.25 {M}-weight full-digital annealing
  processor with a near-memory all-spin-updates-at-once architecture for
  combinatorial optimization with complete spin-spin interactions},\ }in\
  \href@noop {} {\emph {\bibinfo {booktitle} {2020 {IEEE} International
  Solid-State Circuits Conference ({ISSCC})}}}\ (\bibinfo {organization}
  {{IEEE}},\ \bibinfo {year} {2020})\ pp.\ \bibinfo {pages}
  {138--140}\BibitemShut {NoStop}%
\bibitem [{\citenamefont {Kawamura}\ \emph {et~al.}(2023)\citenamefont
  {Kawamura}, \citenamefont {Yu}, \citenamefont {Okonogi}, \citenamefont
  {Jimbo}, \citenamefont {Inoue}, \citenamefont {Hyodo}, \citenamefont
  {Garc\'{i}a-Anas}, \citenamefont {Ando}, \citenamefont {Fukushima-Kimura},
  \citenamefont {Yasudo}, \citenamefont {Van~Chu},\ and\ \citenamefont
  {Motomura}}]{Kawamura2023}%
  \BibitemOpen
  \bibfield  {author} {\bibinfo {author} {\bibfnamefont {K.}~\bibnamefont
  {Kawamura}}, \bibinfo {author} {\bibfnamefont {J.}~\bibnamefont {Yu}},
  \bibinfo {author} {\bibfnamefont {D.}~\bibnamefont {Okonogi}}, \bibinfo
  {author} {\bibfnamefont {S.}~\bibnamefont {Jimbo}}, \bibinfo {author}
  {\bibfnamefont {G.}~\bibnamefont {Inoue}}, \bibinfo {author} {\bibfnamefont
  {A.}~\bibnamefont {Hyodo}}, \bibinfo {author} {\bibfnamefont {{\'{A}}.~L.}\
  \bibnamefont {Garc\'{i}a-Anas}}, \bibinfo {author} {\bibfnamefont
  {K.}~\bibnamefont {Ando}}, \bibinfo {author} {\bibfnamefont {B.~H.}\
  \bibnamefont {Fukushima-Kimura}}, \bibinfo {author} {\bibfnamefont
  {R.}~\bibnamefont {Yasudo}}, \bibinfo {author} {\bibfnamefont
  {T.}~\bibnamefont {Van~Chu}},\ and\ \bibinfo {author} {\bibfnamefont
  {M.}~\bibnamefont {Motomura}},\ }\bibfield  {title} {\bibinfo {title}
  {{Amorphica}: 4-replica 512 fully connected spin {336MHz} metamorphic
  annealer with programmable optimization strategy and compressed-spin-transfer
  multi-chip extension},\ }in\ \href@noop {} {\emph {\bibinfo {booktitle} {2023
  {IEEE} International Solid-State Circuits Conference ({ISSCC})}}}\ (\bibinfo
  {organization} {{IEEE}},\ \bibinfo {year} {2023})\ pp.\ \bibinfo {pages}
  {42--44}\BibitemShut {NoStop}%
\bibitem [{\citenamefont {Lucas}(2014)}]{Lucas2014}%
  \BibitemOpen
  \bibfield  {author} {\bibinfo {author} {\bibfnamefont {A.}~\bibnamefont
  {Lucas}},\ }\bibfield  {title} {\bibinfo {title} {{Ising} formulations of
  many {NP} problems},\ }\href@noop {} {\bibfield  {journal} {\bibinfo
  {journal} {Frontiers in physics}\ }\textbf {\bibinfo {volume} {2}},\ \bibinfo
  {pages} {5} (\bibinfo {year} {2014})}\BibitemShut {NoStop}%
\bibitem [{\citenamefont {Yarkoni}\ \emph {et~al.}(2022)\citenamefont
  {Yarkoni}, \citenamefont {Raponi}, \citenamefont {B{\"a}ck},\ and\
  \citenamefont {Schmitt}}]{Yarkoni2022}%
  \BibitemOpen
  \bibfield  {author} {\bibinfo {author} {\bibfnamefont {S.}~\bibnamefont
  {Yarkoni}}, \bibinfo {author} {\bibfnamefont {E.}~\bibnamefont {Raponi}},
  \bibinfo {author} {\bibfnamefont {T.}~\bibnamefont {B{\"a}ck}},\ and\
  \bibinfo {author} {\bibfnamefont {S.}~\bibnamefont {Schmitt}},\ }\bibfield
  {title} {\bibinfo {title} {Quantum annealing for industry applications:
  Introduction and review},\ }\href@noop {} {\bibfield  {journal} {\bibinfo
  {journal} {Reports on Progress in Physics}\ }\textbf {\bibinfo {volume}
  {85}},\ \bibinfo {pages} {104001} (\bibinfo {year} {2022})}\BibitemShut
  {NoStop}%
\bibitem [{\citenamefont {Patterson}\ and\ \citenamefont
  {Hennessy}(2020)}]{Patterson2020}%
  \BibitemOpen
  \bibfield  {author} {\bibinfo {author} {\bibfnamefont {D.~A.}\ \bibnamefont
  {Patterson}}\ and\ \bibinfo {author} {\bibfnamefont {J.~L.}\ \bibnamefont
  {Hennessy}},\ }\href@noop {} {\emph {\bibinfo {title} {Computer organization
  and design {MIPS} edition: the hardware/software interface}}},\ \bibinfo
  {edition} {6th}\ ed.\ (\bibinfo  {publisher} {{Morgan Kaufmann}},\ \bibinfo
  {address} {Burlington, MA},\ \bibinfo {year} {2020})\BibitemShut {NoStop}%
\bibitem [{\citenamefont {Chaouiya}(2007)}]{Chaouiya2007}%
  \BibitemOpen
  \bibfield  {author} {\bibinfo {author} {\bibfnamefont {C.}~\bibnamefont
  {Chaouiya}},\ }\bibfield  {title} {\bibinfo {title} {{Petri} net modelling of
  biological networks},\ }\href@noop {} {\bibfield  {journal} {\bibinfo
  {journal} {Briefings in Bioinformatics}\ }\textbf {\bibinfo {volume} {8}},\
  \bibinfo {pages} {210} (\bibinfo {year} {2007})}\BibitemShut {NoStop}%
\bibitem [{\citenamefont {Horiuti}(1953)}]{Horiuti1953}%
  \BibitemOpen
  \bibfield  {author} {\bibinfo {author} {\bibfnamefont {J.}~\bibnamefont
  {Horiuti}},\ }\bibfield  {title} {\bibinfo {title} {Stoichiometric number and
  universal kinetic law in the neighbourhood of equilibrium. {I}},\ }\href@noop
  {} {\bibfield  {journal} {\bibinfo  {journal} {Proceedings of the {Japan}
  Academy}\ }\textbf {\bibinfo {volume} {29}},\ \bibinfo {pages} {160}
  (\bibinfo {year} {1953})}\BibitemShut {NoStop}%
\bibitem [{\citenamefont {Zaman}\ \emph {et~al.}(2022)\citenamefont {Zaman},
  \citenamefont {Tanahashi},\ and\ \citenamefont {Tanaka}}]{Zaman2022}%
  \BibitemOpen
  \bibfield  {author} {\bibinfo {author} {\bibfnamefont {M.}~\bibnamefont
  {Zaman}}, \bibinfo {author} {\bibfnamefont {K.}~\bibnamefont {Tanahashi}},\
  and\ \bibinfo {author} {\bibfnamefont {S.}~\bibnamefont {Tanaka}},\
  }\bibfield  {title} {\bibinfo {title} {{PyQUBO}: {Python} library for mapping
  combinatorial optimization problems to {QUBO} form},\ }\href@noop {}
  {\bibfield  {journal} {\bibinfo  {journal} {{IEEE} Transactions on
  Computers}\ }\textbf {\bibinfo {volume} {71}},\ \bibinfo {pages} {838}
  (\bibinfo {year} {2022})}\BibitemShut {NoStop}%
\bibitem [{\citenamefont {{D-Wave Systems Inc.}}(2022)}]{DWaveOcean}%
  \BibitemOpen
  \bibfield  {author} {\bibinfo {author} {\bibnamefont {{D-Wave Systems
  Inc.}}},\ }\href {https://docs.ocean.dwavesys.com/en/stable/} {\bibinfo
  {title} {{D-Wave Ocean Software}}} (\bibinfo {year} {2022}),\ \bibinfo {note}
  {{Accessed: 2022-12-12}}\BibitemShut {NoStop}%
\bibitem [{\citenamefont {Akiba}\ \emph {et~al.}(2019)\citenamefont {Akiba},
  \citenamefont {Sano}, \citenamefont {Yanase}, \citenamefont {Ohta},\ and\
  \citenamefont {Koyama}}]{Akiba2019}%
  \BibitemOpen
  \bibfield  {author} {\bibinfo {author} {\bibfnamefont {T.}~\bibnamefont
  {Akiba}}, \bibinfo {author} {\bibfnamefont {S.}~\bibnamefont {Sano}},
  \bibinfo {author} {\bibfnamefont {T.}~\bibnamefont {Yanase}}, \bibinfo
  {author} {\bibfnamefont {T.}~\bibnamefont {Ohta}},\ and\ \bibinfo {author}
  {\bibfnamefont {M.}~\bibnamefont {Koyama}},\ }\bibfield  {title} {\bibinfo
  {title} {{Optuna}: A next-generation hyperparameter optimization framework},\
  }in\ \href@noop {} {\emph {\bibinfo {booktitle} {Proceedings of the 25th {ACM
  SIGKDD} International Conference on Knowledge Discovery and Data Mining}}}\
  (\bibinfo  {publisher} {{Association for Computing Machinery}},\ \bibinfo
  {year} {2019})\ pp.\ \bibinfo {pages} {2623--2631}\BibitemShut {NoStop}%
\bibitem [{\citenamefont {{Preferred Networks, Inc.}}(2022)}]{Optuna}%
  \BibitemOpen
  \bibfield  {author} {\bibinfo {author} {\bibnamefont {{Preferred Networks,
  Inc.}}},\ }\href {https://optuna.org/} {\bibinfo {title} {{Optuna}}}
  (\bibinfo {year} {2022}),\ \bibinfo {note} {{Accessed:
  2022-12-12}}\BibitemShut {NoStop}%
\bibitem [{\citenamefont {Bergstra}\ \emph {et~al.}(2011)\citenamefont
  {Bergstra}, \citenamefont {Bardenet}, \citenamefont {Bengio},\ and\
  \citenamefont {K\'{e}gl}}]{Bergstra2011}%
  \BibitemOpen
  \bibfield  {author} {\bibinfo {author} {\bibfnamefont {J.}~\bibnamefont
  {Bergstra}}, \bibinfo {author} {\bibfnamefont {R.}~\bibnamefont {Bardenet}},
  \bibinfo {author} {\bibfnamefont {Y.}~\bibnamefont {Bengio}},\ and\ \bibinfo
  {author} {\bibfnamefont {B.}~\bibnamefont {K\'{e}gl}},\ }\bibfield  {title}
  {\bibinfo {title} {Algorithms for hyper-parameter optimization},\ }in\
  \href@noop {} {\emph {\bibinfo {booktitle} {Advances in Neural Information
  Processing Systems}}},\ Vol.~\bibinfo {volume} {24}\ (\bibinfo {year}
  {2011})\BibitemShut {NoStop}%
\bibitem [{\citenamefont {Yang}\ and\ \citenamefont {Shami}(2020)}]{Yang2020}%
  \BibitemOpen
  \bibfield  {author} {\bibinfo {author} {\bibfnamefont {L.}~\bibnamefont
  {Yang}}\ and\ \bibinfo {author} {\bibfnamefont {A.}~\bibnamefont {Shami}},\
  }\bibfield  {title} {\bibinfo {title} {On hyperparameter optimization of
  machine learning algorithms: Theory and practice},\ }\href@noop {} {\bibfield
   {journal} {\bibinfo  {journal} {Neurocomputing}\ }\textbf {\bibinfo {volume}
  {415}},\ \bibinfo {pages} {295} (\bibinfo {year} {2020})}\BibitemShut
  {NoStop}%
\bibitem [{\citenamefont {Ayodele}(2022)}]{Ayodele2022}%
  \BibitemOpen
  \bibfield  {author} {\bibinfo {author} {\bibfnamefont {M.}~\bibnamefont
  {Ayodele}},\ }\href@noop {} {\bibinfo {title} {Comparing the digital annealer
  with classical evolutionary algorithm}} (\bibinfo {year} {2022}),\ \Eprint
  {https://arxiv.org/abs/2205.13586} {arXiv:2205.13586 [cs.NE]} \BibitemShut
  {NoStop}%
\bibitem [{\citenamefont {Goh}\ \emph {et~al.}(2022)\citenamefont {Goh},
  \citenamefont {Bo}, \citenamefont {Gopalakrishnan},\ and\ \citenamefont
  {Lau}}]{Goh2022}%
  \BibitemOpen
  \bibfield  {author} {\bibinfo {author} {\bibfnamefont {S.~T.}\ \bibnamefont
  {Goh}}, \bibinfo {author} {\bibfnamefont {J.}~\bibnamefont {Bo}}, \bibinfo
  {author} {\bibfnamefont {S.}~\bibnamefont {Gopalakrishnan}},\ and\ \bibinfo
  {author} {\bibfnamefont {H.~C.}\ \bibnamefont {Lau}},\ }\bibfield  {title}
  {\bibinfo {title} {Techniques to enhance a {QUBO} solver for
  permutation-based combinatorial optimization},\ }in\ \href@noop {} {\emph
  {\bibinfo {booktitle} {Proceedings of the Genetic and Evolutionary
  Computation Conference Companion}}},\ \bibinfo {series and number} {{GECCO
  '22}}\ (\bibinfo  {publisher} {{Association for Computing Machinery}},\
  \bibinfo {year} {2022})\ pp.\ \bibinfo {pages} {2223--2231}\BibitemShut
  {NoStop}%
\bibitem [{\citenamefont {Lowe}(2017)}]{Lowe2017}%
  \BibitemOpen
  \bibfield  {author} {\bibinfo {author} {\bibfnamefont {D.}~\bibnamefont
  {Lowe}},\ }\href
  {https://figshare.com/articles/dataset/Chemical_reactions_from_US_patents_1976-Sep2016_/5104873}
  {\bibinfo {title} {Chemical reactions from {US} patents (1976-{Sep}2016)}}
  (\bibinfo {year} {2017}),\ \bibinfo {note} {{Accessed:
  2023-4-29}}\BibitemShut {NoStop}%
\bibitem [{\citenamefont {Boothby}\ \emph {et~al.}(2019)\citenamefont
  {Boothby}, \citenamefont {Bunyk}, \citenamefont {Raymond},\ and\
  \citenamefont {Roy}}]{Boothby2019}%
  \BibitemOpen
  \bibfield  {author} {\bibinfo {author} {\bibfnamefont {K.}~\bibnamefont
  {Boothby}}, \bibinfo {author} {\bibfnamefont {P.}~\bibnamefont {Bunyk}},
  \bibinfo {author} {\bibfnamefont {J.}~\bibnamefont {Raymond}},\ and\ \bibinfo
  {author} {\bibfnamefont {A.}~\bibnamefont {Roy}},\ }\href
  {https://www.dwavesys.com/media/jwwj5z3z/14-1026a-c_next-generation-topology-of-dw-quantum-processors.pdf}
  {\emph {\bibinfo {title} {Next-Generation Topology of {D-Wave} Quantum
  Processors}}},\ \bibinfo {type} {Tech. Rep.}\ (\bibinfo  {institution}
  {{D-Wave Systems Inc.}},\ \bibinfo {year} {2019})\BibitemShut {NoStop}%
\bibitem [{\citenamefont {Meurer}\ \emph {et~al.}(2017)\citenamefont {Meurer},
  \citenamefont {Smith}, \citenamefont {Paprocki}, \citenamefont
  {\v{C}ert\'{i}k}, \citenamefont {Kirpichev}, \citenamefont {Rocklin},
  \citenamefont {Kumar}, \citenamefont {Ivanov}, \citenamefont {Moore},
  \citenamefont {Singh}, \citenamefont {Rathnayake}, \citenamefont {Vig},
  \citenamefont {Granger}, \citenamefont {Muller}, \citenamefont {Bonazzi},
  \citenamefont {Gupta}, \citenamefont {Vats}, \citenamefont {Johansson},
  \citenamefont {Pedregosa}, \citenamefont {Curry}, \citenamefont {Terrel},
  \citenamefont {Rou\v{c}ka}, \citenamefont {Saboo}, \citenamefont {Fernando},
  \citenamefont {Kulal}, \citenamefont {Cimrman},\ and\ \citenamefont
  {Scopatz}}]{Meurer2017}%
  \BibitemOpen
  \bibfield  {author} {\bibinfo {author} {\bibfnamefont {A.}~\bibnamefont
  {Meurer}}, \bibinfo {author} {\bibfnamefont {C.~P.}\ \bibnamefont {Smith}},
  \bibinfo {author} {\bibfnamefont {M.}~\bibnamefont {Paprocki}}, \bibinfo
  {author} {\bibfnamefont {O.}~\bibnamefont {\v{C}ert\'{i}k}}, \bibinfo
  {author} {\bibfnamefont {S.~B.}\ \bibnamefont {Kirpichev}}, \bibinfo {author}
  {\bibfnamefont {M.}~\bibnamefont {Rocklin}}, \bibinfo {author} {\bibfnamefont
  {A.}~\bibnamefont {Kumar}}, \bibinfo {author} {\bibfnamefont
  {S.}~\bibnamefont {Ivanov}}, \bibinfo {author} {\bibfnamefont {J.~K.}\
  \bibnamefont {Moore}}, \bibinfo {author} {\bibfnamefont {S.}~\bibnamefont
  {Singh}}, \bibinfo {author} {\bibfnamefont {T.}~\bibnamefont {Rathnayake}},
  \bibinfo {author} {\bibfnamefont {S.}~\bibnamefont {Vig}}, \bibinfo {author}
  {\bibfnamefont {B.~E.}\ \bibnamefont {Granger}}, \bibinfo {author}
  {\bibfnamefont {R.~P.}\ \bibnamefont {Muller}}, \bibinfo {author}
  {\bibfnamefont {F.}~\bibnamefont {Bonazzi}}, \bibinfo {author} {\bibfnamefont
  {H.}~\bibnamefont {Gupta}}, \bibinfo {author} {\bibfnamefont
  {S.}~\bibnamefont {Vats}}, \bibinfo {author} {\bibfnamefont {F.}~\bibnamefont
  {Johansson}}, \bibinfo {author} {\bibfnamefont {F.}~\bibnamefont
  {Pedregosa}}, \bibinfo {author} {\bibfnamefont {M.~J.}\ \bibnamefont
  {Curry}}, \bibinfo {author} {\bibfnamefont {A.~R.}\ \bibnamefont {Terrel}},
  \bibinfo {author} {\bibfnamefont {v.}~\bibnamefont {Rou\v{c}ka}}, \bibinfo
  {author} {\bibfnamefont {A.}~\bibnamefont {Saboo}}, \bibinfo {author}
  {\bibfnamefont {I.}~\bibnamefont {Fernando}}, \bibinfo {author}
  {\bibfnamefont {S.}~\bibnamefont {Kulal}}, \bibinfo {author} {\bibfnamefont
  {R.}~\bibnamefont {Cimrman}},\ and\ \bibinfo {author} {\bibfnamefont
  {A.}~\bibnamefont {Scopatz}},\ }\bibfield  {title} {\bibinfo {title}
  {{SymPy}: symbolic computing in {Python}},\ }\href@noop {} {\bibfield
  {journal} {\bibinfo  {journal} {PeerJ Computer Science}\ }\textbf {\bibinfo
  {volume} {3}},\ \bibinfo {pages} {e103} (\bibinfo {year} {2017})}\BibitemShut
  {NoStop}%
\bibitem [{\citenamefont {{Gurobi Optimization, LLC.}}(2022)}]{Gurobi}%
  \BibitemOpen
  \bibfield  {author} {\bibinfo {author} {\bibnamefont {{Gurobi Optimization,
  LLC.}}},\ }\href {https://www.gurobi.com} {\bibinfo {title} {{Gurobi
  Optimizer}}} (\bibinfo {year} {2022}),\ \bibinfo {note} {{Accessed:
  2022-12-12}}\BibitemShut {NoStop}%
\bibitem [{\citenamefont {Tamura}\ \emph {et~al.}(2021)\citenamefont {Tamura},
  \citenamefont {Shirai}, \citenamefont {Katsura}, \citenamefont {Tanaka},\
  and\ \citenamefont {Togawa}}]{Tamura2021}%
  \BibitemOpen
  \bibfield  {author} {\bibinfo {author} {\bibfnamefont {K.}~\bibnamefont
  {Tamura}}, \bibinfo {author} {\bibfnamefont {T.}~\bibnamefont {Shirai}},
  \bibinfo {author} {\bibfnamefont {H.}~\bibnamefont {Katsura}}, \bibinfo
  {author} {\bibfnamefont {S.}~\bibnamefont {Tanaka}},\ and\ \bibinfo {author}
  {\bibfnamefont {N.}~\bibnamefont {Togawa}},\ }\bibfield  {title} {\bibinfo
  {title} {Performance comparison of typical binary-integer encodings in an
  {Ising} machine},\ }\href@noop {} {\bibfield  {journal} {\bibinfo  {journal}
  {{IEEE} Access}\ }\textbf {\bibinfo {volume} {9}},\ \bibinfo {pages} {81032}
  (\bibinfo {year} {2021})}\BibitemShut {NoStop}%
\bibitem [{\citenamefont {Herr}\ \emph {et~al.}(2017)\citenamefont {Herr},
  \citenamefont {Brown}, \citenamefont {Heim}, \citenamefont {K{\"o}nz},
  \citenamefont {Mazzola},\ and\ \citenamefont {Troyer}}]{Herr2017}%
  \BibitemOpen
  \bibfield  {author} {\bibinfo {author} {\bibfnamefont {D.}~\bibnamefont
  {Herr}}, \bibinfo {author} {\bibfnamefont {E.}~\bibnamefont {Brown}},
  \bibinfo {author} {\bibfnamefont {B.}~\bibnamefont {Heim}}, \bibinfo {author}
  {\bibfnamefont {M.}~\bibnamefont {K{\"o}nz}}, \bibinfo {author}
  {\bibfnamefont {G.}~\bibnamefont {Mazzola}},\ and\ \bibinfo {author}
  {\bibfnamefont {M.}~\bibnamefont {Troyer}},\ }\href@noop {} {\bibinfo {title}
  {Optimizing schedules for quantum annealing}} (\bibinfo {year} {2017}),\
  \Eprint {https://arxiv.org/abs/1705.00420} {arXiv:1705.00420 [quant-ph]}
  \BibitemShut {NoStop}%
\bibitem [{\citenamefont {Chen}\ \emph {et~al.}(2022)\citenamefont {Chen},
  \citenamefont {Chen}, \citenamefont {Lee}, \citenamefont {Zhang},\ and\
  \citenamefont {Hsieh}}]{Chen2022}%
  \BibitemOpen
  \bibfield  {author} {\bibinfo {author} {\bibfnamefont {Y.-Q.}\ \bibnamefont
  {Chen}}, \bibinfo {author} {\bibfnamefont {Y.}~\bibnamefont {Chen}}, \bibinfo
  {author} {\bibfnamefont {C.-K.}\ \bibnamefont {Lee}}, \bibinfo {author}
  {\bibfnamefont {S.}~\bibnamefont {Zhang}},\ and\ \bibinfo {author}
  {\bibfnamefont {C.-Y.}\ \bibnamefont {Hsieh}},\ }\bibfield  {title} {\bibinfo
  {title} {Optimizing quantum annealing schedules with {Monte Carlo} tree
  search enhanced with neural networks},\ }\href@noop {} {\bibfield  {journal}
  {\bibinfo  {journal} {Nature Machine Intelligence}\ }\textbf {\bibinfo
  {volume} {4}},\ \bibinfo {pages} {269} (\bibinfo {year} {2022})}\BibitemShut
  {NoStop}%
\bibitem [{\citenamefont {Ohzeki}(2020)}]{Ohzeki2020}%
  \BibitemOpen
  \bibfield  {author} {\bibinfo {author} {\bibfnamefont {M.}~\bibnamefont
  {Ohzeki}},\ }\bibfield  {title} {\bibinfo {title} {Breaking limitation of
  quantum annealer in solving optimization problems under constraints},\
  }\href@noop {} {\bibfield  {journal} {\bibinfo  {journal} {Scientific
  reports}\ }\textbf {\bibinfo {volume} {10}},\ \bibinfo {pages} {3126}
  (\bibinfo {year} {2020})}\BibitemShut {NoStop}%
\bibitem [{\citenamefont {Hen}\ and\ \citenamefont
  {Spedalieri}(2016)}]{Hen2016}%
  \BibitemOpen
  \bibfield  {author} {\bibinfo {author} {\bibfnamefont {I.}~\bibnamefont
  {Hen}}\ and\ \bibinfo {author} {\bibfnamefont {F.~M.}\ \bibnamefont
  {Spedalieri}},\ }\bibfield  {title} {\bibinfo {title} {Quantum annealing for
  constrained optimization},\ }\href@noop {} {\bibfield  {journal} {\bibinfo
  {journal} {Physical Review Applied}\ }\textbf {\bibinfo {volume} {5}},\
  \bibinfo {pages} {034007} (\bibinfo {year} {2016})}\BibitemShut {NoStop}%
\bibitem [{\citenamefont {Hadfield}\ \emph {et~al.}(2019)\citenamefont
  {Hadfield}, \citenamefont {Wang}, \citenamefont {O'gorman}, \citenamefont
  {Rieffel}, \citenamefont {Venturelli},\ and\ \citenamefont
  {Biswas}}]{Hadfield2019}%
  \BibitemOpen
  \bibfield  {author} {\bibinfo {author} {\bibfnamefont {S.}~\bibnamefont
  {Hadfield}}, \bibinfo {author} {\bibfnamefont {Z.}~\bibnamefont {Wang}},
  \bibinfo {author} {\bibfnamefont {B.}~\bibnamefont {O'gorman}}, \bibinfo
  {author} {\bibfnamefont {E.~G.}\ \bibnamefont {Rieffel}}, \bibinfo {author}
  {\bibfnamefont {D.}~\bibnamefont {Venturelli}},\ and\ \bibinfo {author}
  {\bibfnamefont {R.}~\bibnamefont {Biswas}},\ }\bibfield  {title} {\bibinfo
  {title} {From the quantum approximate optimization algorithm to a quantum
  alternating operator ansatz},\ }\href@noop {} {\bibfield  {journal} {\bibinfo
   {journal} {Algorithms}\ }\textbf {\bibinfo {volume} {12}} (\bibinfo {year}
  {2019})}\BibitemShut {NoStop}%
\bibitem [{\citenamefont {Leipold}\ and\ \citenamefont
  {Spedalieri}(2021)}]{Leipold2021}%
  \BibitemOpen
  \bibfield  {author} {\bibinfo {author} {\bibfnamefont {H.}~\bibnamefont
  {Leipold}}\ and\ \bibinfo {author} {\bibfnamefont {F.~M.}\ \bibnamefont
  {Spedalieri}},\ }\bibfield  {title} {\bibinfo {title} {Constructing driver
  {Hamiltonians} for optimization problems with linear constraints},\
  }\href@noop {} {\bibfield  {journal} {\bibinfo  {journal} {Quantum Science
  and Technology}\ }\textbf {\bibinfo {volume} {7}},\ \bibinfo {pages} {015013}
  (\bibinfo {year} {2021})}\BibitemShut {NoStop}%
\bibitem [{\citenamefont {Zhao}\ \emph {et~al.}(2022)\citenamefont {Zhao},
  \citenamefont {Fan},\ and\ \citenamefont {Han}}]{Zhao2022}%
  \BibitemOpen
  \bibfield  {author} {\bibinfo {author} {\bibfnamefont {Z.}~\bibnamefont
  {Zhao}}, \bibinfo {author} {\bibfnamefont {L.}~\bibnamefont {Fan}},\ and\
  \bibinfo {author} {\bibfnamefont {Z.}~\bibnamefont {Han}},\ }\bibfield
  {title} {\bibinfo {title} {Hybrid quantum {Benders'} decomposition for
  mixed-integer linear programming},\ }in\ \href@noop {} {\emph {\bibinfo
  {booktitle} {2022 {IEEE} Wireless Communications and Networking Conference
  ({WCNC})}}}\ (\bibinfo {organization} {{IEEE}},\ \bibinfo {year} {2022})\
  pp.\ \bibinfo {pages} {2536--2540}\BibitemShut {NoStop}%
\bibitem [{\citenamefont {Ajagekar}\ \emph {et~al.}(2022)\citenamefont
  {Ajagekar}, \citenamefont {Hamoud},\ and\ \citenamefont
  {You}}]{Ajagekar2022}%
  \BibitemOpen
  \bibfield  {author} {\bibinfo {author} {\bibfnamefont {A.}~\bibnamefont
  {Ajagekar}}, \bibinfo {author} {\bibfnamefont {K.~A.}\ \bibnamefont
  {Hamoud}},\ and\ \bibinfo {author} {\bibfnamefont {F.}~\bibnamefont {You}},\
  }\bibfield  {title} {\bibinfo {title} {Hybrid classical-quantum optimization
  techniques for solving mixed-integer programming problems in production
  scheduling},\ }\href@noop {} {\bibfield  {journal} {\bibinfo  {journal}
  {{IEEE} Transactions on Quantum Engineering}\ }\textbf {\bibinfo {volume}
  {3}},\ \bibinfo {pages} {1} (\bibinfo {year} {2022})}\BibitemShut {NoStop}%
\bibitem [{\citenamefont {Bar{\'a}n}\ and\ \citenamefont
  {Villagra}(2016)}]{Baran2016}%
  \BibitemOpen
  \bibfield  {author} {\bibinfo {author} {\bibfnamefont {B.}~\bibnamefont
  {Bar{\'a}n}}\ and\ \bibinfo {author} {\bibfnamefont {M.}~\bibnamefont
  {Villagra}},\ }\bibfield  {title} {\bibinfo {title} {Multiobjective
  optimization in a quantum adiabatic computer},\ }\href@noop {} {\bibfield
  {journal} {\bibinfo  {journal} {Electronic Notes in Theoretical Computer
  Science}\ }\textbf {\bibinfo {volume} {329}},\ \bibinfo {pages} {27}
  (\bibinfo {year} {2016})}\BibitemShut {NoStop}%
\bibitem [{\citenamefont {Kumar}\ \emph {et~al.}(2020)\citenamefont {Kumar},
  \citenamefont {Tomlin}, \citenamefont {Nehrkorn}, \citenamefont {O'Malley},\
  and\ \citenamefont {Dulny~III}}]{Kumar2020}%
  \BibitemOpen
  \bibfield  {author} {\bibinfo {author} {\bibfnamefont {V.}~\bibnamefont
  {Kumar}}, \bibinfo {author} {\bibfnamefont {C.}~\bibnamefont {Tomlin}},
  \bibinfo {author} {\bibfnamefont {C.}~\bibnamefont {Nehrkorn}}, \bibinfo
  {author} {\bibfnamefont {D.}~\bibnamefont {O'Malley}},\ and\ \bibinfo
  {author} {\bibfnamefont {J.}~\bibnamefont {Dulny~III}},\ }\href@noop {}
  {\bibinfo {title} {Achieving fair sampling in quantum annealing}} (\bibinfo
  {year} {2020}),\ \Eprint {https://arxiv.org/abs/2007.08487} {arXiv:2007.08487
  [quant-ph]} \BibitemShut {NoStop}%
\bibitem [{\citenamefont {Mizuno}\ and\ \citenamefont
  {Komatsuzaki}(2021)}]{Mizuno2021}%
  \BibitemOpen
  \bibfield  {author} {\bibinfo {author} {\bibfnamefont {Y.}~\bibnamefont
  {Mizuno}}\ and\ \bibinfo {author} {\bibfnamefont {T.}~\bibnamefont
  {Komatsuzaki}},\ }\href@noop {} {\bibinfo {title} {A note on enumeration by
  fair sampling}} (\bibinfo {year} {2021}),\ \Eprint
  {https://arxiv.org/abs/2104.01941} {arXiv:2104.01941 [quant-ph]} \BibitemShut
  {NoStop}%
\bibitem [{\citenamefont {Kirkpatrick}\ \emph {et~al.}(1983)\citenamefont
  {Kirkpatrick}, \citenamefont {Gelatt},\ and\ \citenamefont
  {Vecchi}}]{Kirkpatrick1983}%
  \BibitemOpen
  \bibfield  {author} {\bibinfo {author} {\bibfnamefont {S.}~\bibnamefont
  {Kirkpatrick}}, \bibinfo {author} {\bibfnamefont {C.~D.}\ \bibnamefont
  {Gelatt}},\ and\ \bibinfo {author} {\bibfnamefont {M.~P.}\ \bibnamefont
  {Vecchi}},\ }\bibfield  {title} {\bibinfo {title} {Optimization by simulated
  annealing},\ }\href@noop {} {\bibfield  {journal} {\bibinfo  {journal}
  {Science}\ }\textbf {\bibinfo {volume} {220}},\ \bibinfo {pages} {671}
  (\bibinfo {year} {1983})}\BibitemShut {NoStop}%
\bibitem [{\citenamefont {Kadowaki}\ and\ \citenamefont
  {Nishimori}(1998)}]{Kadowaki1998}%
  \BibitemOpen
  \bibfield  {author} {\bibinfo {author} {\bibfnamefont {T.}~\bibnamefont
  {Kadowaki}}\ and\ \bibinfo {author} {\bibfnamefont {H.}~\bibnamefont
  {Nishimori}},\ }\bibfield  {title} {\bibinfo {title} {Quantum annealing in
  the transverse {Ising} model},\ }\href@noop {} {\bibfield  {journal}
  {\bibinfo  {journal} {Physical Review E}\ }\textbf {\bibinfo {volume} {58}},\
  \bibinfo {pages} {5355} (\bibinfo {year} {1998})}\BibitemShut {NoStop}%
\bibitem [{\citenamefont {Geman}\ and\ \citenamefont
  {Geman}(1984)}]{Geman1984}%
  \BibitemOpen
  \bibfield  {author} {\bibinfo {author} {\bibfnamefont {S.}~\bibnamefont
  {Geman}}\ and\ \bibinfo {author} {\bibfnamefont {D.}~\bibnamefont {Geman}},\
  }\bibfield  {title} {\bibinfo {title} {Stochastic relaxation, {Gibbs}
  distributions, and the {Bayesian} restoration of images},\ }\href@noop {}
  {\bibfield  {journal} {\bibinfo  {journal} {{IEEE} Transactions on Pattern
  Analysis and Machine Intelligence}\ }\textbf {\bibinfo {volume} {PAMI-6}},\
  \bibinfo {pages} {721} (\bibinfo {year} {1984})}\BibitemShut {NoStop}%
\bibitem [{\citenamefont {Mitra}\ \emph {et~al.}(1986)\citenamefont {Mitra},
  \citenamefont {Romeo},\ and\ \citenamefont
  {Sangiovanni-Vincentelli}}]{Mitra1986}%
  \BibitemOpen
  \bibfield  {author} {\bibinfo {author} {\bibfnamefont {D.}~\bibnamefont
  {Mitra}}, \bibinfo {author} {\bibfnamefont {F.}~\bibnamefont {Romeo}},\ and\
  \bibinfo {author} {\bibfnamefont {A.}~\bibnamefont
  {Sangiovanni-Vincentelli}},\ }\bibfield  {title} {\bibinfo {title}
  {Convergence and finite-time behavior of simulated annealing},\ }\href@noop
  {} {\bibfield  {journal} {\bibinfo  {journal} {Advances in applied
  probability}\ }\textbf {\bibinfo {volume} {18}},\ \bibinfo {pages} {747}
  (\bibinfo {year} {1986})}\BibitemShut {NoStop}%
\bibitem [{\citenamefont {Morita}\ and\ \citenamefont
  {Nishimori}(2007)}]{Morita2007}%
  \BibitemOpen
  \bibfield  {author} {\bibinfo {author} {\bibfnamefont {S.}~\bibnamefont
  {Morita}}\ and\ \bibinfo {author} {\bibfnamefont {H.}~\bibnamefont
  {Nishimori}},\ }\bibfield  {title} {\bibinfo {title} {Convergence of quantum
  annealing with real-time {Schr{\"o}dinger} dynamics},\ }\href@noop {}
  {\bibfield  {journal} {\bibinfo  {journal} {Journal of the Physical Society
  of {Japan}}\ }\textbf {\bibinfo {volume} {76}},\ \bibinfo {pages} {064002}
  (\bibinfo {year} {2007})}\BibitemShut {NoStop}%
\bibitem [{\citenamefont {Morita}\ and\ \citenamefont
  {Nishimori}(2008)}]{Morita2008}%
  \BibitemOpen
  \bibfield  {author} {\bibinfo {author} {\bibfnamefont {S.}~\bibnamefont
  {Morita}}\ and\ \bibinfo {author} {\bibfnamefont {H.}~\bibnamefont
  {Nishimori}},\ }\bibfield  {title} {\bibinfo {title} {Mathematical foundation
  of quantum annealing},\ }\href@noop {} {\bibfield  {journal} {\bibinfo
  {journal} {Journal of Mathematical Physics}\ }\textbf {\bibinfo {volume}
  {49}} (\bibinfo {year} {2008})}\BibitemShut {NoStop}%
\bibitem [{\citenamefont {Cai}\ \emph {et~al.}(2014)\citenamefont {Cai},
  \citenamefont {Macready},\ and\ \citenamefont {Roy}}]{Cai2014}%
  \BibitemOpen
  \bibfield  {author} {\bibinfo {author} {\bibfnamefont {J.}~\bibnamefont
  {Cai}}, \bibinfo {author} {\bibfnamefont {W.~G.}\ \bibnamefont {Macready}},\
  and\ \bibinfo {author} {\bibfnamefont {A.}~\bibnamefont {Roy}},\ }\href@noop
  {} {\bibinfo {title} {A practical heuristic for finding graph minors}}
  (\bibinfo {year} {2014}),\ \Eprint {https://arxiv.org/abs/1406.2741}
  {arXiv:1406.2741 [quant-ph]} \BibitemShut {NoStop}%
\bibitem [{\citenamefont {Sugie}\ \emph {et~al.}(2021)\citenamefont {Sugie},
  \citenamefont {Yoshida}, \citenamefont {Mertig}, \citenamefont {Takemoto},
  \citenamefont {Teramoto}, \citenamefont {Nakamura}, \citenamefont {Takigawa},
  \citenamefont {Minato}, \citenamefont {Yamaoka},\ and\ \citenamefont
  {Komatsuzaki}}]{Sugie2021}%
  \BibitemOpen
  \bibfield  {author} {\bibinfo {author} {\bibfnamefont {Y.}~\bibnamefont
  {Sugie}}, \bibinfo {author} {\bibfnamefont {Y.}~\bibnamefont {Yoshida}},
  \bibinfo {author} {\bibfnamefont {N.}~\bibnamefont {Mertig}}, \bibinfo
  {author} {\bibfnamefont {T.}~\bibnamefont {Takemoto}}, \bibinfo {author}
  {\bibfnamefont {H.}~\bibnamefont {Teramoto}}, \bibinfo {author}
  {\bibfnamefont {A.}~\bibnamefont {Nakamura}}, \bibinfo {author}
  {\bibfnamefont {I.}~\bibnamefont {Takigawa}}, \bibinfo {author}
  {\bibfnamefont {S.}~\bibnamefont {Minato}}, \bibinfo {author} {\bibfnamefont
  {M.}~\bibnamefont {Yamaoka}},\ and\ \bibinfo {author} {\bibfnamefont
  {T.}~\bibnamefont {Komatsuzaki}},\ }\bibfield  {title} {\bibinfo {title}
  {Minor-embedding heuristics for large-scale annealing processors with sparse
  hardware graphs of up to 102,400 nodes},\ }\href@noop {} {\bibfield
  {journal} {\bibinfo  {journal} {Soft Computing}\ }\textbf {\bibinfo {volume}
  {25}},\ \bibinfo {pages} {1731} (\bibinfo {year} {2021})}\BibitemShut
  {NoStop}%
\bibitem [{\citenamefont {Gasteiger}\ and\ \citenamefont
  {Engel}(2003)}]{Gasteiger2003}%
  \BibitemOpen
  \bibinfo {editor} {\bibfnamefont {J.}~\bibnamefont {Gasteiger}}\ and\
  \bibinfo {editor} {\bibfnamefont {T.}~\bibnamefont {Engel}},\ eds.,\
  \href@noop {} {\emph {\bibinfo {title} {Chemoinfomatics: A Textbook}}}\
  (\bibinfo  {publisher} {{Wiley-VCH}},\ \bibinfo {address} {Weinheim,
  Germany},\ \bibinfo {year} {2003})\BibitemShut {NoStop}%
\bibitem [{\citenamefont {Rogers}\ and\ \citenamefont
  {Hahn}(2010)}]{Rogers2010}%
  \BibitemOpen
  \bibfield  {author} {\bibinfo {author} {\bibfnamefont {D.}~\bibnamefont
  {Rogers}}\ and\ \bibinfo {author} {\bibfnamefont {M.}~\bibnamefont {Hahn}},\
  }\bibfield  {title} {\bibinfo {title} {Extended-connectivity fingerprints},\
  }\href@noop {} {\bibfield  {journal} {\bibinfo  {journal} {Journal of
  Chemical Information and Modeling}\ }\textbf {\bibinfo {volume} {50}},\
  \bibinfo {pages} {742} (\bibinfo {year} {2010})}\BibitemShut {NoStop}%
\bibitem [{\citenamefont {Landrum}(2023)}]{RDKit-Morgan}%
  \BibitemOpen
  \bibfield  {author} {\bibinfo {author} {\bibfnamefont {G.}~\bibnamefont
  {Landrum}},\ }\href
  {https://www.rdkit.org/docs/GettingStartedInPython.html#morgan-fingerprints}
  {\bibinfo {title} {{Morgan Fingerprints (Circular Fingerprints)}}} (\bibinfo
  {year} {2023}),\ \bibinfo {note} {{Accessed: 2023-7-23}}\BibitemShut
  {NoStop}%
\end{thebibliography}%

\end{document}